\documentclass[nocopyrightspace]{sigplanconf}
\usepackage{color}
\usepackage{alltt}
\usepackage{mathpartir}
\usepackage{hyperref}
\usepackage{amsmath, amssymb, amsthm}
\usepackage{stmaryrd}
\usepackage{xspace}
\usepackage{graphicx}

\usepackage{listings}

\newcommand{\blankrow}[1]{\multicolumn{#1}{c}{}}

\def\parahead#1{\paragraph{#1.}}

\newcommand{\set}[1]{\{{#1}\}}
\newcommand{\seq}[1]{\overline{#1}}

\newcommand{\ie}{{\emph{i.e.}\ }}
\newcommand{\eg}{{\emph{e.g.}\ }}
\newcommand{\etc}{{\emph{etc.}}}

\newcommand{\langBase}{\ensuremath{\textrm{D}}}
\newcommand{\dtypes}{\text{System}\ \langBase\xspace}

\newcommand{\dntypes}{\text{System}\ \langBase$^{*}$\xspace}

\newcommand{\funAnn}[3]{\ensuremath{\lambda{{#1}\!:\!{#2}}.\ {#3}}}
\newcommand{\funBare}[2]{\ensuremath{\lambda{#1}.\ {#2}}}
\newcommand{\typFun}[2]{\ensuremath{\lambda{#1}.\ {#2}}}
\newcommand{\letinAnn}[4]{\ensuremath{\text{let }\tyBind{#1}{#2} = {#3}\text{ in }{#4}}}
\newcommand{\letinBare}[3]{\ensuremath{\text{let }{#1} = {#2}\text{ in }{#3}}}
\newcommand{\ite}[3]{\ensuremath{\text{if }{#1}\text{ then }{#2}\text{ else }{#3}}}

\newcommand{\dict}[1]{\ensuremath{\set{#1}}}
\newcommand{\emptydict}{\dict{}}
\newcommand{\emptydictR}{\mathit{empty}}

\newcommand{\singledict}[2]{\dict{{#1}\mapsto{#2}}}
\newcommand{\dictextend}[3]{#1\ \text{++}\ \ensuremath{\singledict{#2}{#3}}}

\newcommand{\nullCon}{{\ensuremath{\mathit{Null}}}}
\newcommand{\listCon}{\ensuremath{\mathit{List}}}

\newcommand{\typeCon}[2]{\ensuremath{{#1}[{#2}]}}
\newcommand{\tyList}[1]{\ensuremath{\typeCon{\listCon}{#1}}}

\newcommand{\lookupConName}{\Psi}

\newcommand{\typeConDef}[4]
  {\ensuremath{\text{type}\ {#1}[\seq{#2}]{\{\seq{\tyBind{#3}{#4}}\}}}}
\newcommand{\lookupCon}[4]
  {\ensuremath{\lookupConName(#1)=[#2]{\{\seq{\tyBind{#3}{#4}}\}}}}
\newcommand{\lookupConAnnots}[2]
  {\ensuremath{\lookupConName(#1)=[#2]{\{\ \cdots\ \}}}}
\newcommand{\typeConDefOneTwo}[6]
  {\ensuremath{\text{type}\ {#1}[#2]{\{\tyBind{#3}{#4};
                                       \tyBind{#5}{#6}}\}}}

\newcommand{\varPole}{\ensuremath{\theta}}
\newcommand{\poleCo}{\ensuremath{\texttt{+}}}
\newcommand{\poleContra}{\ensuremath{\texttt{-}}}
\newcommand{\poleBi}{\ensuremath{\texttt{=}}}
\newcommand{\polesName}{\ensuremath{\mathsf{Poles}}}
\newcommand{\poles}[3]{\ensuremath{\polesName({#1},{#2},{#3})}}
\newcommand{\varianceOkName}{\ensuremath{\mathsf{VarianceOk}}}
\newcommand{\varianceOk}[3]{\ensuremath{\varianceOkName({#1},{#2},{#3})}}
\newcommand{\newDataBare}[2]{\ensuremath{{#1}({#2})}}
\newcommand{\newDataAnn}[3]{\ensuremath{{#1}[{#2}]({#3})}}
\newcommand{\inferenceMarker}{\ensuremath{*}}
\newcommand{\unfoldName}{\ensuremath{\mathsf{Unfold}}}
\newcommand{\foldName}{\ensuremath{\mathsf{Fold}}}
\newcommand{\unfold}[2]{\ensuremath{\unfoldName({#1},{#2})}}
\newcommand{\fold}[3]{\ensuremath{\foldName({#1},{#2},{#3})}}

\newcommand{\reduce}{\ensuremath{\hookrightarrow}}
\newcommand{\reducesTo}[2]{{#1}\reduce{#2}}
\newcommand{\reducesToMulti}[2]{{#1}\reduce^{*}{#2}}
\newcommand{\subst}[3]{{#1}[{#3}/{#2}]}

\newcommand{\varVal}{w}
\newcommand{\varLogVal}{\mathit{lw}}
\newcommand{\theV}{\nu}
\newcommand{\varTyp}{T}
\newcommand{\varScm}{S}
\newcommand{\varUnTyp}{U}
\newcommand{\varCon}{C}
\newcommand{\tyvar}{A}
\newcommand{\tyvarB}{B}
\newcommand{\tyVar}{\tyvar}
\newcommand{\tyVarB}{\tyvarB}

\newcommand{\lprimF}{F}
\newcommand{\lprimP}{P}
\newcommand{\varFormOne}{p}
\newcommand{\varFormTwo}{q}
\newcommand{\varFormThr}{r}

\newcommand{\usedBoxes}{\mathcal{U}}
\newcommand{\varTypeDef}{\mathit{td}}

\newcommand{\defeq}{\circeq}
\newcommand{\sepmidsep}{\ensuremath{\hspace{0.02in} | \hspace{0.02in}}}
\newcommand{\refTyp}[1]{\ensuremath{\{ \theV \sepmidsep #1\}}}
\newcommand{\refTypShort}[1]{\ensuremath{\{#1\}}}

\newcommand{\typAll}[2]{\ensuremath{\forall {#1}.\ {#2}}}

\newcommand{\typInst}[2]{\ensuremath{{#1}\ [{#2}]}}
\newcommand{\instName}{\ensuremath{\mathsf{Inst}}}
\newcommand{\inst}[3]{\ensuremath{\mathsf{\instName(}{#1},{#2},{#3}\mathsf{)}}}

\newcommand{\vTrue}{\ensuremath{\texttt{true}}}
\newcommand{\vFalse}{\ensuremath{\texttt{false}}}

\newcommand{\vNull}{\ensuremath{\texttt{null}}}
\newcommand{\vGet}{\texttt{get}}
\newcommand{\vHas}{\texttt{has}}
\newcommand{\vKeys}{\texttt{keys}}
\newcommand{\vUpd}{\texttt{set}}
\newcommand{\vTag}{\texttt{tag}}

\newcommand{\vFix}{\texttt{fix}}

\newcommand{\ttfld}[1]{\ensuremath{\mathtt{``{#1}"}}}
\newcommand{\str}[1]{\ensuremath{\ttfld{#1}}}

\newcommand{\tagInt}{\ttfld{Int}}
\newcommand{\tagStr}{\ttfld{Str}}
\newcommand{\tagBool}{\ttfld{Bool}}
\newcommand{\tagFun}{\ttfld{Fun}}
\newcommand{\tagTFun}{\ttfld{TFun}}
\newcommand{\tagRecd}{\ttfld{Dict}}

\newcommand{\tyTop}{\ensuremath{\mathit{{Top}}}}
\newcommand{\tyInt}{\ensuremath{\mathit{{Int}}}}
\newcommand{\tyIntOrBool}{\ensuremath{\mathit{{IorB}}}}

\newcommand{\tyStr}{\ensuremath{\mathit{Str}}}
\newcommand{\tyBool}{\ensuremath{\mathit{Bool}}}
\newcommand{\tyRecd}{\ensuremath{\mathit{Dict}}}

\newcommand{\tyFld}[3]{\ensuremath{{Fld}(#1, #2, #3)}}

\newcommand{\tySel}[3]{\ensuremath{{Sel}(#1, #2, #3)}}

\newcommand{\tyMApplyF}{\refTyp{\theV = \vNull\ \lor\ 
              \synHastyp{\theV}{\tyInt \ttArr \tyInt}}}

\newcommand{\tyDepNegate}{\tyBind{x}{\tyIntOrBool}\ \ttArr\ {\refTyp{\tagof{\theV} = \tagof{x}}}}

\newcommand{\tyDec}[2]{\ensuremath{{\texttt{#1}}::{#2}}}

\newcommand{\tagofOp}{\ensuremath{\mathit{tag}}}
\newcommand{\tagof}[1]{\tagofOp({#1})}

\newcommand{\formTrue}{\ensuremath{\mathit{true}}}
\newcommand{\formFalse}{\ensuremath{\mathit{false}}}

\newcommand{\logop}[1]{\ensuremath{\mathit{#1}}}
\newcommand{\selR}[2]{\ensuremath{\logop{sel}({#1},{#2})}}
\newcommand{\updR}[3]{\ensuremath{\logop{upd}({#1},{#2},{#3})}}
\newcommand{\hasR}[2]{\ensuremath{\logop{has}({#1}, {#2})}}
\newcommand{\havocR}[3]{\ensuremath{\logop{EqMod}({#1},{#2},{#3})}}

\newcommand{\domR}[1]{\ensuremath{\logop{dom}({#1})}}
\newcommand{\projR}[2]{\ensuremath{\logop{restrict}({#1},{#2})}}

\newcommand{\ttArr}{\ensuremath{{\rightarrow}}}
\newcommand{\tyBind}[2]{\ensuremath{{{#1}\!:\!{#2}}}}

\newcommand{\utArrow}[3]{\ensuremath{\tyBind{#1}{#2}\ \ttArr\ #3}}
\newcommand{\utArrowPlain}[2]{\ensuremath{#1\ \ttArr\ #2}}

\newcommand{\ruleNameFig}[1]{\begin{scriptsize}[#1]\end{scriptsize}}
\newcommand{\ruleName}[1]{\textsc{\begin{normalsize}#1\end{normalsize}}}
\newcommand{\sep}{\hspace{0.06in}}
\newcommand{\sepPremise}{\hspace{0.20in}}
\newcommand{\hsepRule}{\hspace{0.20in}}
\newcommand{\vsepRule}{\vspace{0.12in}}

\newcommand{\miniSepThree}{\hspace{0.03in}}
\newcommand{\setComp}[2]
           {\ensuremath{\{\miniSepThree #1 \hspace{0.02in}
                            \mid\hspace{0.02in} #2 \miniSepThree\}}}
\newcommand{\ttdyn}{\texttt{dynamic}\xspace}

\newcommand{\cnfName}{\ensuremath{\mathsf{Normalize}}}
\newcommand{\cnf}[1]{\ensuremath{\mathsf{\cnfName(}#1\mathsf{)}}}
\newcommand{\boxesName}{\ensuremath{\mathsf{TypeTerms}}}
\newcommand{\boxes}[1]{\ensuremath{\mathsf{\boxesName(}#1\mathsf{)}}}

\newcommand{\typConst}[1]{\ensuremath{ty(#1)}}
\newcommand{\deltaApp}[2]{\ensuremath{\delta({#1},{#2})}}

\newcommand{\valid}[1]{\ensuremath{\mathsf{Valid(}#1\mathsf{)}}}
\newcommand{\embed}[1]{\ensuremath{{\llbracket #1 \rrbracket}}}

\newcommand{\extendGammaName}{\ensuremath{\mathsf{Extend}}}
\newcommand{\extendGamma}[3]
  {\ensuremath{\mathsf{\extendGammaName(}{#1},{#2},{#3}\mathsf{)}}}
\newcommand{\typePredName}{\ensuremath{::}}
\newcommand{\synHastyp}[2]{\ensuremath{{#1}\typePredName{#2}}}
\newcommand{\eraseName}{\ensuremath{\mathsf{erase}}}
\newcommand{\erase}[1]{\ensuremath{\mathsf{\eraseName(}{#1}\mathsf{)}}}
\newcommand{\extractName}{\ensuremath{\mathsf{MustFlow}}}
\newcommand{\extract}[3]
  {\ensuremath{\mathsf{\extractName(}{#1},{#2},{#3}\mathsf{)}}}
\newcommand{\extractNaive}[2]
  {\ensuremath{\mathsf{\extractName(}{#1},{#2}\mathsf{)}}}

\newcommand{\filterAppOneName}{\ensuremath{\mathsf{FilterByArgTyp}}}
\newcommand{\filterAppOne}[3]{\ensuremath{\mathsf{\filterAppOneName(}{#1},{#2},{#3}\mathsf{)}}}
\newcommand{\filterAppTwoName}{\ensuremath{\mathsf{FilterByArgVal}}}
\newcommand{\filterAppTwo}[3]{\ensuremath{\mathsf{\filterAppTwoName(}{#1},{#2},{#3}\mathsf{)}}}
\newcommand{\elimName}{\ensuremath{\mathsf{Elim}}}
\newcommand{\elim}[4]{\ensuremath{\mathsf{\elimName(}{#2},{#3},{#4}\mathsf{)}}}

\newcommand{\intOf}[1]{\ensuremath{\texttt{intOf}}}

\newcommand{\symSub}{\sqsubseteq}
\newcommand{\symSynSub}{<:}

\newcommand{\clauseSym}{\ensuremath{\Rrightarrow}}
\newcommand{\clause}[2]{\ensuremath{#1\clauseSym#2}}
\newcommand{\relImpl}[4]{\ensuremath{#1\vdash_{#4}\clause{#2}{#3}}}
\newcommand{\relAImpl}[4]{\ensuremath{#1;\ {#4} \vdash \clause{#2}{#3}}}

\newcommand{\interp}[1]{\ensuremath{\mathcal{I}_{#1}}}
\newcommand{\formulaSat}[2]{\interp{#1}\models{#2}}

\newcommand{\relTyp}[4]{\ensuremath{#1\vdash_{#4} #2\hspace{0.02in}::\hspace{0.02in}#3}}

\newcommand{\relSub}[4]{\ensuremath{#1\vdash_{#4} #2\hspace{0.02in}\symSub \hspace{0.02in}#3}}
\newcommand{\relASub}[4]
  {\ensuremath{#1;\ {#4}\vdash #2\hspace{0.02in}\symSub \hspace{0.02in}#3}}

\newcommand{\relSynSub}[4]{\ensuremath{#1\vdash_{#4}{#2}\symSynSub{#3}}}

\newcommand{\relASynSub}[4]{\ensuremath{#1;\ {#4} \vdash {#2}\symSynSub{#3}}}

\newcommand{\relSynth}[3]{\ensuremath{#1\vdash #2\hspace{0.02in}\rhd\hspace{0.02in}#3}}
\newcommand{\relConvert}[3]{\ensuremath{#1\vdash #2\hspace{0.02in}\lhd\hspace{0.02in}#3}}

\newcommand{\relWf}[2]{\ensuremath{#1\vdash #2}}

\newcommand{\relDescription}[1]{\ensuremath{\textrm{\textbf{#1}}}}
\newcommand{\judgementHead}[2]{\ensuremath{\relDescription{#1}\hfill\fbox{#2}}}

\newcommand{\ttb}{\ensuremath{\texttt{b}}}
\newcommand{\ttc}{\ensuremath{\texttt{c}}}

\newcommand{\ttf}{\ensuremath{\texttt{f}}}

\newcommand{\ttt}{\ensuremath{\texttt{t}}}

\newcommand{\ttx}{\ensuremath{\texttt{x}}}
\newcommand{\tty}{\ensuremath{\texttt{y}}}

\newif\ifUseIndices
\newcommand{\maybeN}{\ifUseIndices n \else \fi}

\newif\ifLongVersion

\newcommand{\appendixMetatheory}
  {\text{\ifLongVersion \autoref{sec:metatheory}
         \else Appendix~A~\cite{NestedTR}\fi}}
\newcommand{\appendixAlgorithmic}
  {\text{\ifLongVersion \autoref{sec:app-algorithmic}
         \else Appendix~B~\cite{NestedTR}\fi}}
\newcommand{\appendixExamples}
  {\text{\ifLongVersion \autoref{sec:examples}
         \else Appendix~C~\cite{NestedTR}\fi}}

\newcommand{\substwx}[1]{{#1}[{\varVal}/{x}]}
\newcommand{\substlwv}[1]{{#1}[{\varLogVal}/{\theV}]}

\newcommand{\noqedsymbol}{\renewcommand{\qedsymbol}{}}

\newcommand{\spaceCase}{\vspace{0.06in}}

\newcommand{\caseHead}[2]
  {\noindent\sep\makebox[1.10in][l]{Case: \ruleName{#1}.}{#2}}

\newcommand{\multiCaseHead}[1]{\noindent\sep Cases: \ruleName{#1}.}
\newcommand{\multiCaseHeadPlain}[1]{\noindent\sep Cases: {#1}.}
\newcommand{\caseHeadPlain}[1]{\noindent\sep Case: {#1}.}

\newcommand{\subCaseHead}[1]{Subcase: {#1}.}
\newcommand{\subCaseHeadEmpty}{Subcase:}
\newcommand{\subCaseIndentation}{\setlength{\parindent}{0.25in}}

\newcommand{\subSubCaseHead}[1]{Subsubcase: {#1}.}

\newcommand{\subSubCaseIndentation}{\setlength{\parindent}{0.35in}}

\newcommand{\separatingLine}{}

\newcommand{\alignItemTwo}[3]
  {\item \makebox[#1][l]{#2}{#3}}
\newcommand{\alignItemThree}[5]
 {\item \makebox[#1][l]{#3}{\makebox[#2][l]{#4}{#5}}}

\swapnumbers

\newtheorem{ass}{Assumption}

\newtheorem{prop}[ass]{Proposition}
\newtheorem{lemma}[ass]{Lemma}
\newtheorem{mainlemma}[ass]{Main Lemma}
\newtheorem{theorem}[ass]{Theorem}
\newtheorem{corollary}[ass]{Corollary}

\newtheorem*{ass-nonum}{Assumption}
\newtheorem*{definition-nonum}{Definition}
\newtheorem*{prop-nonum}{Proposition}
\newtheorem*{lemma-nonum}{Lemma}
\newtheorem*{mainlemma-nonum}{Main Lemma}
\newtheorem*{theorem-nonum}{Theorem}
\newtheorem*{corollary-nonum}{Corollary}

\newenvironment{sitemize}{\begin{list}
   {$\bullet$}
   {\setlength{\topsep}{0pt}
    \setlength{\itemsep}{0pt}
    \setlength{\leftmargin}{15pt}
    }
    
  }
  {\end{list}}

\definecolor{CconstructorColor}{RGB}{0,112,0}
\newcommand\Cconstructor[1]{\textcolor{CconstructorColor}{#1}}
\definecolor{CcommentColor}{RGB}{153,0,0}

\definecolor{CstringColor}{RGB}{128,0,0}
\newcommand\Cstring[1]{\textcolor{CstringColor}{#1}}

\definecolor{CalphakeywordColor}{RGB}{128,128,128}

\newcommand\Cnonalphakeyword[1]{#1}
\definecolor{CandColor}{RGB}{0,0,112}
\newcommand\Cand[1]{\textcolor{CandColor}{#1}}
\definecolor{CasColor}{RGB}{0,0,112}

\definecolor{CclassColor}{RGB}{0,0,112}

\definecolor{CconstraintColor}{RGB}{0,0,112}

\definecolor{CexceptionColor}{RGB}{0,0,112}

\definecolor{CexternalColor}{RGB}{0,0,112}

\definecolor{CfunColor}{RGB}{0,0,112}
\newcommand\Cfun[1]{\textcolor{CfunColor}{#1}}
\definecolor{CfunctionColor}{RGB}{0,0,112}

\definecolor{CfunctorColor}{RGB}{0,0,112}

\definecolor{CinColor}{RGB}{0,0,112}
\newcommand\Cin[1]{\textcolor{CinColor}{#1}}
\definecolor{CinheritColor}{RGB}{0,0,112}

\definecolor{CinitializerColor}{RGB}{0,0,112}

\definecolor{CletColor}{RGB}{0,0,112}
\newcommand\Clet[1]{\textcolor{CletColor}{#1}}
\definecolor{CmethodColor}{RGB}{0,0,112}

\definecolor{CmoduleColor}{RGB}{0,0,112}

\definecolor{CmutableColor}{RGB}{0,0,112}

\definecolor{CofColor}{RGB}{0,0,112}

\definecolor{CprivateColor}{RGB}{0,0,112}

\definecolor{CrecColor}{RGB}{0,0,112}
\newcommand\Crec[1]{\textcolor{CrecColor}{#1}}
\definecolor{CtypeColor}{RGB}{0,0,112}

\definecolor{CvalColor}{RGB}{0,0,112}

\definecolor{CvirtualColor}{RGB}{0,0,112}

\definecolor{CdoColor}{RGB}{0,0,112}

\definecolor{CdoneColor}{RGB}{0,0,112}

\definecolor{CdowntoColor}{RGB}{0,0,112}

\definecolor{CelseColor}{RGB}{0,0,112}
\newcommand\Celse[1]{\textcolor{CelseColor}{#1}}
\definecolor{CforColor}{RGB}{0,0,112}

\definecolor{CifColor}{RGB}{0,0,112}
\newcommand\Cif[1]{\textcolor{CifColor}{#1}}
\definecolor{ClazyColor}{RGB}{0,0,112}

\definecolor{CmatchColor}{RGB}{0,0,112}

\definecolor{CnewColor}{RGB}{0,0,112}
\newcommand\Cnew[1]{\textcolor{CnewColor}{#1}}
\definecolor{CorColor}{RGB}{0,0,112}

\definecolor{CthenColor}{RGB}{0,0,112}
\newcommand\Cthen[1]{\textcolor{CthenColor}{#1}}
\definecolor{CtoColor}{RGB}{0,0,112}

\definecolor{CtryColor}{RGB}{0,0,112}

\definecolor{CwhenColor}{RGB}{0,0,112}

\definecolor{CwhileColor}{RGB}{0,0,112}

\definecolor{CwithColor}{RGB}{0,0,112}

\definecolor{CassertColor}{RGB}{204,153,0}

\definecolor{CincludeColor}{RGB}{204,153,0}

\definecolor{CopenColor}{RGB}{204,153,0}

\definecolor{CbeginColor}{RGB}{153,0,153}

\definecolor{CendColor}{RGB}{153,0,153}

\definecolor{CobjectColor}{RGB}{153,0,153}

\definecolor{CsigColor}{RGB}{153,0,153}

\definecolor{CstructColor}{RGB}{153,0,153}

\definecolor{CraiseColor}{RGB}{255,0,0}

\definecolor{CasrColor}{RGB}{128,128,128}

\definecolor{ClandColor}{RGB}{128,128,128}

\definecolor{ClorColor}{RGB}{128,128,128}

\definecolor{ClslColor}{RGB}{128,128,128}

\definecolor{ClsrColor}{RGB}{128,128,128}

\definecolor{ClxorColor}{RGB}{128,128,128}

\definecolor{CmodColor}{RGB}{128,128,128}

\definecolor{CbarColor}{RGB}{0,0,112}
\newcommand\Cbar[1]{\textcolor{CbarColor}{#1}}

\title{Nested Refinements for Dynamic Languages}

\authorinfo
  {Ravi Chugh \and Patrick M. Rondon \and Ranjit Jhala}
  {University of California, San Diego}
  {\{rchugh,\ prondon,\ jhala\}@cs.ucsd.edu}

\begin{document}

\lstset{
  numbers=left,
  stepnumber=10,
  numberfirstline=true,
  numberstyle=\tiny,
  escapechar={\@},
  emphstyle=\textbf,
  xleftmargin=.5em,
  columns=fullflexible,
  keepspaces=true,
  language=Python,
  deletekeywords={not}}

\maketitle

\LongVersiontrue  



\begin{abstract}
  Programs written in dynamic languages make heavy use of features ---
  run-time type tests, value-indexed dictionaries, polymorphism, and
  higher-order functions --- that are beyond the reach of type systems
  that employ either purely syntactic or purely semantic reasoning.
  We present a core calculus, \dtypes, that merges these two modes of 
  reasoning into a single powerful mechanism of \emph{nested refinement types} 
  wherein the typing relation is itself a predicate in the refinement logic.
  \dtypes coordinates SMT-based logical implication and syntactic subtyping
  to automatically typecheck sophisticated dynamic language programs. 
  By coupling nested refinements with McCarthy's theory of finite maps, 
  \dtypes can precisely reason about the interaction of higher-order 
  functions, polymorphism, and dictionaries.
  The addition of type predicates to the refinement logic creates a 
  circularity that leads to unique technical challenges in the metatheory,
  which we solve with a novel stratification 
  approach that we use to prove the soundness of \dtypes.
\end{abstract}



\section{Introduction}
\label{sec:intro}

So-called dynamic languages like JavaScript, Python, Racket,
and Ruby are popular as they allow developers to quickly put 
together scripts without having to appease a static type system.
However, these scripts quickly grow into substantial code
bases that would be much easier to maintain, refactor, evolve 
and compile, if only they could be corralled within a suitable 
static type system.

The convenience of dynamic languages comes from their support
of features like run-time type testing, value-indexed
finite maps (\ie dictionaries), and duck typing, a form of
polymorphism where functions operate over any dictionary with
the appropriate keys.
As the empirical study in \cite{lambdajs} shows,
programs written in dynamic languages make heavy use of these
features, and their safety relies on invariants which can only be
established by sophisticated reasoning about the flow of
control, the run-time types of values, and the contents 
of data structures like dictionaries.

The following code snippet, adapted from the popular
Dojo Javascript framework \cite{dojo-js},
illustrates common dynamic features:

\begin{alltt}
\Clet{let} onto callbacks f obj \Cnonalphakeyword{=}
  \Cif{if} f \Cnonalphakeyword{=} null \Cthen{then}
    \Cnew{new} \Cconstructor{List}\Cnonalphakeyword{(}obj\Cnonalphakeyword{,} callbacks\Cnonalphakeyword{)}
  \Celse{else}
    \Clet{let} cb \Cnonalphakeyword{=} \Cif{if} tag f \Cnonalphakeyword{=} \Cstring{"Str"} \Cthen{then} obj\Cnonalphakeyword{[}f\Cnonalphakeyword{]} \Celse{else} f \Cin{in}
      \Cnew{new} \Cconstructor{List}\Cnonalphakeyword{(}\Cfun{fun} \Cnonalphakeyword{(}\Cnonalphakeyword{)} \Cnonalphakeyword{->} cb obj\Cnonalphakeyword{,} callbacks\Cnonalphakeyword{)}
\end{alltt}

The function \verb+onto+ is used to register callback functions
to be called after the DOM and required library modules have 
finished loading. 
The author of \verb+onto+ went to great pains to make it
extremely flexible in the kinds of arguments it takes.
If the \verb+obj+ parameter is provided but \verb+f+ is not,
  then \verb+obj+ is the function to be called after loading.
Otherwise, both \verb+f+ and \verb+obj+ are provided, and either:
(a) \verb+f+ is a string, \verb+obj+ is a dictionary, and
    the (function) value corresponding to key \verb+f+ in 
    \verb+obj+ is called with \verb+obj+ as a parameter 
    after loading; or
(b) \verb+f+ is a function which is called with \verb+obj+
    as a parameter after loading.
To verify the safety of this program, and dynamic code in general,
a type system must reason about dynamic type tests, control flow, higher-order
functions, and heterogeneous, value-indexed dictionaries.

Current type systems are not expressive enough to 
support the full spectrum of reasoning required for 
dynamic languages.
\emph{Syntactic} systems use advanced type-theoretic 
constructs like structural types \cite{Drossopoulou05}, 
row types \cite{Thiemann05}, intersection types \cite{Foster09}, 
and union types \cite{typedracket,lambdajs} to track 
invariants of individual values.
Unfortunately, such techniques cannot reason about 
value-dependent relationships \emph{between} program 
variables, as is required, for example, to determine
the specific types of the variables \verb+f+ and \verb+obj+
in \verb+onto+.
\emph{Semantic} systems like \cite{dminor} support such 
reasoning by using logical predicates to describe invariants 
over program variables. 
Unfortunately, such systems require a clear (syntactic) 
distinction between \emph{complex} values that are typed
with arrows, type variables \etc, and \emph{base} values 
that are typed with predicates~\cite{flanagan06,LiquidPLDI08,Knowles10}. 
Hence, they cannot support the interaction of complex values
and value-indexed dictionaries that is ubiquitous in 
dynamic code, for example in \verb+onto+, which can 
take as a parameter a dictionary containing a function value.

\parahead{Our Approach}
We present \dtypes, a core calculus that supports fully 
automatic checking of dynamic idioms. 
In \dtypes \emph{all} values are described uniformly by 
formulas drawn from a decidable, quantifier-free refinement 
logic.
Our first key insight is that to reason precisely about 
complex values (\eg higher-order functions) nested 
deeply inside structures (\eg dictionaries), we require 
a single new mechanism called \emph{nested refinements} 
wherein syntactic types (resp. the typing relation) may 
be nested as special \emph{type terms} 
(resp. \emph{type predicates}) inside the refinement logic.
Formally, the refinement logic is extended with atomic 
formulas of the form $\synHastyp{x}{\varUnTyp}$ where 
$\varUnTyp$ is a type term,
``$\typePredName$" (read ``has type") 
is a binary, \emph{uninterpreted} predicate in the 
refinement logic, and where the formula states that
the value $x$ ``has the type" described by the term $\varUnTyp$.
This unifying insight allows to us to express the 
invariants in idiomatic dynamic code like \verb+onto+ --- 
including the interaction between higher-order functions 
and dictionaries --- while staying within the boundaries 
of decidability.

\parahead{Expressiveness}
The nested refinement logic underlying \dtypes can 
express complex invariants between base values 
and richer values. For example, we may disjoin 
two tag-equality predicates 
\[
  \refTyp{\tagof{\theV}=\tagInt \vee \tagof{\theV}=\tagStr}
\]
to type a value $\theV$ that is either an integer or a string;
we can then track control flow involving the dynamic type tag-lookup
function $\vTag$ to ensure that the value is safely used at 
either more specific type.
To describe values like the argument \verb+f+ of the \verb+onto+
function we can combine tag-equality predicates
with the type predicate. 
We can give \verb+f+ the type
\[
  \refTyp{\theV=\vNull \vee \tagof{\theV}=\tagStr \vee
          \synHastyp{\theV}{\utArrowPlain{\tyTop}{\tyTop}}}
\]
where $\tyTop$ is an abbreviation for $\refTyp{\formTrue}$,
which is a type that describes all values. Notice the uniformity ---
the types \emph{nested} within this refinement formula are themselves
refinement types.

Our second key insight is that dictionaries are finite maps, and 
so we can precisely type dictionaries with refinement formulas
drawn from the (decidable) theory of finite maps \cite{Mccarthy,KroeningFiniteMaps}.
In particular, McCarthy's two operators ---
$\selR{x}{a}$, which corresponds to the contents of the map
$x$ at the address $a$, and
$\updR{x}{a}{v}$, which corresponds to the new map obtained
by updating $x$ at the address $a$ with the value $v$
--- are precisely what we need to describe reads from and
updates to dictionaries.
For example, we can write 
\[
  \refTyp{\tagof{\theV}=\tagRecd \wedge \tagof{\selR{\theV}{\tty}}=\tagInt}
\]
to type dictionaries $\theV$ that have (at least) an
integer field \verb+y+, where \verb+y+ is a program variable that
dynamically stores the key with which to index the
dictionary.
Even better, since we have nested function types 
into the refinement logic, we can precisely specify, 
for the first time, combinations of dictionaries 
and functions.
For example, we can write the following type for \verb+obj+
\[
  \refTyp{\tagof{\ttf}=\tagStr \Rightarrow
          \synHastyp{\selR{\theV}{\ttf}}{\utArrowPlain{\tyTop}{\tyTop}}}
\]
to describe the second portion of the \verb+onto+ specification, 
all while staying within a decidable refinement logic. In a similar
manner, we show how nested refinements support polymorphism, 
datatypes, and even a form of bounded quantification.

\parahead{Subtyping}
The huge leap in expressiveness yielded by nesting types inside 
refinements is accompanied by some unique technical 
challenges.
The first challenge is that because we nest complex types 
(\eg arrows) as uninterpreted terms in the logic, 
subtyping (\eg between arrows) cannot be carried out
solely via the usual syntactic decomposition into SMT queries
\cite{flanagan06,LiquidPLDI08,dminor}. (A higher-order refinement 
logic would solve this problem, but that would preclude algorithmic
checking; we choose the uninterpreted route precisely to relieve 
the SMT solver of higher-order reasoning!)
We surmount this challenge with a novel decomposition mechanism 
where subtyping between types, syntactic type terms, and 
refinement formulas are defined \emph{inter-dependently}, 
thereby using the logical structure of the refinement 
formulas to divide the labor of subtyping between the 
SMT solver for ground predicates 
(\eg equality, uninterpreted functions, arithmetic, maps, \etc) 
and classical syntactic rules for type terms 
(\eg arrows, type variables, datatypes, \etc).

\parahead{Soundness}
The second challenge is that the inter-dependency 
between the refinement logic and the type system renders 
the standard proof techniques for (refinement) type soundness 
inapplicable. In particular, we illustrate how
uninterpreted type predicates break the usual substitution
property and how nesting makes it difficult to define a type
system that is well-defined and enjoys this property.
We meet this challenge with a new proof technique: we define
an infinite family of \emph{increasingly precise} systems and
prove soundness of the family, of which \dtypes is a member,
thus establishing the soundness of \dtypes.

\parahead{Contributions}
To sum up, we make the following contributions:

\begin{sitemize}
\item We show how nested refinements over the theory of finite maps 
  encode function, polymorphic, dictionary and constructed data 
  types within refinements and permit dependent structural 
  subtyping and a form of bounded quantification.
 
\item We develop a novel algorithmic subtyping mechanism that uses 
  the structure of the refinement formulas to decompose subtyping
  into a collection of SMT and syntactic checks.

\item We illustrate the technical challenges that nesting poses to the
  metatheory of \dtypes and present a novel stratification-based
  proof technique to establish soundness.

\item We define an algorithmic version of
  the type system with local type inference that we implement in
  a prototype checker.
\end{sitemize}

\noindent
Thus, by carefully orchestrating the interplay between syntactic- 
and SMT-based subtyping, the nested refinement types
of \dtypes enable, for the first time, the automatic static checking 
of features found in idiomatic dynamic code.



\section{Overview}
\label{sec:overview}

We start with a series of examples that give an overview of 
our approach.
First, we show how by encoding types using logical refinements, 
\dtypes can reason about control flow and relationships between
program variables.
Next, we demonstrate how nested refinements enable
precise reasoning about values of complex types.
After that, we illustrate how \dtypes uses refinements over
the theory of finite maps to analyze value-indexed 
dictionaries.
We conclude by showing how these features combine to analyze 
the sophisticated invariants in idiomatic dynamic code.

\parahead{Notation} We use the following abbreviations for brevity.
\begin{align*}
\tyTop(x)              \quad \defeq & \quad \formTrue \\
\tyInt(x)              \quad \defeq & \quad \tagof{x} = \tagInt \\
\tyBool(x)             \quad \defeq & \quad \tagof{x} = \tagBool \\
\tyStr(x)              \quad \defeq & \quad \tagof{x} = \tagStr \\
\tyRecd(x)             \quad \defeq & \quad \tagof{x} = \tagRecd \\
\tyIntOrBool(x)        \quad \defeq & \quad \tyInt(x) \vee \tyBool(x) \\
\intertext{We abuse notation to use the above as abbreviations for 
refinement types; for each of the unary abbreviations $T$ 
defined above, an occurrence \emph{without}
the parameter denotes the refinement type
$\refTyp{T(\theV)}$. For example, we write $\tyInt$ 
as an abbreviation for $\refTyp{\tagof{\theV} = \tagInt}$.
Recall that function values are also described by refinement formulas 
(containing type predicates). We often write arrows outside
refinements to abbreviate the following:}
\utArrow{x}{\varTyp_1}{\varTyp_2} \quad \defeq & \quad
     \refTyp{\synHastyp{\theV}{\utArrow{x}{\varTyp_1}{\varTyp_2}}}
\end{align*}
We write $\utArrowPlain{\varTyp_1}{\varTyp_2}$
when the return type $\varTyp_2$ does not refer to $x$.

\subsection{Simple Refinements}\label{sec:overview:controlflow}
To warm up, we show how \dtypes describes all types through 
refinement formulas, and how, by using an SMT solver 
to discharge the subtyping (implication) queries, \dtypes makes
short work of value- and control flow-sensitive reasoning 
\cite{typedracket,lambdajs}.

\paragraph{Ad-Hoc Unions.}
Our first example illustrates the simplest dynamic idiom: 
programs which operate on \textit{ad-hoc} unions. 
The function \verb+negate+ takes an integer or 
boolean and returns its negation:

\begin{alltt}
\Clet{let} negate x \Cnonalphakeyword{=} 
  \Cif{if} tag x \Cnonalphakeyword{=} \Cstring{"Int"} \Cthen{then} 0 \Cnonalphakeyword{-} x \Celse{else} not x
\end{alltt}

In \dtypes we can ascribe to this function the type
$$\tyDec{negate}{\tyIntOrBool\ \ttArr\ {\tyIntOrBool}}$$
which states that the function accepts an integer or boolean argument 
and returns either an integer or boolean result.

To establish this, \dtypes uses
the standard means of reasoning about
control flow in refinement-based systems \cite{LiquidPLDI08},
namely strengthening the environment with the guard
predicate when processing the then-branch of an if-expression
and the negation of the guard predicate for the else-branch.
Thus, in the then-branch, the environment contains
the assumption that $\tagof{\ttx}=\tagInt$, which
allows \dtypes to verify that the expression $0 - \ttx$
is well-typed.
The return value
has the type $\refTyp{\tagof{\theV} = \tagInt \land \theV = 0 - \ttx}$.
This type is a subtype of $\tyIntOrBool$ 
as the SMT solver can prove that 
$\tagof{\theV}=\tagInt$ and $\theV = 0 - \ttx$ implies $\tagof{\theV}=\tagInt \lor \tagof{\theV} = \tagBool$.
Thus, the return value of the then-branch is deduced to have type $\tyIntOrBool$.

On the other hand, in the else-branch, the environment contains the assumption
$\lnot (\tagof{\ttx} = \tagInt)$.
By combining this with the assumption about the type of
\verb+negate+'s input,
$\tagof{\ttx}=\tagInt \lor \tagof{\ttx}=\tagBool$,
the SMT solver can determine that $\tagof{\ttx}=\tagBool$.
This allows our system to type check the call to 
$$\tyDec{not}{\tyBool\ \ttArr\ {\tyBool}},$$
which establishes that the value returned in the else branch
has type \tyIntOrBool.
Thus, our system determines that both branches return a value of type
\tyIntOrBool, and thus that \verb+negate+ meets its specification.

\paragraph{Dependent Unions.}
\dtypes's use of refinements and SMT solvers enable expressive
\emph{relational} specifications that go beyond previous 
techniques~\cite{typedracket,lambdajs}.
While \verb+negate+ takes and returns ad-hoc unions,
there is a relationship between its input and output: the output 
is an integer (resp. boolean) iff the input is an integer (resp. boolean). 
We represent this in \dtypes as 
$$\tyDec{negate}{\tyDepNegate}$$
That is, the refinement for the output states that its 
tag is \emph{the same as} the tag of the input. This function is 
checked through exactly the same analysis as before; the tag test 
ensures that the environment in the then- (resp. else-) branch
implies that \verb+x+ and the returned value are both
$\tyInt$ (resp. $\tyBool$).
That is, in both cases, the output value has the same tag as the input.

\subsection{Nested Refinements}

So far, we have seen how old-fashioned
refinement types (where the predicates refine base values 
\cite{Ou2004,Knowles10,LiquidPLDI08,dminor}) can be used 
to check ad-hoc unions over base values.
However, a type system for dynamic languages must be able
to express invariants about values of base and function
types with equal ease.
We accomplish this in \dtypes by adding types (resp. the 
typing relation) to the refinement logic as nested 
\emph{type terms} (resp. \emph{type predicates}).

However, nesting raises a rather tricky problem: with
the typing relation included in the refinement logic,
subtyping can no longer be carried out entirely via SMT 
implication queries~\cite{dminor}. 
We solve this problem with a new subtyping rule that 
\emph{extracts} type terms from refinements to enable 
syntactic subtyping for nested types.

Consider the function \verb+maybeApply+ which takes an integer
\verb+x+ and a value \verb+f+ which is either $\vNull$ or a function
over integers:

\begin{alltt}
\Clet{let} maybeApply x f \Cnonalphakeyword{=}
  \Cif{if} f \Cnonalphakeyword{=} null \Cthen{then} x \Celse{else} f x
\end{alltt}

In \dtypes, we can use a refinement formula that combines 
a base predicate and a type predicate to assign \verb+maybeApply+ the type
$$\tyDec{maybeApply}{\tyInt\ \ttArr\ \tyMApplyF\ \ttArr\ \tyInt}$$
Note that we have \emph{nested} a function type as a term in the
refinement logic, along with an assertion that a value has
this particular function type.
However, to keep checking algorithmic, we use a simple first-order logic
in which type terms and predicates are completely \emph{uninterpreted}; 
that is, the types can be thought of as constant terms in the logic.
Therefore, we need new machinery to check that
\verb+maybeApply+ actually enjoys the above type, 
\ie to check that 
(a)~\verb+f+ is indeed a function when it is applied,
(b)~it can accept the input \verb+x+, and
(c)~it will return an integer.

\paragraph{Type Extraction.} 
To accomplish the above goals, we \emph{extract} the nested 
function type for \verb+f+ stored in the type environment as 
follows.
Let $\Gamma$ be the type environment at the callsite \verb+(f x)+. 
For each type term $\varUnTyp$ occurring in $\Gamma$, we query 
the SMT solver to determine whether 
$\embed{\Gamma} \Rightarrow \synHastyp{\mathtt{f}}{\varUnTyp}$
holds, where
$\embed{\Gamma}$ is the embedding of $\Gamma$ into the refinement logic
where type terms and predicates are treated in a purely uninterpreted way.
If so, we say that $\varUnTyp$ \emph{must flow to} (or just, flows to) 
the caller expression $\mathtt{f}$. Once we have found the 
type terms that flow to the caller, we map the type terms
to their corresponding type definitions to check the call.

Let us see how this works for \verb+maybeApply+. 
The then-branch is trivial: the assumption that \verb+x+ 
is an integer in the environment allows us to deduce
that the expression \verb#x# is well-typed and has type $\tyInt$.
Next, consider the else-branch.
Let $\varUnTyp_1$ be the type term $\utArrowPlain{\tyInt}{\tyInt}$.
Due to the bindings for \texttt{x} and \texttt{f} and the else-condition,
the environment $\Gamma$ is embedded as
\[
  \embed{\Gamma} \ \defeq \ \tagof{\mathtt{x}} = \tagInt\ \land\
                 ({\mathtt{f}} = \vNull\ \lor\
                 \synHastyp{\mathtt{f}}{\varUnTyp_1}) \land\ 
                 \lnot (\mathtt{f} = \vNull)
\]
Hence, the SMT solver is able to prove that
$\Gamma \Rightarrow \synHastyp{\mathtt{f}}{\varUnTyp_1}$.
This establishes that \verb+f+ is a function on integers
and, since \verb+x+ is known to be an integer,
we can verify that the else-branch has type $\tyInt$ and 
hence check that \verb+maybeApply+ meets its specification.

\parahead{Nested Subtyping}
Next, consider a client of \verb+maybeApply+: 

\begin{alltt}
\Clet{let} \Cnonalphakeyword{_} \Cnonalphakeyword{=} maybeApply 42 negate
\end{alltt}

At the call to \verb+maybeApply+ we must show that 
the actuals are subtypes of the formals, \ie that 
the two subtyping relationships
\begin{align}
\relSub{\Gamma_1}{\refTyp{\theV = 42}}{&\ \tyInt}{} \notag \\
\relSub{\Gamma_1}
       {\refTyp{\theV = \mathtt{negate}}}
       {&\ \refTyp{\theV = \vNull\ \lor\ \synHastyp{\theV}{\varUnTyp_1}}}
       {} \label{applyCallGoal}
\end{align}
hold, where
$\Gamma_1 \defeq
    \tyBind{\mathtt{negate}}{\refTyp{\synHastyp{\theV}{\varUnTyp_0}}},
    \tyBind{\mathtt{maybeApply}}{\cdots}$
and $\varUnTyp_0=\utArrow{x}{\tyIntOrBool}{\refTyp{\tagof{\theV} = \tagof{x}}}$.
Alas, while the SMT solver can make short work of the first
obligation, it cannot be used to discharge the second 
via implication; the ``real" types that must be checked for 
subsumption, namely, $\varUnTyp_0$ and $\varUnTyp_1$, are 
embedded as totally unrelated terms in the refinement logic!

Once again, extraction rides to the rescue. We show that all 
subtyping checks of the form
${\relSub{\Gamma}{\refTyp{\varFormOne}}{\refTyp{\varFormTwo}}{}}$ 
can be reduced to a finite number of sub-goals of the form:
\begin{align*}
\text{\emph{(``type predicate-free")}}\ &\ {\embed{\Gamma'}}\Rightarrow{\varFormOne'} \\
\text{or \emph{(``type predicate")}}   \ &\ {\embed{\Gamma'}}\Rightarrow{\synHastyp{x}{\varUnTyp}}
\end{align*}
The former kind of goal has no type predicates and can be
directly discharged via SMT. For the latter, we use extraction 
to find the finitely many type terms $\varUnTyp_i$ that \emph{flow to} 
$\varFormOne'$. (If there are none, the check fails.)
For each $\varUnTyp_i$ we use \emph{syntactic} subtyping to 
verify that the corresponding type is subsumed by (the type 
corresponding to) $\varUnTyp$ under $\Gamma'$. 

In our example, the goal \ref{applyCallGoal} reduces to proving either
\[
  \embed{\Gamma_1'} \Rightarrow \theV = \vNull \sep\text{ or }\sep
  \embed{\Gamma_1'} \Rightarrow \synHastyp{\theV}{\varUnTyp_1}
\]
where $\Gamma_1' \defeq \Gamma_1, \theV = \mathtt{negate}.$
The former implication contains no type predicates, so we
attempt to prove it by querying the SMT solver. 
The solver tells us that the query is not valid, 
so we turn to the latter implication.
The extraction procedure uses the SMT solver to deduce 
that, under $\Gamma_1'$ the type term $\varUnTyp_0$ flows into $\theV$. 
Thus, all that remains is to retrieve the definition of
$\varUnTyp_0$ and $\varUnTyp_1$ and check 
$$
\relSub{\Gamma_1'}
       {\tyDepNegate}
       {\tyInt\ \ttArr\ \tyInt}
       {}$$
which follows via standard syntactic refinement subtyping~\cite{flanagan06},
thereby checking the client's call.
Thus, by carefully interleaving SMT implication and 
syntactic subtyping, \dtypes enables, for the first
time, the nesting of rich types \emph{within} refinements.

\subsection{Dictionaries}\label{sec:overview:basicdictionaries}
Next, we show how nested refinements allow \dtypes to
precisely check programs that manipulate dynamic dictionaries.
In essence, we demonstrate how structural subtyping 
can be done via nested refinement formulas over the 
theory of finite maps~\cite{Mccarthy,arrayz3}.
We introduce several abbreviations for dictionaries.
\begin{align*}
\tySel{x}{y}{z}         \defeq &\
     \hasR{x}{y} \wedge \selR{x}{y} = z \\
\tyFld{x}{y}{\tyInt}    \defeq &\
     \tyRecd(x) \wedge \tyStr(y) \wedge
     \hasR{x}{y} \wedge \tyInt(\selR{x}{y}) \\
\tyFld{x}{y}{\varUnTyp} \defeq &\
     \tyRecd(x) \wedge \tyStr(y) \wedge
     \hasR{x}{y} \wedge \synHastyp{\selR{x}{y}}{\varUnTyp}
\end{align*}
The last abbreviation states that the type of a
field is a syntactic type term $\varUnTyp$ (\eg an arrow).

\paragraph{Dynamic Lookup.}
SMT-based structural subtyping allows \dtypes to support
the common idiom of dynamic field lookup and update, where 
the field name is a value computed at run-time.
Consider the following function:

\begin{alltt}
\Clet{let} getCount t c \Cnonalphakeyword{=}
  \Cif{if} has t c \Cthen{then} toInt \Cnonalphakeyword{(}t\Cnonalphakeyword{[}c\Cnonalphakeyword{]}\Cnonalphakeyword{)} \Celse{else} 0
\end{alltt}

The function \verb+getCount+ uses the primitive operation
$$\tyDec{\vHas}{\tyBind{d}{\tyRecd}\ \ttArr\ \tyBind{k}{\tyStr}\ \ttArr\  
\refTyp{\tyBool(\theV) \land \theV=\vTrue \Leftrightarrow \hasR{d}{k}}}$$
to check whether the key \verb+c+
exists in \verb+t+.
The refinement for the input $d$ expresses the precondition
that $d$ is a dictionary, while the refinement for the key $k$
expresses the precondition that $k$ is a string.
The refinement of the output expresses the postcondition that
the result is a boolean value which is true if and only if
$d$ has a binding for the key $k$, expressed in our
refinements using $\hasR{d}{k}$, a predicate in the theory of 
maps that is true if and only if there is a binding for key $k$ 
in the map $d$ \cite{Mccarthy}.

The \emph{dictionary lookup} \verb+t[c]+ is desugared 
to $\vGet\ \ttt\ \ttc$ where the primitive operation $\vGet$ has the type
$$\tyDec{\vGet}
{\tyBind{d}{\tyRecd} \ttArr 
  \tyBind{k}{\refTyp{\tyStr(\theV)\ \wedge\ \hasR{d}{k}}} \ttArr
 {\refTyp{\theV=\selR{d}{k}}}}$$
and $\selR{d}{k}$ is an operator in the theory of maps that returns
the binding for key $k$ in the map $d$.
The refinement for the key $k$ expresses the precondition
that it is a string value in the domain of the dictionary
$d$. Similarly, the refinement for the output asserts the
postcondition that the value is the same as the 
contents of the map at the given key.

The function \verb+getCount+
first tests the dictionary \verb+t+ has a binding for the key \verb+c+;
if so, it is read and its contents are 
converted to an integer using the function 
\verb+toInt+, of type $\tyTop \ttArr \tyInt$.
Note that the if-guard strengthens the environment under 
which the lookup appears with the fact $\hasR{\mathtt{t}}{\mathtt{c}}$, 
ensuring the safety of the lookup.
If \verb+t+ does not contain the key \verb+c+, the default value \verb+0+ is returned.
Both branches are thus verified to have type $\tyInt$,
so \dtypes verifies that \verb+getCount+ has the type
$\tyDec{getCount}{\tyRecd\ \ttArr\ \tyStr\ \ttArr\ \tyInt}$.

\paragraph{Dynamic Update.}
Dually, to allow dynamic updates, \dtypes includes a primitive
\begin{align*}
\tyDec{\vUpd}{} 
& \tyBind{d}{\tyRecd}\ \ttArr\ \tyBind{k}{\tyStr}\ \ttArr\ \tyBind{x}{\tyTop}\\ 
\ttArr\ & \refTyp{\havocR{\theV}{d}{k} \wedge \tySel{\theV}{k}{x}}
\end{align*}
where $\havocR{d_1}{d_2}{k}$ abbreviates a predicate that
stipulates that $d_1$ is identical to $d_2$ at all keys \emph{except}
$k$.
Thus, the \vUpd\ primitive returns a 
dictionary that is identical to $d$ everywhere \emph{except} that it maps
the key $k$ to $x$.
The following illustrates how \vUpd\ can be used to 
update (or extend) a dictionary:

\begin{alltt}
\Clet{let} incCount t c \Cnonalphakeyword{=}
  \Clet{let} newcount \Cnonalphakeyword{=} 1 \Cnonalphakeyword{+} getCount t c \Cin{in}
  \Clet{let} res      \Cnonalphakeyword{=} set t c newcount \Cin{in} res
\end{alltt}

We give the function \verb+incCount+ the type
$${\tyBind{d}{\tyRecd}\ \ttArr\ \tyBind{c}{\tyStr}\ \ttArr\
\refTyp{\havocR{\theV}{d}{c} \wedge \tyFld{\theV}{c}{\tyInt}}}$$
The output type of \verb+getCount+ allows \dtypes to 
conclude that $\tyDec{\ensuremath{\mathtt{newcount}}}{\tyInt}$. 
From the type of \vUpd, \dtypes deduces 
$$\tyDec{res}{\refTyp{\havocR{\theV}{\ttt}{\ttc} \wedge
 \tySel{\theV}{\ttc}{\mathtt{newcount}}}}$$
which is a subtype of the output type of \verb+incCount+. Next, consider 

\begin{alltt}
\Clet{let} d0 \Cnonalphakeyword{=} \Cnonalphakeyword{\{}\Cstring{"files"} \Cnonalphakeyword{=} 42 \Cnonalphakeyword{\}} 
\Clet{let} d1 \Cnonalphakeyword{=} incCount d0 \Cstring{"dirs"}
\Clet{let} \Cnonalphakeyword{_}  \Cnonalphakeyword{=} d1\Cnonalphakeyword{[}\Cstring{"files"}\Cnonalphakeyword{]} \Cnonalphakeyword{+} d1\Cnonalphakeyword{[}\Cstring{"dirs"}\Cnonalphakeyword{]}
\end{alltt}

\dtypes verifies that
\begin{align*}
\mathtt{d0}\ & ::\ \refTyp{\tyFld{\theV}{\str{files}}{\tyInt}} \\
\mathtt{d1}\ & ::\ \refTyp{\tyFld{\theV}{\ttfld{files}}{\tyInt} \wedge
                           \tyFld{\theV}{\ttfld{dirs}}{\tyInt}} 
\end{align*}
and, hence, the field lookups return $\tyInt$s that can be safely added.

\subsection{Type Constructors}

Next, we use nesting and extraction to enrich \dtypes with 
data structures, thereby allowing for very expressive 
specifications. 
In general, \dtypes supports arbitrary user-defined datatypes,
but to keep the current discussion simple, let us consider a 
single type constructor $\tyList{T}$ for representing unbounded
sequences of $T$-values. 
Informally, an expression of type $\tyList{T}$ is either a special
\verb+null+ value or a dictionary with a \ttfld{hd} key of type $T$ 
and a \ttfld{tl} key of type $\tyList{T}$.
As for arrows, we use the following notation to write list types 
outside of refinements.
\[
  \tyList{\varTyp} \defeq \refTyp{\synHastyp{\theV}{\tyList{\varTyp}}}
\]

\paragraph{Recursive Traversal.}
Consider a textbook recursive function that takes a list of 
arbitrary values and concatenates the strings:

\begin{alltt}
\Clet{let} \Crec{rec} concat sep xs \Cnonalphakeyword{=} 
  \Cif{if} xs \Cnonalphakeyword{=} null \Cthen{then} \Cstring{""} \Celse{else} 
    \Clet{let} hd \Cnonalphakeyword{=} xs\Cnonalphakeyword{[}\Cstring{"hd"}\Cnonalphakeyword{]} \Cin{in}
    \Clet{let} tl \Cnonalphakeyword{=} xs\Cnonalphakeyword{[}\Cstring{"tl"}\Cnonalphakeyword{]} \Cin{in}
    \Cif{if} tag hd \Cnonalphakeyword{!=} \Cstring{"Str"} \Cthen{then} concat sep tl
    \Celse{else} \Cif{if} tl \Cnonalphakeyword{!=} null \Cthen{then} hd ^ sep ^ concat sep tl
    \Celse{else} hd
\end{alltt}

We ascribe the function the type
${\tyDec{concat}{\tyStr\ \ttArr\ \tyList{\tyTop}\ \ttArr \tyStr}}$.
The null test ensures the safety of the \ttfld{hd} and \ttfld{tl} 
accesses and the tag test ensures the safety of the string concatenation 
using the techniques described above.

\paragraph{Nested Ad-Hoc Unions.}
We can now define ad-hoc unions over constructed types by simply 
nesting $\tyList{\cdot}$ as a type term in the refinement logic. 
The following illustrates a common Python idiom when
an argument is either a single value or a list of values:

\begin{alltt}
\Clet{let} runTest cmd fail_codes \Cnonalphakeyword{=} 
  \Clet{let} status \Cnonalphakeyword{=} syscall cmd \Cin{in}
  \Cif{if} tag fail_codes \Cnonalphakeyword{=} \Cstring{"Int"} \Cthen{then} 
    not \Cnonalphakeyword{(}status \Cnonalphakeyword{=} fail_codes\Cnonalphakeyword{)}
  \Celse{else}
    not \Cnonalphakeyword{(}listMem status fail_codes\Cnonalphakeyword{)}
\end{alltt}

Here, ${\tyDec{listMem}{\tyTop \ttArr \tyList{\tyTop} \ttArr \tyBool}}$ 
and $\tyDec{syscall}{\tyStr \ttArr \tyInt}$.
The input \verb+cmd+ is a string, and \verb+fail_codes+ is either a
single integer or a list of integer failure codes. 
Because we nest $\tyList{\cdot}$ as a type term in our
logic, we can use the same kind of type extraction
reasoning as we did for \verb+maybeApply+ to ascribe
\verb+runTest+ the type
$$\tyDec{runTest}{\tyStr\ \ttArr\ \refTyp{\tyInt(\theV) \vee
\synHastyp{\theV}{\tyList{\tyInt}}}\ \ttArr\ \tyBool}$$

\subsection{Parametric Polymorphism}

Similarly, we can add parametric polymorphism to
\dtypes by simply treating type variables $\tyVar,\tyVarB,\etc$\ as 
(uninterpreted) type terms in the logic.
As before, we use the following notation
to write type variables outside of refinements.
\[
  \tyVar \defeq \refTyp{\synHastyp{\theV}{\tyVar}}
\]

\paragraph{Generic Containers.}
We can compose the type constructors in the ways we all know and
love. Here is list \verb+map+ in \dtypes:

\begin{alltt}
\Clet{let} \Crec{rec} map f xs \Cnonalphakeyword{=}
  \Cif{if} xs \Cnonalphakeyword{=} null \Cthen{then} null
  \Celse{else} \Cnew{new} \Cconstructor{List}\Cnonalphakeyword{(}f xs\Cnonalphakeyword{[}\Cstring{"hd"}\Cnonalphakeyword{]}\Cnonalphakeyword{,} map f xs\Cnonalphakeyword{[}\Cstring{"tl"}\Cnonalphakeyword{]}\Cnonalphakeyword{)}
\end{alltt}

(Of course, pattern matching would improve matters, but we are merely 
trying to demonstrate how much can be --- and is! --- achieved with
dictionaries.) By combining extraction with the reasoning used for
\verb+concat+, it is easy to check that
$$\tyDec{map}{\typAll{\tyVar,\tyVarB}{(\tyVar \ttArr \tyVarB)\ \ttArr\
\tyList{\tyVar}\ \ttArr{\tyList{\tyVarB}}}}$$
Note that type abstractions are automatically inserted
where a function is ascribed a polymorphic type.

\paragraph{Predicate Functions.}
Consider the list filter function:

\begin{alltt}
\Clet{let} \Crec{rec} filter f xs \Cnonalphakeyword{=}
  \Cif{if} xs \Cnonalphakeyword{=} null \Cthen{then} null
  \Celse{else} \Cif{if} not \Cnonalphakeyword{(}f xs\Cnonalphakeyword{[}\Cstring{"hd"}\Cnonalphakeyword{]}\Cnonalphakeyword{)} \Cthen{then} filter f \Cnonalphakeyword{(}xs\Cnonalphakeyword{[}\Cstring{"tl"}\Cnonalphakeyword{]}\Cnonalphakeyword{)}
  \Celse{else} \Cnew{new} \Cconstructor{List}\Cnonalphakeyword{(}xs\Cnonalphakeyword{[}\Cstring{"hd"}\Cnonalphakeyword{]}\Cnonalphakeyword{,} filter f xs\Cnonalphakeyword{[}\Cstring{"tl"}\Cnonalphakeyword{]}\Cnonalphakeyword{)}
\end{alltt}

In \dtypes, we can ascribe \verb+filter+ the type
$$\typAll{\tyVar,\tyVarB}
    {{(\tyBind{x}{\tyVar}\ \ttArr\ \refTyp{\theV = \vTrue \Rightarrow \synHastyp{x}{\tyVarB}})}
    \ \ttArr\ {\tyList{\tyVar}}\ \ttArr\ \tyList{\tyVarB}},$$
Note that the return type of the predicate, \verb+f+, tells us what
type is satisfied by values \verb+x+ for which \verb+f+
returns $\vTrue$,
and the return type of \verb+filter+
states that the items \verb+filter+ returns all
have the type implied by the predicate \verb+f+.
Thus, the general mechanism of nested refinements
subsumes the kind of reasoning performed by specialized
techniques like latent predicates~\cite{typedracket}.

\paragraph{Bounded Quantification.}
Nested refinements enable a form of bounded quantification. 
Consider the function

\begin{alltt}
\Clet{let} dispatch d f \Cnonalphakeyword{=} d\Cnonalphakeyword{[}f\Cnonalphakeyword{]} d
\end{alltt}

The function \verb+dispatch+ works for any dictionary
\verb+d+ of type $\tyVar$ that has a key \verb+f+ 
bound to a function that maps values of type $\tyVar$
to values of type $\tyVarB$. We can specify this via the dependent signature
$$
\forall \tyVar, \tyVarB.\ \tyBind{d}{\refTyp{\tyRecd(\theV) \land \synHastyp{\theV}{\tyVar}}}\ \ttArr\ 
   \refTyp{\tyFld{d}{\theV}{\tyVar \ttArr \tyVarB}}\ \ttArr\ 
   \tyVarB
$$
Note that there is no need for explicit type bounds; all that 
is required is the conjunction of the appropriate nested refinements.

\subsection{All Together Now}

With the tools we've developed in this section, \dtypes is now
capable of type checking sophisticated code from the wild.
The original source code for the following can be
found in \appendixExamples.

\paragraph{Unions, Generic Dispatch, and Polymorphism.}
We now have everything we need to type the 
motivating example from the introduction, \verb+onto+,
which combined multiple dynamic idioms: dynamic 
fields, tag-tests, and the dependency between 
nested dictionary functions and their arguments.
Nested refinements let us formalize the flexible 
interface for \verb+onto+ given in the introduction:
\begin{align*}
\forall \tyvar. &\
  \tyBind{callbacks}{\tyList{\utArrowPlain{\tyTop}{\tyTop}}} \\
\ttArr &\
  \tyBind{f}{\refTyp{\theV = \vNull \lor \tyStr(\theV) \lor
                      \synHastyp{\theV}{\utArrowPlain{\tyVar}{\tyTop}}}} \\
\ttArr &\
  \tyBind{obj}{\{ \theV \sepmidsep
                 \synHastyp{\theV}{\tyvar} \land
                 (f = \vNull \Rightarrow
                 \synHastyp{\theV}{\utArrowPlain{\tyTop}{\tyTop}}) } \\
       &\
  \hspace{0.55in}
          \land (\tyStr(f) \Rightarrow \tyFld{\theV}{f}{\utArrowPlain{\tyVar}{\tyTop}}) \} \\
\ \ttArr &\ \tyList{\utArrowPlain{\tyTop}{\tyTop}}
\end{align*}
Using reasoning similar to that used in the 
previous examples, \dtypes checks that 
\verb+onto+ enjoys the above type, where 
the specification for $obj$ is enabled by 
the kind of bounded quantification described earlier.

\paragraph{Reflection.}

Finally, to round off the overview, we present one last example that shows
how all the features presented combine to allow \dtypes to statically type
programs that introspect on the contents of dictionaries.
The function \verb+toXML+ shown below is adapted
from the Python 3.2 standard library's 
\verb+plistlib.py+ \cite{python-32}: 

\begin{alltt}
\Clet{let} \Crec{rec} toXML x \Cnonalphakeyword{=}
  \Cif{if} tag x \Cnonalphakeyword{=} \Cstring{"Bool"} \Cthen{then}
    \Cif{if} x \Cthen{then} element \Cstring{"true"} null
    \Celse{else}      element \Cstring{"false"} null
  \Celse{else} \Cif{if} tag x \Cnonalphakeyword{=} \Cstring{"Int"} \Cthen{then}
    element \Cstring{"integer"} \Cnonalphakeyword{(}intToStr x\Cnonalphakeyword{)}
  \Celse{else} \Cif{if} tag x \Cnonalphakeyword{=} \Cstring{"Str"} \Cthen{then}
    element \Cstring{"string"} x
  \Celse{else} \Cif{if} tag x \Cnonalphakeyword{=} \Cstring{"Dict"} \Cthen{then}
    \Clet{let} ks \Cnonalphakeyword{=} keys x \Cin{in}
    \Clet{let} vs \Cnonalphakeyword{=} map \Cnonalphakeyword{\{}v\Cbar{|} \Cconstructor{Str}\Cnonalphakeyword{(}v\Cnonalphakeyword{)} \Cand{and} has\Cnonalphakeyword{(}x\Cnonalphakeyword{,}v\Cnonalphakeyword{)}\Cnonalphakeyword{\}} \Cconstructor{Str}
      \Cnonalphakeyword{(}\Cfun{fun} k \Cnonalphakeyword{->} element \Cstring{"key"} k ^ toXML x\Cnonalphakeyword{[}k\Cnonalphakeyword{]}\Cnonalphakeyword{)} ks \Cin{in}
    \Cstring{"<data>"} ^ concat \Cstring{"\(\backslash\)n"} vs ^ \Cstring{"</data>"}
  \Celse{else} element \Cstring{"function"} null
\end{alltt}

The function 
takes an arbitrary value and renders it as an XML string,
and illustrates several idiomatic uses of dynamic features.
If we give the auxiliary
function \verb+intToStr+ the type
$\tyInt\ \ttArr\ \tyStr$
and \verb+element+ the type
$\tyStr\ \ttArr\ \refTyp{\theV = \vNull \lor \tyStr(\theV)}\ \ttArr \tyStr$,
we can verify that
\begin{align*}
\mathtt{toXML} & ::\ {\tyTop\ \ttArr\ \tyStr} \\
\intertext{Of especial interest is the dynamic field lookup 
$\mathtt{x[k]}$ used in the function passed to $\mathtt{map}$ 
to recursively convert each binding of the dictionary to XML.
The primitive operation \vKeys\ has the type}
\vKeys & ::\ \tyBind{d}{\tyRecd}\ \ttArr\ {\tyList{\refTyp{\tyStr(\theV) \wedge \hasR{d}{\theV}}}}
\end{align*}
that is, it returns a list of string keys that belong to the
input dictionary. Thus, $\mathtt{ks}$ has type
$\tyList{\refTyp{\tyStr(\theV) \wedge \hasR{\mathtt{x}}{\theV}}}$,
which enables the call to $\mathtt{map}$ to typecheck, 
since the body of the argument is checked in an 
environment where
$\tyDec{k}{\refTyp{\tyStr(\theV) \wedge \hasR{\mathtt{x}}{\theV}}}$,
which is the type that $\tyVar$ is instantiated with.
This binding suffices to prove the safety of the dynamic 
field access. The control flow reasoning described previously
uses the tag tests guarding the other cases to prove each of them safe.



\section{Syntax and Semantics}
\label{sec:syntax-and-semantics}


\newcommand{\vv}{\varVal}

\begin{figure}[t]
  \centering
  $$
  \begin{array}{rcll}
  \vv & ::= & \hspace{1.10in}   
                                                    & \textbf{Values} \\
      &   | & x                                     & \text{variable} \\
      &   | & c                                     & \text{constant} \\
      &   | & \dictextend{\vv_1}{\vv_2}{\vv_3}      & \text{dictionary extension} \\
      &   | & \funBare{x}{e}                        & \text{function} \\
      &   | & \typFun{\tyvar}{e}                    & \text{type function} \\
      &   | & \newDataBare{\varCon}{\seq{\varVal}}  & \text{constructed data} \\
    \blankrow{4} \\
    e & ::= &                                     & \textbf{Expressions} \\
      &   | & \vv                                 & \text{value} \\
      &   | & \vv_1\ \vv_2                        & \text{function application} \\
      &   | & \typInst{\vv}{\varTyp}              & \text{type function application} \\
      &   | & \ite{\vv}{e_1}{e_2}                 & \text{if-then-else} \\
      &   | & \letinBare{x}{e_1}{e_2}             & \text{let-binding} \\
    \blankrow{4} \\
  \varTypeDef
      & ::= & \typeConDef{\varCon}
                         {\varPole\tyvar}
                         {f}
                         {\varTyp}                & \textbf{Datatype Definitions} \\
    \blankrow{4} \\
  \mathit{prg} & ::= & \seq{\varTypeDef}; e       & \textbf{Programs} \\
    \blankrow{4} \\
    \varLogVal
      & ::= &                                     & \textbf{Logical Values} \\
      &   | & \vv                                 & \text{value} \\
      &   | & \lprimF(\seq{\varLogVal})           & \text{logical function application} \\
    \blankrow{4} \\
    \varFormOne, \varFormTwo, \varFormThr
      & ::= &                                     & \textbf{Refinement Formulas} \\
      &   | & \lprimP(\seq{\varLogVal})           & \text{predicate} \\
      &   | & \synHastyp{\varLogVal}{\varUnTyp}   & \text{type predicate} \\
      &   | & \varFormOne \land \varFormTwo
                \sep | \sep
                \varFormOne \lor \varFormTwo
                \sep | \sep
                \neg \varFormOne                  & \text{logical connective} \\
    \blankrow{4} \\
    \varTyp
      & ::= & \refTyp{\varFormOne}                & \textbf{Refinement Types} \\
    \blankrow{4} \\
    \varUnTyp
      & ::= &                                     & \textbf{Type Terms} \\
      &   | & \utArrow{x}{\varTyp_1}{\varTyp_2}   & \text{arrow} \\
      &   | & \tyvar                              & \text{type variable} \\
      &   | & \typeCon{\varCon}{\seq{\varTyp}}    & \text{constructed type} \\
      &   | & \nullCon                            & \text{null} \\
    \blankrow{4} \\
    \varScm
      & ::= & \varTyp \sep | \sep
              \typAll{\tyvar}{\varScm}            & \textbf{Type Schemes}
  \end{array}
  $$
  \caption{Syntax of \dtypes}
  \label{fig:syntax}
\end{figure}

We begin with the syntax and evaluation semantics of \dtypes.
\autoref{fig:syntax} shows the syntax of values, expressions, and types.


\parahead{Values}
Values $\varVal$ include variables
constants, functions, type functions, dictionaries,
and records created by type constructors.
The set of constants $c$ include base values like integer, boolean, and
string constants, the empty dictionary $\emptydict$, and $\vNull$.
Logical values $\varLogVal$ are all values and applications of
primitive function symbols $\lprimF$, such as addition $+$ and
dictionary selection $\logop{sel}$, to logical values.
The constant $\vTag$ allows introspection on the type tag of a 
value at run-time. For example,
$$\begin{array}{rclrcl}
\vTag(3)                          & \!\defeq\! & \tagInt
&
\vTag(\vTrue)                     & \!\defeq\! & \tagBool 
\\
\vTag(``\text{joe}")                     & \!\defeq\! & \tagStr 
& 
\vTag(\funBare{x}{e})             & \!\defeq\! & \tagFun 
\\
\vTag(\emptydict)                 & \!\defeq\! & \tagRecd 
&
\vTag(\typFun{\tyVar}{e})         & \!\defeq\! & \tagTFun 
\end{array}$$

\parahead{Dictionaries}
A dictionary $\dictextend{\varVal_1}{\varVal_2}{\varVal_3}$ extends
the dictionary $\varVal_1$ with the binding from string $\varVal_2$
to value $\varVal_3$.
For example, the dictionary mapping \ttfld{x} to 3 and
\ttfld{y} to $\vTrue$ is written
$$\dictextend{\dictextend{\emptydict}{\mathtt{``x"}}{3}}
             {\mathtt{``y"}}
             {\vTrue}.$$
The set of constants also includes operations 
for extending dictionaries and accessing their fields.
The function $\vGet$ is used to access dictionary fields 
and is defined
\begin{align*}
  \vGet\ (\dictextend{\varVal}{\str{x}}{\varVal_x})\ \str{x} &
      \defeq \varVal_x \\
  \vGet\ (\dictextend{\varVal}{\str{y}}{\varVal_y})\ \str{x} &
      \defeq \vGet\ \varVal\ \str{x}
\intertext{The function $\vHas$ tests for the presence of a field
and is defined}
  \vHas\ (\dictextend{\varVal}{\str{y}}{\varVal_y})\ \str{x} &
      \defeq \vHas\ \varVal\ \str{x} \\
  \vHas\ (\dictextend{\varVal}{\str{x}}{\varVal_x})\ \str{x} &
      \defeq \vTrue \\
  \vHas\ \emptydict\ \str{x}
      & \defeq \vFalse
\end{align*}
The function $\vUpd$ updates the value bound to a key and is defined
\begin{align*}
  \vUpd\ d\ k\ \varVal & \defeq \dictextend{d}{k}{\varVal}
\end{align*}


\parahead{Expressions}
The set of expressions $e$ consists of values,
function applications, type instantiations,
if-then-else expressions, and let-bindings.
We use an A-normal presentation so that we need
only define substitution of values (not arbitrary expressions) 
into types. 


\parahead{Types}
We stratify types into monomorphic types $\varTyp$
and polymorphic type schemes $\typAll{\tyVar}{\varScm}$.
In \dtypes, a type $\varTyp$ is a \emph{refinement type} of the
form $\refTyp{\varFormOne}$, where $\varFormOne$ is a
\emph{refinement formula}, and is read ``$\theV$ such that $\varFormOne$.''
The values of this type are all values $\varVal$ such that the
formula $\subst{\varFormOne}{\theV}{\varVal}$ ``is true.''
What this means, formally, is core to our approach and will be
considered in detail in \autoref{sec:soundness}.

\parahead{Refinement Formulas}
The language of \emph{refinement formulas} includes predicates
$\lprimP$, such as the equality predicate and
dictionary predicates $\logop{has}$ and $\logop{sel}$,
and the usual logical connectives.
For example, the type of integers is $\refTyp{\tagof{\theV}=\tagInt}$,
which we abbreviate to $\tyInt$. The type of positive integers is 
$$\refTyp{\tagof{\theV}=\tagInt \wedge \theV > 0}$$ 
and the type of dictionaries with an integer field $\ttfld{f}$ is 
\[
 \refTyp{\tagof{\theV}=\tagRecd \wedge
         \hasR{\theV}{\ttfld{f}} \wedge
         \tagof{\selR{\theV}{\ttfld{f}}} = \tagInt}.
\]
We refer to the binder $\theV$ in refinement types as
``the value variable."

\parahead{Nesting: Type Predicates and Terms}
To express the types of values like functions and dictionaries containing
functions, \dtypes permits types to be nested within refinement 
formulas. 
Formally, the language of refinement formulas includes a form, 
$\synHastyp{\varLogVal}{\varUnTyp},$ called a \emph{type predicate}, 
where $\varUnTyp$ is a \emph{type term}.
The type term $\utArrow{x}{\varTyp_1}{\varTyp_2}$ 
describes values that have a dependent function type, \ie 
functions that accept arguments $\varVal$ of type $\varTyp_1$
and return values of type $\varTyp_2[\varVal/x]$, where
$x$ is bound in $\varTyp_2$.
We write $\utArrowPlain{\varTyp_1}{\varTyp_2}$ when $x$ does
not appear in $\varTyp_2$.
Type terms $\tyVar,\tyVarB,\etc$ correspond to type parameters to
polymorphic functions.
The type term $\nullCon$ corresponds to the type of the constant value $\vNull$.
The type term $\typeCon{\varCon}{\seq{\varTyp}}$ corresponds to
records constructed with the $\varCon$ type constructor instantiated
with the sequence of type arguments $\seq{\varTyp}$.
For example, the type of the (integer) successor function is 
\[
\refTyp{\synHastyp{\theV}{\utArrow{x}{\tyInt}{\refTyp{\tagof{\theV}=\tagInt
\land \theV = x + 1}}}},
\]
dictionaries where the value at key $\ttfld{f}$ maps $\tyInt$ to
$\tyInt$ have type
\[
 \refTyp{\tagof{\theV}=\tagRecd \wedge
         \hasR{\theV}{\ttfld{f}} \wedge
         \synHastyp{\selR{\theV}{\ttfld{f}}}
                   {\utArrowPlain{\tyInt}{\tyInt}}},
\]
and the constructed record $\newDataBare{\listCon}{\texttt{1},\vNull}$
can be assigned the type
$\refTyp{\synHastyp{\theV}{\typeCon{\listCon}{\tyInt}}}$.


\parahead{Datatype Definitions}
A datatype definition of $\varCon$ defines a named, possibly recursive type.
A datatype definition includes a sequence $\seq{\varPole\tyVar}$ of
type parameters $\tyVar$ paired with \emph{variance annotations} $\varPole$.
A variance annotation is either $\poleCo$ (covariant),
$\poleContra$ (contravariant), or $\poleBi$ (bivariant).
The rest of the definition specifies a sequence $\seq{\tyBind{f}{\varTyp}}$
of field names and their types. The types of the fields may refer
to the type parameters of the declaration. A well-formedness check,
which will be described in \autoref{sec:type-checking}, ensures that
occurrences of type parameters in the field types respect their declared
variance annotations.
By convention, we will use the subscript $i$ to index into the sequence
$\seq{\varPole\tyVar}$ and $j$ for $\seq{\tyBind{f}{\varTyp}}$.
For example, $\varPole_i$ refers to the variance annotation of the $i^{th}$
type parameter, and $f_j$ refers to the name of the $j^{th}$ field.


\parahead{Programs}
A program is a sequence of datatype definitions $\seq{\varTypeDef}$ followed
by an expression $e$.
Requiring all datatype definitions to appear first simplifies
the subsequent presentation.



\parahead{Semantics}
The small-step operational semantics of \dtypes is standard for a
call-by-value, polymorphic lambda calculus;
we provide the formal definition in \appendixMetatheory.
Following standard practice, the semantics is parametrized by
a function $\delta$ that assigns meaning to primitive functions $c$,
including dictionary operations like $\vHas$, $\vGet$, and $\vUpd$.



\section{Type Checking}
\label{sec:type-checking}

\begin{figure}[t]
\centering

\judgementHead{Well-Formed Type Schemes}{$\relWf{\Gamma}{\varScm}$}

\vsepRule

$\inferrule*
  {x \textrm{ fresh} \sepPremise
   \relWf{\Gamma,\tyBind{x}{\tyTop}}{\subst{\varFormOne}{\theV}{x}}}
  {\relWf{\Gamma}{\refTyp{\varFormOne}}}$
\hsepRule
$\inferrule*
  {\relWf{\Gamma,\tyVar}{\varScm}}
  {\relWf{\Gamma}{\typAll{\tyVar}{\varScm}}}$

\vsepRule

\judgementHead{Well-Formed Formulas}{$\relWf{\Gamma}{\varFormOne}$}

\vsepRule

$\inferrule*
  {
   \relWf{\Gamma}{\varLogVal} \sepPremise
   \relWf{\Gamma}{\varUnTyp}}
  {\relWf{\Gamma}{\synHastyp{\varLogVal}{\varUnTyp}}}$
\hsepRule
$\inferrule*
  {
   \forall i.\ \relWf{\Gamma}{\varLogVal_i}
  }
  {\relWf{\Gamma}{\lprimP(\seq{\varLogVal})}}$
\hsepRule
$\inferrule*
  {
   \relWf{\Gamma}{\varFormOne} \sepPremise
   \relWf{\Gamma}{\varFormTwo}
  }
  {\relWf{\Gamma}{\varFormOne\wedge\varFormTwo}}$

\vsepRule

\judgementHead{Well-Formed Type Terms}{$\relWf{\Gamma}{\varUnTyp}$}

\vsepRule

$\inferrule*
  {\relWf{\Gamma}{\varTyp_1} \\\\
   \relWf{\Gamma,\tyBind{x}{\varTyp_1}}{\varTyp_2}}
  {\relWf{\Gamma}{\utArrow{x}{\varTyp_1}{\varTyp_2}}}$
\hsepRule
$\inferrule*
  {\tyVar\in\Gamma}
  {\relWf{\Gamma}{\tyVar}}$
\hsepRule
$\inferrule*
  { }
  {\relWf{\Gamma}{\nullCon}}$
\hsepRule
$\inferrule*
  {\varCon\in Dom(\lookupConName) \\\\
   \forall i.\ \relWf{\Gamma}{\varTyp_i}}
  {\relWf{\Gamma}{\typeCon{\varCon}{\seq{\varTyp}}}}$

\vsepRule

\judgementHead{Well-Formed Type Environments}{$\relWf{}{\Gamma}$}

\vsepRule

$\inferrule*
  { }
  {\relWf{}{\emptyset}}$
\hsepRule
$\inferrule*
  {x\notin Dom(\Gamma) \\\\
   \relWf{}{\Gamma} \sepPremise
   \relWf{\Gamma}{\varScm}}
  {\relWf{}{\Gamma,\tyBind{x}{\varScm}}}$
\hsepRule
$\inferrule*
  {\relWf{}{\Gamma} \\\\
   \tyVar\notin\Gamma}
  {\relWf{}{\Gamma,\tyVar}}$
\hsepRule
$\inferrule*
  {\relWf{}{\Gamma} \sepPremise
   \relWf{\Gamma}{\varFormOne}}
  {\relWf{}{\Gamma,\varFormOne}}$

\vsepRule

\judgementHead{Well-Formed Type Definitions}{$\relWf{}{\varTypeDef}$}

\vsepRule

$\inferrule*
  {\forall j.\ \relWf{\seq{\tyVar}}{\varTyp_j} \sepPremise
   \forall i.\ \varianceOk{\tyVar_i}{\varPole_i}{\seq{\varTyp}}
  }
  {\relWf{}{\typeConDef{\varCon}{\varPole\tyVar}{f}{\varTyp}}}
$

\vsepRule

\caption{Well-formedness for \dtypes}
\label{fig:well-formedness}
\end{figure}

In this section, we present the \dtypes type system, comprising 
several well-formedness relations,
an expression typing relation, and,
at the heart of our approach, 
a novel subtyping relation which discharges obligations involving
nested refinements through a combination of syntactic and semantic,
SMT-based reasoning.
We first define environments for type checking.

\parahead{Environments}
Type environments $\Gamma$ are of the form
\begin{displaymath}
\begin{array}{rcl}
  \Gamma & ::= &   \emptyset
       \sep | \sep \Gamma,\tyBind{x}{\varScm}
       \sep | \sep \Gamma,\tyVar
       \sep | \sep \Gamma,\varFormOne
\end{array} 
\end{displaymath}
where bindings either record the derived type $\varScm$ for a
variable $x$, a type variable $\tyVar$ introduced in the scope
of a type function, or a formula $\varFormOne$ that is recorded
to track the control flow along branches of an if-expression.
A type definition environment $\lookupConName$ 
records the definition of each constructor type $\varCon$.
As type definitions appear at the beginning of a program, 
we assume for clarity that $\lookupConName$ is fixed and 
globally visible, and elide it from the judgments.
In the sequel, we assume that $\lookupConName$ contains 
at least the definition
\[
  \typeConDefOneTwo
    {\listCon}{\poleCo\tyVar}
    {\ttfld{hd}}{\refTyp{\synHastyp{\theV}{\tyVar}}}
    {\ttfld{tl}}{\refTyp{\synHastyp{\theV}{\tyList{\tyVar}}}}.
\]
%


\subsection{Well-formedness}

\autoref{fig:well-formedness} defines the well-formedness relations.

\parahead{Formulas, Types and Environments}
We require that types be \emph{well-formed} within the current type
environment, which means that formulas used in types
are boolean propositions and mention only variables that are currently
in scope.
By convention, we assume that variables used as binders throughout
the program are distinct and different from the special value
variable $\theV$, which is reserved for types. Therefore, $\theV$
is never bound in $\Gamma$.
When checking the well-formedness of
a refinement formula $\varFormOne$, we substitute a fresh variable $x$
for $\theV$ and check that $\subst{\varFormOne}{\theV}{x}$ is well-formed
in the environment extended with $\tyBind{x}{\tyTop}$,
to the environment, where $\tyTop=\refTyp{\formTrue}$.
We use fresh variables to prevent duplicate bindings of $\theV$.
 
Note that the well-formedness of formulas does \emph{not} depend
on type checking; all that is needed is the ability to syntactically 
distinguish between terms and propositions.
Checking that formulas are well-formed is straightforward;
the important point is that a variable $x$ may be used only 
if it is bound in $\Gamma$.

\parahead{Datatype Definitions}
To check that a datatype definition is well-formed, we first check
that the types of the fields are well-formed in an environment
containing the declared type parameters.
Then, to enable a sound subtyping rule for constructed types in the
sequel, we check that the declared variance annotations are
respected within the type definition.
For this, we use a procedure $\varianceOkName$
(defined in \appendixMetatheory) that recursively walks
formulas to record whether type variables occur in positive
or negative positions within the types of the fields.
%


\subsection{Expression Typing}

The expression typing judgment $\relTyp{\Gamma}{e}{\varScm}{}$,
defined in \autoref{fig:dec-typing}, verifies
that expression $e$ has type scheme $\varScm$ 
in environment $\Gamma$.
We highlight the important aspects of the typing rules.


\parahead{Constants}
Each primitive constant $c$ has a type, denoted by $\typConst{c}$, that is
used by \ruleName{T-Const}.
Basic values like integers, booleans, \etc\ are given
singleton types stating that their value equals the corresponding constant
in the refinement logic. For example:
$$\begin{array}{rlrl}
  \texttt{1}       \ :: & \refTyp{\theV = \texttt{1}}
  &
  \vTrue           \ :: & \refTyp{\theV = \vTrue} 
  \\
  \text{``joe''}   \ :: & \refTyp{\theV = \text{``joe''}}
  &
  \vFalse          \ :: & \refTyp{\theV = \vFalse} 
\end{array}$$
Arithmetic and boolean operations have types
that reflect their semantics.
Equality on base values is defined in the standard way, 
while equality on function values is physical equality.
\begin{align*}
\texttt{+}     \ ::\ & \utArrow{x}{\tyInt}
                         {\utArrow{y}{\tyInt}{\refTyp{\tyInt(\theV) \wedge \theV=x+y}}} \\
\texttt{not}   \ ::\ & \utArrow{x}{\tyBool}
                         {\refTyp{\tyBool(\theV) \wedge
                                       x=\vTrue\Leftrightarrow\theV=\vFalse}} \\
\texttt{=}     \ ::\ & \utArrow{x}{\tyTop}
                         {\utArrow{y}{\tyTop}
                            {\refTyp{\tyBool(\theV) \wedge
                                     \theV=\vTrue \Leftrightarrow x=y}}} \\
\vFix \ ::\ & \typAll{\tyVar}{\utArrowPlain{(\utArrowPlain{\tyVar}{\tyVar})}{\tyVar}} \\
\texttt{tag}   \ ::\ & \utArrow{x}{\tyTop}{\refTyp{\theV=\tagof{x}}}
\end{align*}
The constant $\vFix$ is used to encode recursion, 
and the type for the tag-test operation uses an
axiomatized function in the logic.

The operations on dictionaries are given refinement types 
over the theory of finite maps.
\begin{align*}
  \emptydict    \ ::\ & \refTyp{\theV = \emptydictR} \\
  \vHas         \ ::\ & \utArrow{d}{\tyRecd}
                          {\utArrow{k}{\tyStr}
                             {\refTyp{\tyBool(\theV) \wedge
                                      \theV=\vTrue\Leftrightarrow\hasR{d}{k}}}} \\
  \vGet         \ ::\ & \tyBind{d}{\tyRecd}\ \ttArr\
                        \tyBind{k}{\refTyp{\tyStr(\theV) \wedge\ 
                                           \hasR{d}{\theV}}} \ \ttArr\
                        \refTyp{\theV=\selR{d}{k}} \\
  \vUpd         \ ::\ & \tyBind{d}{\tyRecd}\ \ttArr\ 
                        \tyBind{k}{\tyStr}\ \ttArr\ \tyBind{x}{\tyTop} \\
                      & \ttArr\ \refTyp{\havocR{\theV}{d}{k} \wedge
                                        \hasR{d}{k} \wedge
                                        \selR{d}{k}=x}\\
  \vKeys        \ ::\ & \utArrow{d}{\tyRecd}{\tyList{\refTyp{\tyStr(\theV)\wedge\hasR{d}{\theV}}}}
\end{align*}
In the theory of finite maps, the operator $\domR{d}$ denotes the 
domain of the map $d$, and $\projR{d}{y}$ restricts $d$ to the set 
of keys $y$. (These primitives can all be reduced to McCarthy's
select and update operators \cite{Mccarthy,KroeningFiniteMaps};
we define these in \appendixMetatheory).
Thus, we define $\emptydictR$ as a special constant such that 
$\domR{\emptydictR} = \emptyset$. The refinements for the other
operators use $\hasR{d}{k}$, which abbreviates $k \in \domR{d}$, 
and $\havocR{d_1}{d_2}{a}$, which abbreviates
$$\projR{d_1}{\domR{d_1}\setminus\set{a}}=\projR{d_2}{\domR{d_2}{\setminus\set{a}}}$$
The predicate $\hasR{d}{k}$ checks that a key $k$ is defined in a 
map $d$, and is used as a precondition for $\vGet$.
%
The predicate $\havocR{d_1}{d_2}{k}$ states that the dictionaries 
$d_1$ and $d_2$ are identical \emph{except} at the key $k$. This is
useful for dictionary updates where we do not know the \emph{exact} 
value being stored, but do know some abstraction thereof, \eg its type. 
For example, in \verb+incCounter+ (from \autoref{sec:overview}) 
we do not know what value is stored in the count field $\mathtt{c}$,
only that it is an integer. Thus, we say 
that the new dictionary is the same as the old except at 
$\mathtt{c}$, where the binding is an integer. 
A more direct approach would be to use an existentially quantified
variable to represent the stored value and say that the resulting
dictionary is the original dictionary updated to contain this quantified
value.
Unfortunately, that would take the formulas outside the decidable 
quantifier-free fragment of the logic, thereby precluding SMT-based
logical subtyping. 


\UseIndicesfalse

\begin{figure}[t]
\centering


\judgementHead{Type Checking}{$\relTyp{\Gamma}{e}{\varScm}{\maybeN}$}

\vsepRule

$\inferrule*[right=\ruleNameFig{T-Const}]
  { }
  {\relTyp{\Gamma}{c}{\typConst{c}}{\maybeN}}
$

\vsepRule

$\inferrule*[right=\ruleNameFig{T-Var}]
  {\Gamma(x)=\varTyp}
  {\relTyp{\Gamma}{x}{\refTyp{\theV=x}}{\maybeN}}
$
\sep\sep
$\inferrule*[right=\ruleNameFig{T-VarPoly}]
  {\Gamma(x)=\typAll{\tyvar}{\varScm}}
  {\relTyp{\Gamma}{x}{\typAll{\tyvar}{\varScm}}{\maybeN}}
$

\vsepRule

$\inferrule*[right=\ruleNameFig{T-Extend}]
  {\relTyp{\Gamma}{\varVal_1}{\tyRecd}{\maybeN} \sepPremise
   \relTyp{\Gamma}{\varVal_2}{\tyStr}{\maybeN} \sepPremise
   \relTyp{\Gamma}{\varVal_3}{\varScm}{\maybeN}}
  {\relTyp{\Gamma}{\dictextend{\varVal_1}{\varVal_2}{\varVal_3}}
                  {\refTyp{\theV=\dictextend{\varVal_1}{\varVal_2}{\varVal_3}}}
                  {\maybeN}}
$

\vsepRule

$\inferrule*[right=\ruleNameFig{T-If}]
  {\relTyp{\Gamma}{\varVal}{\tyBool}{\maybeN} \\\\
   \relTyp{\Gamma,\varVal=\vTrue}{e_1}{\varScm}{\maybeN} \sepPremise
   \relTyp{\Gamma,\varVal=\vFalse}{e_2}{\varScm}{\maybeN}
  }
  {\relTyp{\Gamma}{\ite{\varVal}{e_1}{e_2}}{\varScm}{\maybeN}}
$

\vsepRule

$\inferrule*[right=\ruleNameFig{T-Fun}]
  {\relWf{\Gamma}{\varTyp_1} \sepPremise
   \relTyp{\Gamma,\tyBind{x}{\varTyp_1}}{e}{\varTyp_2}{\maybeN}}
  {\relTyp{\Gamma}
          {\funBare{x}{e}}
          {\refTyp{\theV=\funBare{x}{e}\wedge
                   \synHastyp{\theV}{\utArrow{x}{\varTyp_1}{\varTyp_2}}}}{\maybeN}}
$

\vsepRule

$\inferrule*[right=\ruleNameFig{T-App}]
  {\relTyp{\Gamma}{\varVal_1}
          {\refTyp{\synHastyp{\theV}{\utArrow{x}
                                             {\varTyp_{11}}
                                             {\varTyp_{12}}}}}{\maybeN} \sepPremise
   \relTyp{\Gamma}{\varVal_2}{\varTyp_{11}}{\maybeN}}
  {\relTyp{\Gamma}{\varVal_1\ \varVal_2}{\varTyp_{12}[\varVal_2/x]}{\maybeN}}
$

\vsepRule

$\inferrule*[right=\ruleNameFig{T-TFun}]
  {\tyvar\notin\Gamma \sepPremise
   \relTyp{\Gamma, \tyvar}{e}{\varScm}{\maybeN}}
  {\relTyp{\Gamma}{\typFun{\tyvar}{e}}{\typAll{\tyvar}{\varScm}}{\maybeN}}
$

\vsepRule

$\inferrule*[right=\ruleNameFig{T-TApp}]
  {\relWf{\Gamma}{\varTyp} \sepPremise
   \relTyp{\Gamma}{\varVal}{\typAll{\tyvar}{S}}{\maybeN}}
  {\relTyp{\Gamma}{\typInst{\varVal}{\varTyp}}
          {\inst{\varScm}{\tyVar}{\varTyp}}{\maybeN}}
$

\vsepRule

$\inferrule*[right=\ruleNameFig{T-Fold}]
  {\forall i.\ \relWf{\Gamma}{\varTyp_i} \sepPremise
   \lookupCon{\varCon}{\seq{\varPole\tyvar}}{f}{\varTyp'} \\\\
   \forall j.\ \relTyp{\Gamma}
                      {\varVal_j}
                      {\inst{\varTyp'_j}{\seq{\tyvar}}{\seq{\varTyp}}}
                      {\maybeN}
  }
  {\relTyp{\Gamma}{\newDataBare{\varCon}{\seq{\varVal}}}
          {\refTyp{\fold{\varCon}{\seq{\varTyp}}{\seq{\varVal}}}}{\maybeN}}
$

\vsepRule

$\inferrule*[right=\ruleNameFig{T-Unfold}]
  {\relTyp{\Gamma}{e}
          {\refTyp{\synHastyp{\theV}{\typeCon{\varCon}{\seq{\varTyp}}}}}{\maybeN}}
  {\relTyp{\Gamma}{e}
          {\refTyp{\unfold{\varCon}{\seq{\varTyp}}}}{\maybeN}}
$

\vsepRule

$\inferrule*[right=\ruleNameFig{T-Let}]
  {\relWf{\Gamma}{\varScm_1} \sepPremise
   \relTyp{\Gamma}{e_1}{\varScm_1}{\maybeN} \sepPremise
   \relTyp{\Gamma, \tyBind{x}{\varScm_1}}{e_2}{\varScm_2}{\maybeN} \sepPremise
   \relWf{\Gamma}{\varScm_2}}
  {\relTyp{\Gamma}{\letinBare{x}{e_1}{e_2}}{\varScm_2}{\maybeN}}
$

\vsepRule

$\inferrule*[right=\ruleNameFig{T-Sub}]
  {\relTyp{\Gamma}{e}{\varScm'}{\maybeN} \sepPremise
   \relSub{\Gamma}{\varScm'}{\varScm}{\maybeN} \sepPremise
   \relWf{\Gamma}{\varScm}}
  {\relTyp{\Gamma}{e}{\varScm}{\maybeN}}
$

\vsepRule

\ifUseIndices
  \caption{Type checking for \dntypes}
  \label{fig:dec-typing-indexed}
\else
  \caption{Type checking for \dtypes}
  \label{fig:dec-typing}
\fi

\end{figure}


\parahead{Standard Rules}
We briefly identify several typing rules that are standard for lambda calculi
with dependent refinements.
\ruleName{T-Var} and \ruleName{T-VarPoly} assign types to variable
expressions $x$.
If $x$ is bound to a (monomorphic) refinement type in $\Gamma$, then
\ruleName{T-Var} assigns $x$ the singleton type that says that the
expression $x$ evaluates to the same value as the variable $x$.
\ruleName{T-If} assigns the type scheme $\varScm$ to an if-expression
if the condition $\varVal$ is a boolean-valued expression, the
then-branch expression $e_1$ has type scheme $\varScm$ under the assumption
that $\varVal$ evaluates to $\vTrue$, and the else-branch expression $e_2$
has type scheme $\varScm$ under the assumption that $\varVal$ evaluates to
$\vFalse$.
The \ruleName{T-App} rule is standard, but notice that the arrow type
of $\varVal_1$ is nested inside a refinement type.
In \ruleName{T-Let}, the type scheme $\varScm_2$ must be 
well-formed in $\Gamma$, which prevents the variable $x$ 
from escaping its scope.
\ruleName{T-Sub} allows expression $e$ to be used with
type $\varScm$ if $e$ has type $\varScm'$ and $\varScm'$ 
is a subtype of $\varScm$.


\parahead{Type Instantiation}
The \ruleName{T-TApp} rules uses the procedure $\instName$ to instantiate
a type variable with a (monomorphic) type.
$\instName$ is defined recursively on formulas, type terms, and types,
where the only non-trivial case involves type predicates with type variables:
\begin{align*}
\inst{\synHastyp{\varLogVal}{\tyvar}}{\tyvar}{\refTyp{\varFormOne}} &=
  \subst{\varFormOne}{\theV}{\varLogVal} \\
\inst{\synHastyp{\varLogVal}{\tyvarB}}{\tyvar}{\varTyp} &=
  \synHastyp{\varLogVal}{\tyVarB}
\end{align*}
We write $\inst{\varScm}{\seq{\tyVar}}{\seq{\varTyp}}$ to mean the result
of applying $\instName$ to $\varScm$ with the type variables and type arguments 
in succession.
%


\parahead{Fold and Unfold}
The \ruleName{T-Fold} rule is used for records of data created with
the datatype constructor $\varCon$ and type arguments $\seq{\varTyp}$.
The rule succeeds if the argument $\varVal_j$ provided for each
field $f_j$ has the required type $\varTyp'_j$ after instantiating
all type parameters $\seq{\tyVar}$ with the type arguments $\seq{\varTyp}$.
If these conditions are satisfied, the formula returned by
$\fold{\varCon}{\seq{\varTyp}}{\seq{\varVal}}$, defined as
\[
  \theV\neq\vNull \wedge
  \tagof{\theV}=\tagRecd \wedge
  \synHastyp{\theV}{\typeCon{\varCon}{\seq{\varTyp}}} \wedge
  (\wedge_j\ \selR{\theV}{f_j}=\varVal_j)
\]
records that the value is non-null, that the values stored in
the fields are precisely the values used to construct the record,
and that the value has a type corresponding to
the specific constructor used to create the value.
\ruleName{T-Unfold} exposes the fields of non-null constructed 
data as a dictionary, using $\unfold{\varCon}{\seq{\varTyp}}$, 
defined as
\begin{align*}
  \theV\neq\vNull \Rightarrow & 
  (\tagof{\theV}=\tagRecd \wedge
  (\wedge_j \embed{\varTyp''_j}(\selR{\theV}{f_j}))) \\
\intertext{where $\lookupCon{\varCon}{\seq{\varPole\tyVar}}{f}{\varTyp'}$,
$\embed{\refTyp{\varFormOne}}(\varLogVal) \defeq
\subst{\varFormOne}{\theV}{\varLogVal}$,
and for all $j$, $\varTyp''_j=\inst{\varTyp'_j}{\seq{\tyVar}}{\seq{\varTyp}}$.
For example, $\unfold{\listCon}{\tyInt}$ is}
\theV\neq\vNull \Rightarrow & (\tagof{\theV}=\tagRecd \wedge \tagof{\selR{\theV}{\ttfld{hd}}}=\tagInt \\
   & \wedge \synHastyp{\selR{\theV}{\ttfld{tl}}}{\tyList{\tyInt}})
\end{align*}


\subsection{Subtyping}

\UseIndicesfalse

\begin{figure}[t]
\centering

\judgementHead{Subtyping}
              {$\relSub{\Gamma}{\varScm_1}{\varScm_2}{\maybeN}$}

\vsepRule


$\inferrule*[right=\ruleNameFig{S-Mono}]
  {x \text{ fresh} \sepPremise
   \varFormOne_1' = \subst{\varFormOne_1}{\theV}{x} \sepPremise
   \varFormOne_2' = \subst{\varFormOne_2}{\theV}{x} \\\\
   \cnf{\varFormOne_2'} = \wedge_i(\clause{\varFormTwo_i}{\varFormThr_i}) \sepPremise
   \forall i.\
     \relImpl{\Gamma,\varFormOne_1'}{\varFormTwo_i}{\varFormThr_i}{\maybeN}}
  {\relSub{\Gamma}{\refTyp{\varFormOne_1}}{\refTyp{\varFormOne_2}}{\maybeN}}
$

\vsepRule

$\inferrule*[right=\ruleNameFig{S-Poly}]
  {\relSub{\Gamma}{\varScm_1}{\varScm_2}{\maybeN}}
  {\relSub{\Gamma}{\typAll{\tyVar}{\varScm_1}}{\typAll{\tyVar}{\varScm_2}}{\maybeN}}
$

\vsepRule

\judgementHead{Clause Implication}
              {$\relImpl{\Gamma}{\varFormTwo}{\varFormThr}{\maybeN}$}

\vsepRule

$\inferrule*[right=\ruleNameFig{C-Valid}]
  {\valid{\embed{\Gamma}\wedge\varFormTwo\Rightarrow\varFormThr}}
  {\relImpl{\Gamma}{\varFormTwo}{\varFormThr}{\maybeN}}
$
\ifUseIndices
  \quad
  $\inferrule*[right=\ruleNameFig{C-Valid-n}]
    {\formulaSat{n}{\embed{\Gamma}\wedge\varFormTwo\Rightarrow\varFormThr}}
    {\relImpl{\Gamma}{\varFormTwo}{\varFormThr}{n}}
  $
\fi

\vsepRule


$\inferrule*[right=\ruleNameFig{C-ImpSyn}]
  {\exists j. \sepPremise
   \valid{\embed{\Gamma} \wedge \varFormTwo \Rightarrow
          \synHastyp{\varLogVal_j}{\varUnTyp}} \sepPremise
   \relSynSub{\Gamma, \varFormTwo}{\varUnTyp}{\varUnTyp_j}{\maybeN}
  }
  {\relImpl{\Gamma}
           {\varFormTwo}
           {\vee_i\ \synHastyp{\varLogVal_i}{\varUnTyp_i}}
           {\maybeN}}
$

\vsepRule

\judgementHead{Syntactic Subtyping}
              {$\relSynSub{\Gamma}{\varUnTyp_1}{\varUnTyp_2}{\maybeN}$}

\vsepRule

$\inferrule*[right=\ruleNameFig{U-Arrow}]
  {
   \relSub{\Gamma}{\varTyp_{21}}{\varTyp_{11}}{\maybeN} \sepPremise
   \relSub{\Gamma,\tyBind{x}{\varTyp_{21}}}{\varTyp_{12}}{\varTyp_{22}}{\maybeN}
  }
  {\relSynSub{\Gamma}{\utArrow{x}{\varTyp_{11}}{\varTyp_{12}}}
                     {\utArrow{x}{\varTyp_{21}}{\varTyp_{22}}}{\maybeN}}
$

\vsepRule

$\inferrule*[right=\ruleNameFig{U-Var}]
  { }
  {\relSynSub{\Gamma}{\tyvar}{\tyvar}{\maybeN}}
$
\hsepRule
$\inferrule*[right=\ruleNameFig{U-Null}]
  { }
  {\relSynSub{\Gamma}{\nullCon}{\typeCon{\varCon}{\seq{\varTyp}}}{\maybeN}}
$

\vsepRule

$\inferrule*[right=\ruleNameFig{U-Datatype}]
  {\lookupConAnnots{\varCon}{\seq{\varPole\tyvar}} \\\\
   \forall i.\
      \text{if } \varPole_i\in\set{\poleCo,\poleBi}
      \text{ then } \relSub{\Gamma}{\varTyp_{1i}}{\varTyp_{2i}}{\maybeN} \\\\
   \forall i.\ 
      \text{if } \varPole_i\in\set{\poleContra,\poleBi}
      \text{ then } \relSub{\Gamma}{\varTyp_{2i}}{\varTyp_{1i}}{\maybeN}
  }
  {\relSynSub{\Gamma}{\typeCon{\varCon}{\seq{\varTyp_1}}}
                     {\typeCon{\varCon}{\seq{\varTyp_2}}}{\maybeN}}
$

\vsepRule

\ifUseIndices
  \caption{Subtyping for \dntypes}
  \label{fig:dec-subtyping-indexed}
\else
  \caption{Subtyping for \dtypes}
  \label{fig:dec-subtyping}
\fi

\end{figure}

In traditional refinement type systems, there is a 
two-level hierarchy between types and refinements 
that allows a syntax-directed reduction of subtyping
obligations to SMT implications
\cite{flanagan06,LiquidPLDI08,Knowles10}.
In contrast, \dtypes's refinements include 
uninterpreted type predicates that are beyond the 
scope of (first-order) SMT solvers.

Let us consider the problem of establishing the subtyping judgment
${\relSub{\Gamma}{\refTyp{\varFormOne_1}}{\refTyp{\varFormOne_2}}{\maybeN}}$. 
We cannot use the SMT query 
\begin{equation}
  \embed{\Gamma}\wedge\varFormOne_1\Rightarrow\varFormOne_2
  \label{eq:plain-impl}
\end{equation}
as the presence of (uninterpreted) type-predicates 
may conservatively render the implication invalid. 
Instead, our strategy is to massage the refinements 
into a normal form that makes it easy to factor 
the implication in (\ref{eq:plain-impl}) into a collection of subgoals 
whose consequents are either simple (non-type) predicates 
or type predicates. 
The former can be established via SMT and the latter 
by recursively invoking syntactic subtyping. 
Next, we show how this strategy is realized by the rules in
\autoref{fig:dec-subtyping}.

\parahead{Step 1: Split query into subgoals} 
We start by converting $\varFormOne_2$ into a \emph{normalized 
conjunction} $\wedge_i (\clause{\varFormTwo_i}{\varFormThr_i})$.
Each conjunct, or \emph{clause}, $\clause{\varFormTwo_i}{\varFormThr_i}$
is normalized such that its consequent is a \emph{disjunction} 
of type predicates.
We use the symbol $\clauseSym$ instead of the usual implication
arrow $\Rightarrow$ to emphasize the normal structure of
each clause.
By splitting $\varFormOne_2$ into its normalized clauses, 
rule \ruleName{S-Mono} reduces the goal ($\ref{eq:plain-impl}$)
to the equivalent collection of subgoals 
\[
  \forall i.\
    \relImpl{\Gamma,\varFormOne_1}{\varFormTwo_i}{\varFormThr_i}{}
\]

\parahead{Step 2: Discharge subgoals} 
The normalization ensures that the consequent of each subgoal above
is a disjunction of type predicates. When the disjunction of a clause is 
\emph{empty}, the subgoal is  
\begin{align}
\text{\emph{(``type predicate-free")}}\quad &\
  \relImpl{\Gamma,\varFormOne_1}{\varFormTwo_i}{\formFalse}{}
\notag \\
\intertext{which rule \ruleName{C-Valid} handles by SMT. 
Otherwise, the subgoal is}
\text{\emph{(``type predicate")}}\quad &\
  \relImpl{\Gamma,\varFormOne_1}
          {\varFormTwo_i}
          {\synHastyp{\varLogVal_j}{\varUnTyp_j}}{}
\notag
\end{align}
which rule \ruleName{C-ImpSyn} handles via 
type extraction followed by an invocation 
of syntactic subtyping. 
In particular, the rule tries to establish 
\emph{one of} the disjuncts $\synHastyp{\varLogVal_j}{\varUnTyp_j}$, 
by searching for a type term $\varUnTyp$ that occurs in $\Gamma$ 
that 1) \emph{flows to} $\varLogVal_j$, \ie for which we can deduce via SMT that
\[
  \embed{\Gamma}\wedge\varFormOne_1\wedge\varFormTwo_i\Rightarrow
    \synHastyp{\varLogVal_j}{\varUnTyp}
\]
is valid and,
2) is a syntactic subtype of $\varUnTyp_j$ in an appropriately
strengthened environment
(written $\relSynSub{\Gamma,\varFormOne_1,\varFormTwo_i}{\varUnTyp}{\varUnTyp_j}{}$). 
The rules \ruleName{U-Datatype} and \ruleName{U-Arrow} establish 
syntactic (refinement) subtyping, by (recursively) establishing 
that subtyping holds for the
matching components~\cite{flanagan06, LiquidPLDI08, dminor}.
Because syntactic subtyping recursively refers to subtyping,
the \ruleName{S-Mono} rule uses fresh variables to avoid
duplicate bindings of $\theV$ in the environment.

\parahead{Formula Normalization}
Procedure $\cnfName$ converts
a formula $\varFormOne$ into a conjunction of clauses
$\wedge_i(\clause{\varFormTwo_i}{\varFormThr_i})$
as described above.
The conversion is carried out by translating $\varFormOne$ to 
conjunctive normal form (CNF), and then for each CNF clause,
rearranging literals and adding negations as necessary.
For example,
\begin{align*}
  \cnf{\theV=\vNull} & \defeq \neg(\theV=\vNull) \clauseSym \formFalse \\
  \cnf{\theV=\vNull \vee \synHastyp{\theV}{\varUnTyp}} & \defeq
    \neg(\theV=\vNull) \clauseSym \synHastyp{\theV}{\varUnTyp}
\end{align*}

\parahead{Formula Implication}
In each SMT implication query
$\embed{\Gamma}\wedge\varFormOne\Rightarrow\varFormTwo$, the operator 
$\embed{\cdot}$ describes the embedding of environments and types
into the logic as follows:
\[
\begin{array}{rclrcl}
  \embed{\refTyp{\varFormOne}} & \!\defeq\! & \varFormOne &
  \embed{\Gamma, \tyBind{x}{\varTyp}} & \!\defeq\! &
              \embed{\Gamma} \land \subst{\embed{\varTyp}}{\theV}{x} \\
  \embed{\emptyset} & \!\defeq\! & \formTrue &
  \embed{\Gamma, \tyBind{x}{\typAll{\tyVar}{\varScm}}} & \!\defeq\! &
        \embed{\Gamma} \\
  \embed{\Gamma, \varFormOne}  & \!\defeq\! & \embed{\Gamma} \land \varFormOne &
  \embed{\Gamma, \tyVar}  & \!\defeq\! & \embed{\Gamma}
\end{array}
\]
%



\parahead{Recap}
Recall that our goal is to typecheck programs which use value-indexed
dictionaries which may contain functions as values.
On the one hand, the theory of finite maps allows us to use logical refinements
to express and verify complex invariants about the contents of dictionaries.
On the other, without resorting to higher-order logic,
such theories cannot express that a dictionary maps a key to a value of function
type.

To resolve this tension, we introduced the novel concept of
\emph{nested refinements},
where types are nested into the logic as uninterpreted terms and
the typing relation is nested as an uninterpreted predicate.
The logical validity queries arising in typechecking
are discharged by rearranging the formula in question into an implication
between a purely logical formula and a disjunction of type predicates.
This implication is discharged using a novel
combination of logical queries, discharged by an SMT solver, and syntactic subtyping.
This approach enables the efficient, automatic type checking of
sophisticated dynamic language programs that manipulate complex data,
including dictionaries which map keys to function values.



\section{Soundness}
\label{sec:soundness}

At this point in the proceedings, it is customary to make a claim about 
the soundness of the type system by asserting that it enjoys the standard
preservation and progress properties. Unfortunately, the presence of nested 
refinements means this route is unavailable to us, as the usual substitution 
property does not hold! 
Next, we describe why substitution is problematic and define
a \emph{stratified} system \dntypes for which we establish the 
preservation and progress properties. The soundness of $\dtypes$ follows,
as it is a special case of the stratified \dntypes.

\subsection{The Problems}

The key insight in $\dtypes$ is that we can use uninterpreted functions to
nest types inside refinements, thereby unlocking the door to expressive
SMT-based reasoning for dynamic languages. However, this very strength 
precludes the usual substitution lemma upon which preservation 
proofs rest.

\parahead{Substitution} 
The standard substitution
property requires that if 
${\relTyp{\tyBind{x}{\varScm},\Gamma}{e}{\varScm'}{}}$  
and 
${\relTyp{\ }{\varVal}{\varScm}{}}$,
then
${\relTyp{\substwx{\Gamma}}{\substwx{e}}{\substwx{\varScm'}}{}}$.
The following snippet shows why \dtypes lacks this property:

\begin{alltt}
\Clet{let} foo f \Cnonalphakeyword{=} 0 \Cin{in} foo \Cnonalphakeyword{(}\Cfun{fun} x \Cnonalphakeyword{->} x \Cnonalphakeyword{+} 1\Cnonalphakeyword{)}
\end{alltt}

Suppose that we ascribe to \verb+foo+ the type 
\[
  \tyDec{foo}{\tyBind{f}{(\utArrowPlain{\tyInt}{\tyInt})}\ \ttArr\
              \refTyp{\synHastyp{f}{\utArrowPlain{\tyInt}{\tyInt}}}}.
\]
The return type of the function states that its argument $f$ 
is a function from integers to integers and does not impose any
constraints on the return value itself.
To check that \texttt{foo} does indeed have this type,
by $\ruleName{T-Fun}$, the following judgment must be derivable:
\begin{equation}
 \relTyp{\tyBind{f}{\utArrowPlain{\tyInt}{\tyInt}}}
        {\texttt{0}}
        {\refTyp{\synHastyp{f}{\utArrowPlain{\tyInt}{\tyInt}}}}{}
        \label{eq:footype}
\end{equation}
By $\ruleName{T-Const}$, $\ruleName{T-Sub}$, $\ruleName{S-Mono}$ and 
$\ruleName{C-Valid}$ the judgment reduces to the implication
\[
  \formTrue\wedge
  \synHastyp{f}{\utArrowPlain{\tyInt}{\tyInt}}\wedge
  \subst{\embed{\typConst{\texttt{0}}}}{\theV}{\texttt{0}}\Rightarrow
  \synHastyp{f}{\utArrowPlain{\tyInt}{\tyInt}}.
\]
which is trivially valid, thereby deriving (\ref{eq:footype}), and showing
that \texttt{foo} does indeed have the ascribed type.

Next, consider the call to \texttt{foo}. By $\ruleName{T-App}$, the result 
has type
\[
  \refTyp{\synHastyp{\texttt{(fun x -> x + 1)}}
                         {\utArrowPlain{\tyInt}{\tyInt}}}.
\]
The expression
\texttt{foo (fun x -> x + 1)} evaluates in one step to \texttt{0}.
Thus, if the substitution property is to hold, \texttt{0} 
should also have the above type. In other words, \dtypes must 
be able to derive 
\[
  \relTyp{}{\texttt{0}}
           {\refTyp{\synHastyp{\texttt{(fun x -> x + 1)}}
                              {\utArrowPlain{\tyInt}{\tyInt}}}}{}.
                              \label{eq:footypestep}
\]
By $\ruleName{T-Const}$, $\ruleName{T-Sub}$,
$\ruleName{S-Mono}$, and $\ruleName{C-Valid}$, 
the judgment reduces to the implication
\begin{equation}
  \formTrue\wedge
  \subst{\embed{\typConst{\texttt{0}}}}{\theV}{\texttt{0}}\Rightarrow
  \synHastyp{\texttt{(fun x -> x + 1)}}
            {\utArrowPlain{\tyInt}{\tyInt}}
\label{eqn:problem}
\end{equation}
which is \emph{invalid} as type predicates are \emph{uninterpreted} 
in our refinement logic! Thus, the call to \verb+foo+ and the
reduced value do not have the same type in \dtypes, which 
illustrates the crux of the problem: the $\ruleName{C-Valid}$ 
rule is not closed under substitution.

\parahead{Circularity}
Thus, it is clear that the substitution lemma will require that we 
define an interpretation for type predicates.
As a first attempt, we can define an interpretation $\interp{}$ that
interprets type predicates involving arrows as:
\[
\formulaSat{}{\synHastyp{\funBare{x}{e}}
             {\utArrow{x}{\varTyp_{1}}{\varTyp_{2}}}}
\quad \textit{ iff } \quad
\relTyp{\tyBind{x}{\varTyp_{1}}}{e}{\varTyp_{2}}{}.
\]
Next, let us replace $\ruleName{C-Valid}$ with the following rule
that restricts the antecedent to the above interpretation:
\[
\inferrule*[right=\ruleNameFig{C-Valid-Interpreted}]
  {\formulaSat{}{\embed{\Gamma}\wedge\varFormOne\Rightarrow\varFormTwo}}
  {\relImpl{\Gamma}{\varFormOne}{\varFormTwo}{}}
\]
Notice that the new rule requires the implication be valid in
the particular interpretation $\interp{}$ instead of in all interpretations.
This allows the logic to ``hook back" into the type system to derive
types for closed lambda expressions, thereby discharging the problematic 
implication query in (\ref{eqn:problem}).
While the rule solves the problem with substitution, it does
not take us safely to the shore --- it introduces a circular 
dependence between the typing judgments and the interpretation
$\interp{}$.
Since our refinement logic includes negation, the type system
corresponding the set of rules outlined earlier combined with
$\ruleName{C-Valid-Interpreted}$ is not necessarily well-defined.

\subsection{The Solution: Stratified \dntypes}

Thus, to prove soundness, we require a well-founded means of 
interpreting type predicates. We achieve this by \emph{stratifying} 
the interpretations and type derivations, requiring that type 
derivations at each level refer to interpretations at the same level,
and that interpretations at each level refer to derivations at strictly
lower levels.
Next, we formalize this intuition and state the important lemmas and
theorems. The full proofs may be found in \appendixMetatheory.

Formally, we make the following changes. 
First, we index typing judgments ($\vdash_{n}$) and 
interpretations ($\interp{n}$) with a natural number $n$. 
We call these the level-$n$ judgments and interpretations,
respectively.
Second, we allow level-$n$ judgments to use the rule
\[
\inferrule*[right=\ruleNameFig{C-Valid-n}]
  {\formulaSat{n}{\embed{\Gamma}\wedge\varFormOne\Rightarrow\varFormTwo}}
  {\relImpl{\Gamma}{\varFormOne}{\varFormTwo}{n}}
\]
and the level-$n$ interpretations to use lower-level type derivations:
\[
\formulaSat{n}{\synHastyp{\funBare{x}{e}}
              {\utArrow{x}{\varTyp_{1}}{\varTyp_{2}}}}
\quad \textit{ iff } \quad
\relTyp{\tyBind{x}{\varTyp_{1}}}{e}{\varTyp_{2}}{n-1}.
\]
Finally, we write 
\[
\relTyp{\Gamma}{e}{\varScm}{*} 
\quad \textit{ iff } \quad
\exists n.\ \relTyp{\Gamma}{e}{\varScm}{n}.
\]

\noindent
The derivations in \dntypes consist of the derivations 
at all levels. The following ``lifting" lemma states 
that the derivations at each level include the 
derivations at all lower levels:

\begin{lemma-nonum}[Lifting Derivations]
\mbox{}
\begin{enumerate}
\item If $\relTyp{\Gamma}{e}{\varScm}{}$,
      then $\relTyp{\Gamma}{e}{{\varScm}}{*}$.
\item If $\relTyp{\Gamma}{e}{\varScm}{n}$,
      then $\relTyp{\Gamma}{e}{{\varScm}}{n+1}$.
\end{enumerate}
\end{lemma-nonum}

\noindent
The first clause holds since the original \dtypes derivations
cannot use the \ruleName{C-Valid-n} rule, 
\ie ${\relTyp{\Gamma}{e}{\varScm}{}}$ exactly 
when ${\relTyp{\Gamma}{e}{\varScm}{0}}$.
The second clause follows from the definitions 
of $\vdash_n$ and $\interp{n}$.
Stratification snaps the circularity knot and enables the 
proof of the following stratified substitution lemma:
\begin{lemma-nonum}[Stratified Substitution]
\mbox{}

If $\relTyp{\tyBind{x}{\varScm},\Gamma}{e}{\varScm'}{n}$
and $\relTyp{\ }{\varVal}{\varScm}{n}$,

then $\relTyp{\substwx{\Gamma}}{\substwx{e}}{\substwx{\varScm'}}{n+1}$.
\end{lemma-nonum}
\noindent
The proof of the above depends on the following lemma, which
captures the connection between our typing rules and the 
logical interpretation of formulas in our refinement logic:

\begin{lemma-nonum}[Satisfiable Typing]
\mbox{}

If $\relTyp{\ }{\varVal}{\varTyp}{n}$,
then $\formulaSat{n+1}{\substwx{\embed{\varTyp}}}$.
\end{lemma-nonum}

\noindent
Stratified substitution enables the following
preservation result:

\begin{theorem-nonum}[Stratified Preservation]
\mbox{}

If $\relTyp{\ }{e}{\varScm}{n}$,
and $\reducesTo{e}{e'}$
then $\relTyp{\ }{e'}{\varScm}{n+1}$.
\end{theorem-nonum}

\noindent
From this, and a separate progress result, we establish
the type soundness of \dntypes:

\begin{theorem-nonum}[\dntypes Type Soundness]
\mbox{}

If $\relTyp{\ }{e}{\varScm}{*}$,
then either $e$ is a value
or $\reducesTo{e}{e'}$ and $\relTyp{\ }{e'}{\varScm}{*}$.
\end{theorem-nonum}

\noindent
By coupling this with Lifting, we obtain the soundness of \dtypes
as a corollary.



\section{Algorithmic Typing}
\label{sec:algorithmic}

Having established the expressiveness and soundness of
\dtypes, we establish its practicality by implementing a type checker
and applying it to several interesting examples.
The declarative rules for type checking \dtypes programs, shown in
\autoref{sec:type-checking}, are not syntax-directed and thus
unsuitable for implementation.
We highlight the problematic rules and
sketch an \emph{algorithmic}
version of the type system
that also performs local type inference \cite{pierce-turner}.
The algorithmic system is sound with respect to the declarative
one and, modulo a restriction to ensure that subtyping terminates,
is as precise.
Our prototype implementation \cite{NestedTR}
verifies all of the examples in this paper and 
in \cite{typedracket},
using Z3 \cite{z3} to discharge SMT obligations.
A more detailed discussion of the algorithmic system
may be found in \appendixAlgorithmic.


\subsection{Algorithmic Subtyping}

Nearly all the declarative subtyping rules presented in
\autoref{fig:dec-subtyping} are non-overlapping and directed by the
structure of the judgment being derived.
The sole exception is \ruleName{C-ImpSyn}, whose first premise
requires us to synthesize a type term $\varUnTyp$ such that
the SMT solver can prove $\synHastyp{\varLogVal_j}{\varUnTyp}$ for some
$j$, where $\varUnTyp$ is used in the second premise.
We note that, since type predicates are uninterpreted,
the only type terms $\varUnTyp$ that can satisfy this criterion
must come from the environment $\Gamma$.
Thus, we define a procedure
$\extractNaive{\Gamma}{\varTyp}$ that uses the SMT solver to compute
the set of type terms $\varUnTyp'$, out of \emph{all} possible type terms
mentioned in $\Gamma$, such that for all values $x$,
$\tyBind{x}{\varTyp}$ implies that
$\synHastyp{x}{\varUnTyp'}$.
To implement \ruleName{C-ImpSyn}, we call
$\extractNaive{\Gamma}{\refTyp{\theV=\varLogVal_j}}$ to compute the set
$\usedBoxes$ of type terms that might be needed by the second premise.
Since the declarative rule cannot possibly refer to a type
term $\varUnTyp$ that is not in $\Gamma$, this strategy guarantees that
$\varUnTyp\in\usedBoxes$ and, thus, does not forfeit precision.

\paragraph{Ensuring Termination.}
An important concern remains:
because we extract type terms from the environment and recursively
invoke the subtyping relation on them,
we do not have the usual guarantee that subtyping is recursively invoked
on strictly syntactically smaller terms, and thus
it is not clear whether subtyping checks
will terminate.
Indeed, they may not!
\appendixAlgorithmic\ presents an example
obligation that, although unlikely to appear in practice,
leads to non-termination when subtyping is implemented directly.
The crux of the matter is that an inner subtyping obligation may be
isomorphic to an outer one, triggering an infinitely repeating
derivation.
Fortunately, we can cut the loop as follows:
along any branch of a subtyping derivation, we allow a type term
to be returned by $\extractName$ at most once.
Since there are only finitely many type terms in the environment,
this is enough to ensure termination.
The price we pay is that algorithmic subtyping is not complete with respect
to declarative subtyping; we have not found and do not expect
this to be a problem in practice.


\subsection{Bidirectional Type Checking}

We extend the syntax of \dtypes with optional type annotations
for binding constructs and constructed data, and, following work on local type
inference \cite{pierce-turner},
we define a \emph{bidirectional}
type checking algorithm.
In the remainder of this section, we highlight
the novel aspects of our bidirectional type system.


\parahead{Function Applications}
To typecheck an application $\varVal_1\ \varVal_2$,
we must synthesize a type
$\varTyp_1$ for the function $\varVal_1$ and use type extraction to
convert $\varTyp_1$ to a syntactic arrow.
Since the procedure $\extractName$ can return an arbitrary number of
type terms, we must decide how to proceed in the event
that $\varTyp_1$ can be extracted to multiple different arrow types.
To avoid the need for backtracking in the type checker, and to
provide a semantics
that is simple for the programmer to understand, we
synthesize a type for $\varVal_1$
only if there is \emph{exactly one} syntactic arrow that is applicable
to the given argument $\varVal_2$.
%



\parahead{Remaining Rules}
We will now briefly summarize some of the other algorithmic rules presented
in \appendixAlgorithmic.
Uses of \ruleName{T-Sub} can be factored into other typing rules.
However, uses of \ruleName{T-Unfold} cannot,
since we cannot syntactically predict where it is
needed. Since we do not have pattern matching to determine exactly when to
unfold type definitions, as in languages like ML, we eagerly
unfold type definitions to anticipate all situations in which unfolding
might be required.
For let-expressions, to handle the fact that synthesized types might refer
to variables that are about to go out of scope, making them ill-formed,
we use several simple
heuristics to eliminate occurrences of local variables.
In all of the examples we have tested, the annotations
provided on top-level let-bindings are sufficient to allow synthesizing
well-formed types for all unannotated inner let-expressions.
Precise types are synthesized for if-expressions by synthesizing the types
of both branches, guarding them by the appropriate branch conditions,
and conjoining them.
For constructed data expressions, we allow the programmer to provide hints
in type definitions that help the type checker decide how to infer type
parameters that are omitted.
For example, suppose the $\listCon$ definition is updated as follows:
\[
  \typeConDefOneTwo
    {\listCon}
    {\poleCo\tyVar}
    {\ttfld{hd}}{\refTyp{\synHastyp{\theV}{\tyVar}}}
    {\ttfld{tl}}{\refTyp{\synHastyp{\theV}
                                   {\tyList{\inferenceMarker\tyVar}}}}
\]
Due to the presence of the marker $\inferenceMarker$ in the type of the \ttfld{tl}
field, local type inference will use the type of $\varVal_2$ to infer
the omitted type parameter in $\newDataBare{\listCon}{\varVal_1,\varVal_2}$.
Finally, although the techniques in \cite{pierce-turner} would allow us
to, for simplicity we do not attempt to synthesize parameters to
type functions.


\parahead{Soundness}
We write $\relConvert{\Gamma}{e}{\varScm}$ for the algorithmic type
checking judgment, which verifies $e$ against the given type $\varScm$, and
$\relSynth{\Gamma}{e}{\varScm}$ for the algorithmic type
synthesis judgment, which produces a type $\varScm$ for expression $e$.
Each of the techniques employed in this section are sound with respect
to the declarative system, so we can show the following property,
where we use a procedure $\eraseName$ to remove type annotations from functions,
let-bindings, and constructed data because the syntax of the declarative
system does not permit them:

\begin{prop-nonum}[Sound Algorithmic Typing]
\mbox{}

  {If   $\relSynth{\Gamma}{e}{\varScm}$ or}
  {     $\relConvert{\Gamma}{e}{\varScm}$,}
  {then $\relTyp{\Gamma}{\erase{e}}{\varScm}{}$.}
\end{prop-nonum}



\section{Related Work}
\label{sec:related}

In this section, we highlight related approaches to statically
verifying features of dynamic languages. For a thorough
introduction to contract-based and other hybrid approaches,
see \cite{Findler02,Sie06,Knowles10}.

\parahead{Dynamic Unions and Control Flow}
Among the earliest attempts at mixing static and dynamic 
typing was adding the special type \ttdyn to a statically-typed
language like ML~\cite{Abadi89}. In this approach, an arbitrary 
value can be injected into \ttdyn, and a typecase construct allows
inspecting its precise type at run-time. However, one cannot 
guarantee that a particular \ttdyn value is of one of a subset 
of types (cf. \verb+negate+ from \autoref{sec:overview}). 
Several researchers have used union types and tag-test 
sensitive control-flow analyses to support such idioms. 
Most recently, $\lambda_{TR}$~\cite{typedracket} and 
$\lambda_S$~\cite{lambdajs} feature values of (untagged) 
union types that can be used at more precise types based 
on control flow. 
In the former, each expression is assigned two propositional
formulas that hold when the expression evaluates to either 
true or false; these propositions are strengthened by recording 
the guard of an if-expression in the typing environment when typing 
its branches. Typechecking proceeds by solving propositional 
constraints to compute, for each value at each program point, 
the set of tags it may correspond to.
The latter shows how a similar strategy can be developed 
in an imperative setting, by coupling a type system with 
a data flow analysis. However, both systems are limited to 
ad-hoc unions over basic and function values.
In contrast, \dtypes shows how, by pushing all the information 
about the value (resp. reasoning about flow) into expressive, 
but decidable refinement predicates (resp. into SMT solvers),
one can statically reason about significantly richer 
idioms (related tags, dynamic dictionaries, polymorphism, \etc).

\parahead{Records and Objects}
There is a large body of work
on type systems for objects \cite{PalsbergBook,Kennedy06}.
Several early advances incorporate records into 
ML~\cite{Rem89}, but the use of records
in these systems are unfortunately unlikely to be 
flexible enough for dynamic dictionaries. In particular, 
record types cannot be joined when they disagree on the 
type of a common field, which is crucially enabled by the 
use of the theory of finite maps in our setting.
Recent work includes type systems for JavaScript and Ruby.
\cite{Drossopoulou05} presents a rich type system and 
inference algorithm for JavaScript, which uses row-types 
and width subtyping to model dictionaries (objects). 
The system does not support unions, and uses fixed 
field names. This issue is addressed in \cite{Thiemann05},
which models dictionaries using row types labeled by
singletons indexed by string constants, and depth subtyping. 
A recent proposal \cite{Zhao10} incorporates an
initialization phase during which object types can be updated.
However, these systems preclude truly dynamic dictionaries, 
which require dependent types, and moreover lack the 
control flow analysis required to support ad-hoc unions.
DRuby \cite{Foster09} is a powerful type system designed
to support Ruby code that mixes intersections, unions, 
classes, and parametric polymorphism. DRuby supports 
``duck typing," by converting from nominal
to structural types appropriately. However, it does not
support ad-hoc unions or dynamic dictionary accesses.

\parahead{Dependent Types and SMT Solvers} 
The observation that ad-hoc unions can be checked 
via dependent types is not new.
\cite{Komondoor05} develops a dependent type system called
guarded types that is used to describe records and ad-hoc 
unions in legacy Cobol programs that make extensive use of 
tag-tests, where the ``tag" is simply the first few bytes 
of a structure. 
\cite{JMX07} presents an SMT-based system for statically 
inferring dependent types that verify the safety of ad-hoc 
unions in legacy C programs. 
\cite{Condit09} describes how type-checking and property
verification are two sides of the same coin for C (which 
is essentially uni-typed.) It develops a precise logic-based
type system for C and shows how SMT solvers can be used 
for type-checking.
\cite{dminor} uses refinement types to formalize similar 
ideas in the context of Dminor, a first-order functional
data description language with fixed-key records and 
run-time tag-tests. The authors show how unions and 
intersections can be expressed in refinements (and 
even collections, via recursive functions), and hence
how SMT solvers can wholly discharge all subtyping 
obligations.
However, the above techniques apply only to first-order languages,
with static keys and dictionaries over base values.

\paragraph{Combining Decision Procedures.}
Our approach of combining logical reasoning by SMT solvers
and syntactic reasoning by subtyping is reminiscent
of work on combining decision procedures \cite{NelsonOppen, Shostak84}.
However, such techniques require the theories being combined to be disjoint;
since our logic includes type terms which
themselves contain arbitrary terms, our theory of
syntactic types cannot be separated from the other theories in our
system, so these techniques cannot be directly applied.



\section{Conclusions and Future Work}

We have shown how, by nesting type predicates within refinement
formulas and carefully interleaving syntactic-  and SMT-based 
subtyping, \dtypes can statically type check dynamic programs that 
manipulate dictionaries, polymorphic higher-order 
functions and containers.
Thus, we believe that \dtypes can be a foundation for two distinct
avenues of research: the addition of heterogeneous dictionaries 
to static languages like C\#, Java, OCaml and Haskell, or dually, 
the addition of expressive static typing to dynamic languages 
like Clojure, JavaScript, Racket, and Ruby.

We anticipate several concrete lines of work 
that are needed to realize the above goals.
First, we need to add support for references and imperative
update, features common to most popular dynamic languages.
Since every dictionary operation in an imperative language
goes through a reference, we will need to
extend the type system with flow-sensitive analyses,
as in \cite{LiquidPOPL10} and \cite{lambdajs},
to precisely track the values stored in reference
cells at each program point.
Furthermore, to precisely track updates to dictionaries in
the imperative setting, we will likely need to introduce some
flow-sensitivity to the type system itself, adopting strong update
techniques as in \cite{Thiemann10} and \cite{Zhao10}.
Second, our system treats strings as atomic constants. 
Instead, it should be possible to incorporate modern 
decision procedures for strings \cite{HooimeijerV11} 
to support logical operations on keys, which would give 
even more precise support for reflective metaprogramming.
Third, we plan to extend our local inference techniques
to automatically derive polymorphic instantiations
\cite{pierce-turner} and use Liquid Types \cite{LiquidPLDI08} to
globally infer refinement types. Finally, for dynamic languages,
it would be useful to incorporate some form of staged 
analysis to support dynamic code generation \cite{sif,An11}.


\bibliographystyle{abbrvnat}
\bibliography{sw}

\begin{thebibliography}{35}
\providecommand{\natexlab}[1]{#1}
\providecommand{\url}[1]{\texttt{#1}}
\expandafter\ifx\csname urlstyle\endcsname\relax
  \providecommand{\doi}[1]{doi: #1}\else
  \providecommand{\doi}{doi: \begingroup \urlstyle{rm}\Url}\fi

\bibitem[Nes()]{NestedTR}
Supplemental materials.

\bibitem[Abadi et~al.(1989)Abadi, Cardelli, Pierce, and Plotkin]{Abadi89}
M.~Abadi, L.~Cardelli, B.~C. Pierce, and G.~Plotkin.
\newblock Dynamic typing in a statically-typed language.
\newblock In \emph{POPL}, 1989.

\bibitem[An et~al.(2011)An, Chaudhuri, Foster, and Hicks]{An11}
J.-h.~D. An, A.~Chaudhuri, J.~S. Foster, and M.~Hicks.
\newblock Dynamic inference of static types for ruby.
\newblock In \emph{POPL}, 2011.

\bibitem[Anderson et~al.(2005)Anderson, Drossopoulou, and
  Giannini]{Drossopoulou05}
C.~Anderson, S.~Drossopoulou, and P.~Giannini.
\newblock {Towards Type Inference for JavaScript}.
\newblock In \emph{ECOOP}, pages 428--452, June 2005.

\bibitem[Bierman et~al.(2010)Bierman, Gordon, Hritcu, and Langworthy]{dminor}
G.~M. Bierman, A.~D. Gordon, C.~Hritcu, and D.~E. Langworthy.
\newblock Semantic subtyping with an smt solver.
\newblock In \emph{ICFP}, 2010.

\bibitem[Chugh et~al.(2009)Chugh, Meister, Jhala, and Lerner]{sif}
R.~Chugh, J.~A. Meister, R.~Jhala, and S.~Lerner.
\newblock Staged information flow for javascript.
\newblock In \emph{Proceedings of PLDI 2009}, pages 50--62, 2009.

\bibitem[Condit et~al.(2009)Condit, Hackett, Lahiri, and Qadeer]{Condit09}
J.~Condit, B.~Hackett, S.~K. Lahiri, and S.~Qadeer.
\newblock Unifying type checking and property checking for low-level code.
\newblock In \emph{POPL}, 2009.

\bibitem[de~Moura and Bj{\o}rner(2008)]{z3}
L.~de~Moura and N.~Bj{\o}rner.
\newblock Z3: An efficient {SMT} solver.
\newblock In \emph{TACAS}, 2008.

\bibitem[de~Moura and Bj{\o}rner(2009)]{arrayz3}
L.~de~Moura and N.~Bj{\o}rner.
\newblock Generalized, efficient array decision procedures.
\newblock In \emph{FMCAD}, pages 45--52, 2009.

\bibitem[Findler and Felleisen(2002)]{Findler02}
R.~B. Findler and M.~Felleisen.
\newblock Contracts for higher-order functions.
\newblock In \emph{ICFP}, pages 48--59, 2002.

\bibitem[Flanagan(2006)]{flanagan06}
C.~Flanagan.
\newblock Hybrid type checking.
\newblock In \emph{POPL}. ACM, 2006.

\bibitem[Furr et~al.(2009)Furr, hoon (David)~An, Foster, and Hicks]{Foster09}
M.~Furr, J.~hoon (David)~An, J.~S. Foster, and M.~W. Hicks.
\newblock Static type inference for ruby.
\newblock In \emph{SAC}, pages 1859--1866, 2009.

\bibitem[Guha et~al.(2011)Guha, Softoiu, and Krishnamurthi]{lambdajs}
A.~Guha, C.~Softoiu, and S.~Krishnamurthi.
\newblock Typing local control and state using flow analysis.
\newblock In \emph{ESOP}, 2011.

\bibitem[Heidegger and Thiemann(2010)]{Thiemann10}
P.~Heidegger and P.~Thiemann.
\newblock Recency types for analyzing scripting languages.
\newblock In \emph{ECOOP}, pages 200--224, 2010.

\bibitem[Hooimeijer and Veanes(2011)]{HooimeijerV11}
P.~Hooimeijer and M.~Veanes.
\newblock An evaluation of automata algorithms for string analysis.
\newblock In \emph{VMCAI}, pages 248--262, 2011.

\bibitem[Jhala et~al.(2007)Jhala, Majumdar, and Xu]{JMX07}
R.~Jhala, R.~Majumdar, and R.-G. Xu.
\newblock State of the union: Type inference via craig interpolation.
\newblock In \emph{TACAS}, 2007.

\bibitem[Kennedy and Pierce(2007)]{Kennedy06}
A.~J. Kennedy and B.~C. Pierce.
\newblock On decidability of nominal subtyping with variance.
\newblock In \emph{FOOL-WOOD}, 2007.

\bibitem[Knowles and Flanagan(2010)]{Knowles10}
K.~Knowles and C.~Flanagan.
\newblock Hybrid type checking.
\newblock \emph{ACM TOPLAS}, 32\penalty0 (2), 2010.

\bibitem[Komondoor et~al.(2005)Komondoor, Ramalingam, Chandra, and
  Field]{Komondoor05}
R.~Komondoor, G.~Ramalingam, S.~Chandra, and J.~Field.
\newblock Dependent types for program understanding.
\newblock In \emph{TACAS}, pages 157--173, 2005.

\bibitem[Kr{\"o}ning et~al.(2009)Kr{\"o}ning, R{\"u}mmer, and
  Weissenbacher]{KroeningFiniteMaps}
D.~Kr{\"o}ning, P.~R{\"u}mmer, and G.~Weissenbacher.
\newblock A proposal for theory of finite sets, lists, and maps for the smt-lib
  standard.
\newblock In \emph{SMT}, 2009.

\bibitem[Mc{C}arthy(1962)]{Mccarthy}
J.~Mc{C}arthy.
\newblock Towards a mathematical science of computation.
\newblock In \emph{In IFIP Congress}, pages 21--28. North-Holland, 1962.

\bibitem[Nelson and Oppen(1979)]{NelsonOppen}
G.~Nelson and D.~C. Oppen.
\newblock Simplification by cooperating decision procedures.
\newblock \emph{TOPLAS}, 1979.

\bibitem[Ou et~al.(2004)Ou, Tan, Mandelbaum, and Walker]{Ou2004}
X.~Ou, G.~Tan, Y.~Mandelbaum, and D.~Walker.
\newblock Dynamic typing with dependent types.
\newblock In \emph{IFIP TCS}, pages 437--450, 2004.

\bibitem[Palsberg and Schwartzbach(1994)]{PalsbergBook}
J.~Palsberg and M.~I. Schwartzbach.
\newblock \emph{OO Type Systems}.
\newblock Wiley, 1994.

\bibitem[Pierce and Turner(1998)]{pierce-turner}
B.~C. Pierce and D.~N. Turner.
\newblock Local type inference.
\newblock In \emph{POPL}, pages 252--265, 1998.

\bibitem[R\'{e}my(1989)]{Rem89}
D.~R\'{e}my.
\newblock Type checking records and variants in a natural extension of ml.
\newblock In \emph{POPL}, 1989.

\bibitem[Rondon et~al.(2008)Rondon, Kawaguchi, and Jhala]{LiquidPLDI08}
P.~Rondon, M.~Kawaguchi, and R.~Jhala.
\newblock Liquid types.
\newblock In \emph{PLDI}, 2008.

\bibitem[Rondon et~al.(2010)Rondon, Kawaguchi, and Jhala]{LiquidPOPL10}
P.~Rondon, M.~Kawaguchi, and R.~Jhala.
\newblock Low-level liquid types.
\newblock In \emph{POPL}, pages 131--144, 2010.

\bibitem[Shostak(1984)]{Shostak84}
R.~Shostak.
\newblock Deciding combinations of theories.
\newblock \emph{Journal of the ACM}, 31\penalty0 (1):\penalty0 1--12, 1984.

\bibitem[Siek and Taha(2006)]{Sie06}
J.~Siek and W.~Taha.
\newblock Gradual typing for functional languages.
\newblock In \emph{Scheme and Functional Programming Workshop}, 2006.

\bibitem[{The Dojo Foundation}()]{dojo-js}
{The Dojo Foundation}.
\newblock Dojo toolkit.
\newblock \url{http://dojotoolkit.org/}.

\bibitem[{The Python Software Foundation}()]{python-32}
{The Python Software Foundation}.
\newblock Python 3.2 standard library.
\newblock \url{http://python.org/}.

\bibitem[Thiemann(2005)]{Thiemann05}
P.~Thiemann.
\newblock Towards a type system for analyzing javascript programs.
\newblock In \emph{ESOP}, 2005.

\bibitem[Tobin-Hochstadt and Felleisen(2010)]{typedracket}
S.~Tobin-Hochstadt and M.~Felleisen.
\newblock Logical types for untyped languages.
\newblock In \emph{ICFP}, pages 117--128, 2010.

\bibitem[Zhao(2010)]{Zhao10}
T.~Zhao.
\newblock Type inference for scripting languages with implicit extension.
\newblock In \emph{FOOL}, 2010.

\end{thebibliography}

\ifLongVersion
{ \clearpage
  \appendix



\section{Metatheory}
\label{sec:metatheory}



This section deals with the formal properties of \dntypes.
First, we provide some definitions that were omitted from
the presentation of \dtypes in
Sections~\ref{sec:syntax-and-semantics} and \ref{sec:type-checking}.
Next, we provide the complete definitions of stratified \dntypes.
Finally, we specify the assumptions and definitions specific to our
refinement logic, and present the details of the proof.
%
%
Compared to the proof outline in \autoref{sec:soundness}, we prove
the progress and preservation parts of \dntypes Type Soundness
together, rather than with separate progress and Stratified Preservation
theorems.


\subsection{Additional \dtypes Definitions}

\subsubsection{Operational Semantics}

The small-step operational semantics of
\dtypes expressions is parametrized on a function $\delta$ that defines the
behavior of constants $c$ that are functions.
Dictionary operations like $\vHas$, $\vGet$, and $\vUpd$ are factored
into the $\delta$ function.
As terms are in A-normal form, there is a single congruence rule, $\ruleName{E-Compat}$.

{\centering
  \vsepRule
  \judgementHead{}{$\reducesTo{e}{e'}$}
  

$\inferrule*[right=\ruleNameFig{E-Delta}]
  {\textrm{if } \deltaApp{c}{\varVal} \textrm{ is defined}}
  {\reducesTo{c\ \varVal}{\deltaApp{c}{\varVal}}}
$

\vsepRule

$\inferrule*[right=\ruleNameFig{E-App}]
  {}
  {\reducesTo{(\funBare{x}{e})\ \varVal}{\subst{e}{x}{\varVal}}}
$

\vsepRule

$\inferrule*[right=\ruleNameFig{E-Let}]
  {}
  {\reducesTo{\letinBare{x}{\varVal}{e}}{\subst{e}{x}{\varVal}}}
$

\vsepRule

$\inferrule*[right=\ruleNameFig{E-TApp}]
  {}
  {\reducesTo{\typInst{(\typFun{\tyVar}{e})}{\varTyp}}{e}}
$

\vsepRule

$\inferrule*[right=\ruleNameFig{E-IfTrue}]
  {}
  {\reducesTo{\ite{\vTrue}{e_1}{e_2}}{e_1}}
$

\vsepRule

$\inferrule*[right=\ruleNameFig{E-IfFalse}]
  {}
  {\reducesTo{\ite{\vFalse}{e_1}{e_2}}{e_2}}
$

\vsepRule

$\inferrule*[right=\ruleNameFig{E-Compat}]
  {\reducesTo{e_1}{e_1'}}
  {\reducesTo{\letinBare{x}{e_1}{e_2}}
             {\letinBare{x}{e_1'}{e_2}}}
$


}

\subsubsection{Well-formedness}

We briefly supplement our discussion in \autoref{sec:type-checking}.

\parahead{Refinement Types}
The well-formedness of formulas does \emph{not} depend
on type checking; all that is needed is the ability to syntactically 
distinguish between terms and propositions.
We omit the straightforward rules for well-formed values. 
The important point is that a variable $x$ may be used only 
if it is (bound) in $\Gamma$.
Since our refinement logic is unsorted,
all logical predicate and function symbols must be defined for
\emph{all} values in any model of the logic.
Thus,
ill-typed expressions like $\formTrue + \formFalse$ may evaluate
to nonstandard ``error'' values in such models.
This means that, for example,
$\refTyp{\theV > 0}$ is not the same as
$\refTyp{\tagof{\theV} = \tagInt \wedge \theV > 0}$
since the former may also include non-integer values.
Such values never arise at run-time, as the types of our primitive
operations and constants guarantee that they only consume and produce
standard, non-error values.

\parahead{Datatype Definitions}
To enable a sound subtyping rule for constructed types in the
sequel, we check that the declared variance annotations are
respected within the type definition. The $\varianceOkName$
predicate is defined as
\begin{align*}
\varianceOk{\tyVar}{\poleCo}{\seq{\varTyp}} \quad & \text{iff} \quad
   (\cup_j\ \poles{\tyVar}{\poleCo}{\varTyp_j}) \subseteq \set{\poleCo} \\
\varianceOk{\tyVar}{\poleContra}{\seq{\varTyp}} \quad & \text{iff} \quad
   (\cup_j\ \poles{\tyVar}{\poleCo}{\varTyp_j}) \subseteq \set{\poleContra} \\
\varianceOk{\tyVar}{\poleBi}{\seq{\varTyp}} \quad & \text{always}
\end{align*}

\noindent
where $\polesName$ is a helper procedure that
recursively walks formulas, type terms, and types to record
where type variables occur within the types of the fields.
$\poles{\tyVar}{\poleCo}{\varTyp}$ computes a subset of
$\set{\poleCo,\poleContra}$ that includes $\poleCo$
(resp. $\poleContra$) if $\tyVar$ occurs in at least one
positive (resp. negative) position inside $\varTyp$.
For each type variable, these polarities are computed across
all field types in the definition and then checked against its
variance annotation.
After successfully checking that a type definition is well-formed,
it is added to the globally-available type definition environment 
$\lookupConName$.
For example, when checking the well-formedness of the type term
$\typeCon{\varCon}{\seq{\varTyp}}$, we make sure that $\varCon$ is defined
by testing for its presence in $\lookupConName$.

\[
\neg\varPole =
  \left\{
     \begin{array}{lr}
       \poleContra & \textrm{if } \varPole=\poleCo \\
       \poleCo     & \textrm{if } \varPole=\poleContra
     \end{array}
   \right.
\]
\begin{align*}
\poles{\tyVar}{\varPole}{\refTyp{\varFormOne}} &=
  \poles{\tyVar}{\varPole}{\varFormOne} \\[7pt]
\poles{\tyVar}{\varPole}{\lprimP(\seq{\varLogVal})} &= \emptyset \\
\poles{\tyVar}{\varPole}{\synHastyp{\varLogVal}{\varUnTyp}} &=
  \poles{\tyVar}{\varPole}{\varUnTyp} \\
\poles{\tyVar}{\varPole}{\varFormOne\wedge\varFormTwo} &=
  \poles{\tyVar}{\varPole}{\varFormOne} \cup
  \poles{\tyVar}{\varPole}{\varFormTwo} \\
\poles{\tyVar}{\varPole}{\varFormOne\vee\varFormTwo} &=
  \poles{\tyVar}{\varPole}{\varFormOne} \cup
  \poles{\tyVar}{\varPole}{\varFormTwo} \\
\poles{\tyVar}{\varPole}{\neg\varFormOne} &=
  \poles{\tyVar}{\neg\varPole}{\varFormOne} \\[7pt]
\poles{\tyVar}{\varPole}{\tyVar} &= \set{\varPole} \\
\poles{\tyVar}{\varPole}{\tyVarB} &= \emptyset \\
\poles{\tyVar}{\varPole}{\utArrow{x}{\varTyp_1}{\varTyp_2}} &=
   \poles{\tyVar}{\neg\varPole}{\varTyp_1} \cup
   \poles{\tyVar}{\varPole}{\varTyp_2} \\
\poles{\tyVar}{\varPole}{\nullCon} &= \emptyset \\
\poles{\tyVar}{\varPole}{\typeCon{\varCon}{\seq{\varTyp}}} &= \cup_i
  \left\{
     \begin{array}{ll}
       \poles{\tyVar}{\varPole}{\varTyp_i} &
          \textrm{if } \varPole_i=\poleCo \\
       \poles{\tyVar}{\neg\varPole}{\varTyp_i} &
          \textrm{if } \varPole_i=\poleContra \\ 
       \poles{\tyVar}{\poleCo}{\varTyp_i} & \\
       \hspace{0.07in} \cup\ \poles{\tyVar}{\poleContra}{\varTyp} &
          \textrm{if } \varPole_i=\poleBi
     \end{array}
   \right.
\end{align*}
In the last case of this definition,
$\lookupConAnnots{\varCon}{\seq{\varPole\tyVarB}}$.


\UseIndicestrue

\UseIndicestrue

\subsection{Stratified \dntypes}

The complete definition of the \dntypes typing and subtyping relations
in Figures~\ref{fig:dec-typing-indexed} and \ref{fig:dec-subtyping-indexed}.
The only differences compared to the base system are that all typing and
subtyping derivations are now indexed with an integer $n$, and the clause
implication relation contains the new $\ruleName{C-Valid-n}$ rule.
The well-formedness relations remain unchanged.


\subsection{Definitions and Assumptions}

We often use the following abbreviations for
types and substitution into types.
\begin{align*}
  \refTypShort{\varFormOne} \defeq &\ \refTyp{\varFormOne} \\
  \varFormOne(\varLogVal) \defeq &\ \subst{\varFormOne}{\theV}{\varLogVal} \\
  \embed{\varTyp}(\varLogVal) \defeq &\ \subst{\embed{\varTyp}}{\theV}{\varLogVal} \\
\end{align*}

\begin{prop-nonum}[Refinement Logic]
The refinement logic underlying the type system at level zero
is the quantifier-free fragment of
first-order logic with equality and the decidable theories listed
below.
Logical terms of a universal sum sort called Val include integers,
booleans, strings, and dictionaries (finite maps from strings to values).
Expressions, formulas and type terms can be encoded in the logic as
uninterpreted constructed terms.
Function and type function terms are pairs of
formal parameters and expression terms.

\begin{itemize}

\item (Theory: Uninterpreted Functions)
\spaceCase

\item (Theory: Linear Arithmetic)
\spaceCase

\spaceCase
\item (Theory: Dictionaries)
\spaceCase

We use the following axiomatization of dictionaries that can be
reduced to the theory of finite maps \cite{KroeningFiniteMaps}.
\vsepRule


$\forall \varVal.$

\begin{itemize}
\item $\neg\hasR{\emptydictR}{\varVal}$
\end{itemize}

$\forall \varVal_1,\varVal_2,\varVal_3.$

\begin{itemize}
\item
$\hasR{\dictextend{\varVal_1}{\varVal_2}{\varVal_3}}{\varVal_2}$
\item
$\selR{\dictextend{\varVal_1}{\varVal_2}{\varVal_3}}{\varVal_2}=\varVal_3$
\item
$\havocR{\dictextend{\varVal_1}{\varVal_2}{\varVal_3}}{\varVal_1}{\varVal_2}$
\end{itemize}

$\forall \varVal_1,\varVal_2,x,y.$

\begin{itemize}
\item
$\havocR{\varVal_1}{\varVal_2}{x}\wedge x\neq y \Rightarrow 
 (\hasR{\varVal_1}{y}\Leftrightarrow\hasR{\varVal_2}{y})$
\item
$\havocR{\varVal_1}{\varVal_2}{x}\wedge x\neq y \Rightarrow 
 (\selR{\varVal_1}{y}=\selR{\varVal_2}{y})$
\end{itemize}


\spaceCase
\item (Assumption: Tag Function)
\spaceCase

We assume the presence of a unary function symbol $\tagofOp$ that maps
values to strings.

\begin{displaymath}
\begin{array}{rcl}
    \tagof{\vTrue}              & = & \tagBool \\
    \tagof{\vFalse}             & = & \tagBool \\
    \tagof{n}                   & = & \tagInt \\
    \tagof{\funBare{x}{e}}      & = & \tagFun \\
    \tagof{\typFun{\tyVar}{e}}  & = & \tagTFun \\
    \tagof{\dictextend{\varVal_1}{\varVal_2}{\varVal_3}}
                                & = & \tagRecd \\
    \tagof{\newDataBare{\varCon}{\seq{\varVal}}}
                                & = & \tagRecd \\
    \tagof{c}                   & = & \tagFun \textrm{ if } c \textrm{ is a function}
\end{array}
\end{displaymath}

\spaceCase
\item (Fact: Validity)
\spaceCase

We write $\valid{\varFormOne}$ to mean that, as usual,
$\varFormOne$ is satisfiable in all interpretations.
In the $\ruleName{C-Valid}$ rule, we appeal to a decision
procedure to check whether $\valid{\varFormOne}$.

\spaceCase
\item (Assumption: Boolean Values)
\spaceCase

We assume
$\valid{\tagof{\varVal}=\tagBool}$ iff $\varVal\in\set{\vTrue,\vFalse}$.

\spaceCase
\item (Fact: Free Variable Substitution)
\spaceCase

If $\theV$ appears free in $\varFormOne$ and $\varFormTwo$,

then $\varFormOne\Rightarrow\varFormTwo$ implies
$\substwx{\varFormOne}\Rightarrow\substwx{\varFormTwo}$
for all $\varVal$.

\spaceCase
\item (Fact: Uninterpreted Predicate Substitution)
\spaceCase

If $\lprimP$ is an uninterpreted predicate symbol in
$\varFormOne$ and $\varFormTwo$,

then $\varFormOne\Rightarrow\varFormTwo$ implies
$\subst{\varFormOne}{\lprimP}{\lprimP'}\Rightarrow
 \subst{\varFormTwo}{\lprimP}{\lprimP'}$ for all $\lprimP'$.

\end{itemize}

\end{prop-nonum}


\begin{ass-nonum}[Constant Types]
For every constant $c\in Dom(ty)$, the following properties hold.
\begin{enumerate}

\item (Well-formed).
      \hspace{0.2in}
      $\relWf{}{\typConst{c}}$.

\spaceCase
\item (Normal).

      \hspace{0.7in}
      $\typConst{c}={\refTyp{\theV=c\wedge p}}$ where either

      \hspace{0.7in}
      $p=\formTrue$ or

      \hspace{0.7in}
      $p=\synHastyp{\theV}{\utArrow{x}{\varTyp_{1}}{\varTyp_{2}}}$.

\spaceCase
\item (App).

      \hspace{0.7in}
      if $\typConst{c}=
            \refTyp{\theV=c\wedge
                    \synHastyp{\theV}{\utArrow{x}{\varTyp_{1}}{\varTyp_{2}}}}$,

      \hspace{0.7in}
      then for all $\varVal'$ and $n$ such that
        $\relTyp{}{\varVal'}{\varTyp_{1}}{n}$,

      \hspace{0.7in}
      $\deltaApp{c}{\varVal'}$ is defined and
      $\relTyp{}{\deltaApp{c}{\varVal'}}{\subst{\varTyp_{2}}{x}{\varVal'}}{n}$.

\spaceCase
\item (Valid).

      \hspace{0.7in}
      $\valid{\subst{\typConst{c}}{\theV}{c}}$.

      \hspace{0.7in}
      In other words, we add these to the initial

      \hspace{0.9in}
      typing environment.

\end{enumerate}
\end{ass-nonum}


\begin{definition-nonum}[Type Predicate Interpretation]
The System D Interpretation at level $n$ interprets type predicates
as follows.

\begin{itemize}

\item

$\formulaSat{n}{\synHastyp{\varVal}{\utArrow{x}{\varTyp_{11}}{\varTyp_{12}}}}$
  iff $\relWf{\ }{\utArrow{x}{\varTyp_{11}}{\varTyp_{12}}}$ and either:

  \begin{enumerate}

  \item
    $\varVal=\funBare{x}{e}$ and

    $\relTyp{\tyBind{x}{\varTyp_{11}}}{e}{\varTyp_{12}}{n-1}$; or

  \item

    $\varVal=c$,

    $\typConst{c}=
       \refTyp{\theV=c\wedge
               \synHastyp{\theV}{\utArrow{x}{\varTyp_{01}}{\varTyp_{02}}}}$, and

    $\relSynSub{}{\utArrow{x}{\varTyp_{01}}{\varTyp_{02}}}
                 {\utArrow{x}{\varTyp_{11}}{\varTyp_{12}}}{n-1}$.

  \end{enumerate}

\spaceCase
\item

$\formulaSat{n}{\synHastyp{\varVal}{\tyVar}}$ never.

\spaceCase
\item

$\formulaSat{n}{\synHastyp{\varVal}{\nullCon}}$ iff $\varVal=\vNull$.

\spaceCase
\item

$\formulaSat{n}{\synHastyp{\varVal}{\typeCon{\varCon}{\seq{\varTyp}}}}$ 
  iff $\relWf{\ }{\seq{\varTyp}}$,
      $\lookupCon{\varCon}{\seq{\varPole\tyVar}}{f}{\varTyp'}$,
      and either:

  \begin{enumerate}

  \item $\varVal=\vNull$; or

  \item
    $\varVal=\newDataBare{\varCon}{\seq{\varVal}}$ and

    for all j,
    $\relTyp{\ }{\varVal_j}{\inst{\varTyp'_j}{\seq{\tyVar}}{\seq{\varTyp}}}{n-1}$.

  \end{enumerate}

\end{itemize}

\end{definition-nonum}


\begin{ass-nonum}[Datatype Representation]
This assumption requires that the implementation treats constructed
data just like ordinary dictionaries.
Let $\lookupCon{\varCon}{\seq{\varPole\tyVar}}{f}{\varTyp'}$.

\vsepRule
 
\makebox[0.25in][l]{If}
   $\formulaSat{n}{\synHastyp{\varVal}{\typeCon{\varCon}{\seq{\varTyp}}}}$,

\makebox[0.25in][l]{then}
   $\formulaSat{n}{\tagof{\varVal}=\tagRecd}$

\makebox[0.25in][l]{and}
   $\formulaSat{n}
       {\wedge_j \embed{\inst{\varTyp'_j}{\seq{\tyVar}}{\seq{\varTyp}}}
                       (\selR{\varVal}{f_j})}$.

\end{ass-nonum}


\subsection{Formal Properties}

\separatingLine

\noindent
To reduce clutter, we elide the well-formedness requirements of
all expressions, formulas, types, type terms, typing environments, and
type definitions mentioned in the lemmas and theorems that follow.


\begin{lemma}[Inversion]
\mbox{}

\begin{enumerate}
\item If   $\relSynSub{\Gamma}{\utArrow{x}{\varTyp_{11}}{\varTyp_{12}}}
                              {\utArrow{x}{\varTyp_{21}}{\varTyp_{22}}}{n}$,

      then $\relSub{\Gamma}{\varTyp_{21}}{\varTyp_{11}}{n}$
      and  $\relSub{\Gamma,\tyBind{x}{\varTyp_{21}}}
                   {\varTyp_{12}}{\varTyp_{22}}{n}$.

\item If   $\relTyp{\Gamma}{\funBare{x}{e}}{\varScm}{n}$,

      then $\relTyp
            {\Gamma}
            {\funBare{x}{e}}
            {\refTypShort{\theV=\funBare{x}{e}\wedge
                                 \synHastyp{\theV}
                                           {\utArrow{x}{\varTyp_1}{\varTyp_2}}}}
            {n}$

\end{enumerate}
\end{lemma}

\begin{proof}
By induction.
Note that we have only listed the properties we will need to use.
\end{proof}


\begin{lemma}[Reflexive Subtyping]
\mbox{}
\begin{enumerate}
\item $\relImpl{\Gamma}{\varFormOne}{\varFormOne}{n}$
\item $\relSynSub{\Gamma}{\varUnTyp}{\varUnTyp}{n}$
\item $\relSub{\Gamma}{\varScm}{\varScm}{n}$
\end{enumerate}
\end{lemma}

\begin{proof}
By mutual induction.
\end{proof}


\begin{lemma}[Transitive Subtyping]
\mbox{}
\begin{enumerate}
\alignItemThree{0.95in}{1.10in}
  {If   $\relImpl{\Gamma}{\varFormOne}{\varFormTwo}{n}$}
  {and  $\relImpl{\Gamma}{\varFormTwo}{\varFormThr}{n}$,}
  {then $\relImpl{\Gamma}{\varFormOne}{\varFormThr}{n}$.}
\alignItemThree{0.95in}{1.10in}
  {If   $\relSynSub{\Gamma}{\varUnTyp_1}{\varUnTyp_2}{n}$}
  {and  $\relSynSub{\Gamma}{\varUnTyp_2}{\varUnTyp_3}{n}$,}
  {then $\relSynSub{\Gamma}{\varUnTyp_1}{\varUnTyp_3}{n}$.}
\alignItemThree{0.95in}{1.10in}
  {If   $\relSub{\Gamma}{\varScm_1}{\varScm_2}{n}$}
  {and  $\relSub{\Gamma}{\varScm_2}{\varScm_3}{n}$,}
  {then $\relSub{\Gamma}{\varScm_1}{\varScm_3}{n}$.}
\end{enumerate}
\end{lemma}

\begin{proof}
By mutual induction.
\end{proof}


\begin{lemma}[Narrowing]
Suppose $\relSub{\Gamma}{\varScm}{\varScm'}{n}$.
\begin{enumerate}
\alignItemTwo{1.30in}
  {If   $\relImpl{\Gamma,\tyBind{x}{\varScm'}}{\varFormOne}{\varFormTwo}{n}$,}
  {then $\relImpl{\Gamma,\tyBind{x}{\varScm}}{\varFormOne}{\varFormTwo}{n}$.}
\alignItemTwo{1.30in}
  {If   $\relSynSub{\Gamma,\tyBind{x}{\varScm'}}{\varUnTyp_1}{\varUnTyp_2}{n}$,}
  {then $\relSynSub{\Gamma,\tyBind{x}{\varScm}}{\varUnTyp_1}{\varUnTyp_2}{n}$.}
\alignItemTwo{1.30in}
  {If   $\relSub{\Gamma,\tyBind{x}{\varScm'}}{\varScm_1}{\varScm_2}{n}$,}
  {then $\relSub{\Gamma,\tyBind{x}{\varScm}}{\varScm_1}{\varScm_2}{n}$.}
\alignItemTwo{1.30in}
  {If   $\relTyp{\Gamma,\tyBind{x}{\varScm'}}{e}{\varScm_1}{n}$,}
  {then $\relTyp{\Gamma,\tyBind{x}{\varScm}}{e}{\varScm_1}{n}$.}
\end{enumerate}
\end{lemma}

\begin{proof}
By mutual induction.
\end{proof}


\begin{lemma}[Weakening]
Suppose $\Gamma=\Gamma_1,\Gamma_2$ and $\Gamma'$ is such that
$\relWf{}{\Gamma'}$ and either

\begin{center}
$\Gamma'=\Gamma_1,\tyBind{x}{\varScm},\Gamma_2$
\sep or \sep
$\Gamma'=\Gamma_1,\varFormOne,\Gamma_2$
\sep or \sep
$\Gamma'=\Gamma_1,\tyVar,\Gamma_2$.
\end{center}

\begin{enumerate}
\alignItemTwo{1.00in}
  {If   $\relImpl{\Gamma}{\varFormTwo_1}{\varFormTwo_2}{n}$,}
  {then $\relImpl{\Gamma'}{\varFormTwo_1}{\varFormTwo_2}{n}$.}
\alignItemTwo{1.00in}
  {If   $\relSynSub{\Gamma}{\varUnTyp_1}{\varUnTyp_2}{n}$,}
  {then $\relSynSub{\Gamma'}{\varUnTyp_1}{\varUnTyp_2}{n}$.}
\alignItemTwo{1.00in}
  {If   $\relSub{\Gamma}{\varScm_1}{\varScm_2}{n}$,}
  {then $\relSub{\Gamma'}{\varScm_1}{\varScm_2}{n}$.}
\alignItemTwo{1.00in}
  {If   $\relTyp{\Gamma}{e}{\varScm}{n}$,}
  {then $\relTyp{\Gamma'}{e}{\varScm}{n}$.}
\end{enumerate}
\end{lemma}

\begin{proof}
By mutual induction.
\end{proof}



\begin{lemma}[Free Variables in Subtyping]
Recall that the variable $\theV$ can appear free in the formulas, type terms,
and types mentioned in the outputs of the following derivations.
Suppose $\varLogVal$ is a closed, well-formed value.
\begin{enumerate}
\alignItemTwo{1.00in}
  {If   $\relImpl{\Gamma}{\varFormOne}{\varFormTwo}{n}$,}
  {then $\relImpl{\Gamma}{\substlwv{\varFormOne}}{\substlwv{\varFormTwo}}{n}$.}
\alignItemTwo{1.00in}
  {If   $\relSynSub{\Gamma}{\varUnTyp_1}{\varUnTyp_2}{n}$,}
  {then $\relSynSub{\Gamma}{\substlwv{\varUnTyp_1}}{\substlwv{\varUnTyp_2}}{n}$.}
\alignItemTwo{1.00in}
  {If   $\relSub{\Gamma}{\varScm_1}{\varScm_2}{n}$,}
  {then $\relSub{\Gamma}{\substlwv{\varScm_1}}{\substlwv{\varScm_2}}{n}$.}
\end{enumerate}
\end{lemma}

\begin{proof}
By mutual induction. The premise of the $\ruleName{C-Valid-n}$ case is
$\formulaSat{n}{\embed{\Gamma}\wedge\varFormOne\Rightarrow\varFormTwo}$.
Since $\theV$ appears free in the implication,
by Free Variable Substitution,
$\formulaSat{n}{\embed{\Gamma}\wedge\substlwv{\varFormOne}\Rightarrow
                                    \substlwv{\varFormTwo}}$.
Thus, by $\ruleName{C-Valid-n}$,
$\relImpl{\Gamma}{\substlwv{\varFormOne}}{\substlwv{\varFormTwo}}{n}$.
The rest of the proof is a straightforward induction.
\end{proof}


\begin{lemma}[Sound Variance]
\mbox{}
\begin{enumerate}

\item Suppose $\relSub{\Gamma}{\varTyp_1}{\varTyp_2}{n}$.

  \begin{enumerate}

  \item
    If   $\tyVarB$ appears only positively in $\varTyp$,

    then $\relSub{\Gamma}{\inst{\varTyp}{\tyVarB}{\varTyp_1}}
                         {\inst{\varTyp}{\tyVarB}{\varTyp_2}}{n}$.
  \item
    If   $\tyVarB$ appears only positively in $\varFormOne$,

    then $\relSub{\Gamma}{\refTypShort{\inst{\varFormOne}{\tyVarB}{\varTyp_1}}}
                         {\refTypShort{\inst{\varFormOne}{\tyVarB}{\varTyp_2}}}{n}$.

  \end{enumerate}

\item Suppose $\relSub{\Gamma}{\varTyp_1}{\varTyp_2}{n}$.

  \begin{enumerate}

  \item
    If   $\tyVarB$ appears only negatively in $\varTyp$,

    then $\relSub{\Gamma}{\inst{\varTyp}{\tyVarB}{\varTyp_2}}
                         {\inst{\varTyp}{\tyVarB}{\varTyp_1}}{n}$.

  \item
    If   $\tyVarB$ appears only negatively in $\varFormOne$,

    then $\relSub{\Gamma}{\refTypShort{\inst{\varFormOne}{\tyVarB}{\varTyp_2}}}
                         {\refTypShort{\inst{\varFormOne}{\tyVarB}{\varTyp_1}}}{n}$.
  \end{enumerate}

\item Suppose $\relSub{\Gamma}{\varTyp_1}{\varTyp_2}{n}$
         and  $\relSub{\Gamma}{\varTyp_2}{\varTyp_1}{n}$.

  \begin{enumerate}

  \item
    Then $\relSub{\Gamma}{\inst{\varTyp}{\tyVarB}{\varTyp_2}}
                         {\inst{\varTyp}{\tyVarB}{\varTyp_1}}{n}$

    and  $\relSub{\Gamma}{\inst{\varTyp}{\tyVarB}{\varTyp_1}}
                         {\inst{\varTyp}{\tyVarB}{\varTyp_2}}{n}$.

  \item
    Then $\relSub{\Gamma}{\refTypShort{\inst{\varFormOne}{\tyVarB}{\varTyp_2}}}
                         {\refTypShort{\inst{\varFormOne}{\tyVarB}{\varTyp_1}}}{n}$

    and  $\relSub{\Gamma}{\refTypShort{\inst{\varFormOne}{\tyVarB}{\varTyp_1}}}
                         {\refTypShort{\inst{\varFormOne}{\tyVarB}{\varTyp_2}}}{n}$.
  \end{enumerate}

\end{enumerate}
\end{lemma}

\begin{proof}
The proofs of (1) and (2) are by mutual induction on types and formulas.
The proof of (3) is a stand-alone induction on types and formulas.

\begin{proof}[Proof of (1a)]\noqedsymbol
\mbox{}

\spaceCase

Let $\varTyp=\refTyp{\varFormOne}$.

The goal follows by IH (1b), since

  \indent\indent
  $\inst{\refTyp{\varFormOne}}{\tyVarB}{\varTyp_1}=
   \refTyp{\inst{\varFormOne}{\tyVarB}{\varTyp_1}}$ and

  \indent\indent
  $\inst{\refTyp{\varFormOne}}{\tyVarB}{\varTyp_2}=
   \refTyp{\inst{\varFormOne}{\tyVarB}{\varTyp_2}}$.
\end{proof}

\begin{proof}[Proof of (1b)]\noqedsymbol
\mbox{}

\spaceCase
\caseHeadPlain{
  $\varFormOne=\varLogVal$}
\spaceCase

Trivial, since $\inst{\varFormOne}{\tyVarB}{\varTyp_1}=
                \inst{\varFormOne}{\tyVarB}{\varTyp_2}=
                \varFormOne$.

\spaceCase
\multiCaseHeadPlain{
  $\varFormOne=\varFormTwo_1\wedge\varFormTwo_2$,
  $\varFormOne=\varFormTwo_1\vee\varFormTwo_2$}
\spaceCase

By IH (1b), $\ruleName{C-Valid}$, and $\ruleName{S-Mono}$.

\spaceCase
\caseHeadPlain{
  $\varFormOne=\neg\varFormTwo$}
\spaceCase

By IH (2b), $\ruleName{C-Valid}$, and $\ruleName{S-Mono}$.

\spaceCase
\caseHeadPlain{
  $\varFormOne=\synHastyp{\varLogVal}{\varUnTyp}$}
\spaceCase

\subCaseHead{$\varUnTyp=\tyVarB$}
\spaceCase

  {\subCaseIndentation

  By definition,
  $\inst{\varFormOne}{\tyVarB}{\varTyp_1}=\embed{\varTyp_1}(\varLogVal)$.

  By definition,
  $\inst{\varFormOne}{\tyVarB}{\varTyp_2}=\embed{\varTyp_2}(\varLogVal)$.

  By Free Variables in Subtyping,
  $\relSub{\ }{\subst{\varTyp_1}{\theV}{\varLogVal}}
              {\subst{\varTyp_2}{\theV}{\varLogVal}}{n}$.

  That is,
  $\relSub{\ }{\refTypShort{\embed{\varTyp_1}(\varLogVal)}}
              {\refTypShort{\embed{\varTyp_2}(\varLogVal)}}{n}$.

  }

\spaceCase
\subCaseHead{$\varUnTyp=\utArrow{x}{\varScm_1}{\varScm_2}$}
\spaceCase

  {\subCaseIndentation
  (Note that we are using $\varScm_1$ and $\varScm_2$ for types.)

  Let
  $\varScm_{11}=\inst{\varScm_1}{\tyVarB}{\varTyp_1}$ and
  $\varScm_{12}=\inst{\varScm_1}{\tyVarB}{\varTyp_2}$.

  Let
  $\varScm_{21}=\inst{\varScm_2}{\tyVarB}{\varTyp_1}$ and
  $\varScm_{22}=\inst{\varScm_2}{\tyVarB}{\varTyp_2}$.

  Since $\tyVarB$ appears only pos in $\varTyp$, it appears only neg in $\varScm_1$,
  
    \indent\indent
    by the well-formedness of the type definition and

    \indent\indent
    the definition of $\polesName$.

  By IH (2a), $\relSub{\ }{\varScm_{12}}{\varScm_{11}}{n}$.

  Since $\tyVarB$ appears only pos in $\varTyp$, it appears only pos in $\varScm_2$.

  By IH (1a), $\relSub{\ }{\varScm_{21}}{\varScm_{22}}{n}$.

  By Weakening, $\relSub{\tyBind{x}{\varScm_{21}}}{\varScm_{21}}{\varScm_{22}}{n}$.

  By $\ruleName{U-Arrow}$,

  \indent\indent
  $\relSynSub{}{\utArrow{x}{\varScm_{11}}{\varScm_{21}}}
               {\utArrow{x}{\varScm_{12}}{\varScm_{22}}}{n}$.

  By $\ruleName{C-Valid}$,

  \indent\indent
  $\relImpl{\synHastyp{\varFormOne}{\utArrow{x}{\varScm_{11}}{\varScm_{21}}}}{\formTrue}
           {\synHastyp{\varFormOne}{\utArrow{x}{\varScm_{11}}{\varScm_{21}}}}{n}$.

  By $\ruleName{C-ImpSyn}$,

  \indent\indent
  $\relImpl{}{\synHastyp{\varFormOne}{\utArrow{x}{\varScm_{11}}{\varScm_{21}}}}
             {\synHastyp{\varFormOne}{\utArrow{x}{\varScm_{12}}{\varScm_{22}}}}{n}$.

  By $\ruleName{S-Mono}$,

  \indent\indent
  $\relSub
    {}
    {\refTypShort{\synHastyp{\varFormOne}{\utArrow{x}{\varScm_{11}}{\varScm_{21}}}}}
    {\refTypShort{\synHastyp{\varFormOne}{\utArrow{x}{\varScm_{12}}{\varScm_{22}}}}}
    {n}$.

  That is, $\relSub{\ }{\inst{\varTyp}{\tyVarB}{\varTyp_1}}
                       {\inst{\varTyp}{\tyVarB}{\varTyp_2}}{n}$.
  }

\spaceCase
\subCaseHead{$\varUnTyp=\typeCon{\varCon}{\varScm}$,
      where $\lookupConAnnots{\varCon}{\varPole\tyVar}$}
\spaceCase

  {\subCaseIndentation
  (Note that we are using $\varScm$ for a type.)

    \spaceCase
    \subSubCaseHead{$\varPole=\poleCo$}
    \spaceCase

    {\subSubCaseIndentation
     Let $\varScm_1=\inst{\varScm}{\tyVarB}{\varTyp_1}$
     and $\varScm_2=\inst{\varScm}{\tyVarB}{\varTyp_2}$.
 
     Since $\tyVarB$ appears only pos in $\varTyp$, it appears only pos in $\varScm$.
 
     By IH (1a),
     $\relSub{\ }{\varScm_1}{\varScm_2}{n}$.
 
     By $\ruleName{U-Datatype}$,
     $\relSynSub{\ }{\typeCon{\varCon}{\varScm_1}}{\typeCon{\varCon}{\varScm_2}}{n}$.

     By $\ruleName{C-Valid}$, $\ruleName{C-ImpSyn}$, and $\ruleName{S-Mono}$,
 
       \indent\indent
       $\relSub{\ }{\refTypShort{\synHastyp{\varFormOne}{\typeCon{\varCon}{\varScm_1}}}}
                   {\refTypShort{\synHastyp{\varFormOne}{\typeCon{\varCon}{\varScm_2}}}}{n}$.
 
    }

    \spaceCase
    \subSubCaseHead{$\varPole=\poleContra$}
    \spaceCase

    {\subSubCaseIndentation
    Similar.
    }

    \spaceCase
    \subSubCaseHead{$\varPole=\poleBi$}
    \spaceCase

    {\subSubCaseIndentation
    Since $\tyVarB$ appears only pos in $\varTyp$, it cannot appear in $\varScm$.

    Thus,
      $\inst{\varTyp}{\tyVarB}{\varTyp_1}=
       \inst{\varTyp}{\tyVarB}{\varTyp_2}=
       \varTyp$.
    }
  }
\end{proof}
\begin{proof}[Proof of (2a) and (2b)]\noqedsymbol
\mbox{}
Similar.
\end{proof}
\begin{proof}[Proof of (3)]\noqedsymbol
\mbox{}
Straightforward induction.
\end{proof}
\end{proof}



\begin{lemma}[Lifting]
\mbox{}
\begin{enumerate}
\alignItemTwo{1.00in}
  {If   $\relImpl{\Gamma}{\varFormOne}{\varFormTwo}{n}$,}
  {then $\relImpl{\Gamma}{\varFormOne}{\varFormTwo}{n+1}$.}
\alignItemTwo{1.00in}
  {If   $\relSynSub{\Gamma}{\varUnTyp_1}{\varUnTyp_2}{n}$,}
  {then $\relSynSub{\Gamma}{\varUnTyp_1}{\varUnTyp_2}{n+1}$.}
\alignItemTwo{1.00in}
  {If   $\relSub{\Gamma}{\varScm_1}{\varScm_2}{n}$,}
  {then $\relSub{\Gamma}{\varScm_1}{\varScm_2}{n+1}$.}
\alignItemTwo{1.00in}
  {If   $\relTyp{\Gamma}{e}{\varScm}{n}$,}
  {then $\relTyp{\Gamma}{e}{\varScm}{n+1}$.}
\alignItemTwo{1.00in}
  {If   $\formulaSat{n}{p}$,}
  {then $\formulaSat{n+1}{p}$.}
\end{enumerate}
Furthermore, for each of the first four properties, the size of the
output derivation is the same size as the original.
\end{lemma}

\begin{proof}
By mutual induction. In the \ruleName{C-Valid-n} case of (1), the conclusion
follows by \ruleName{C-Valid-n} after applying IH (5). The type predicate case for
(5) follows from IH (4).
\end{proof}


\begin{lemma}[Strengthening]
Suppose $\formulaSat{n}{p}$.
\begin{enumerate}
\alignItemTwo{1.12in}
  {If   $\relImpl{p,\Gamma}{\varFormTwo_1}{\varFormTwo_2}{n}$,}
  {then $\relImpl{  \Gamma}{\varFormTwo_1}{\varFormTwo_2}{n}$.}
\alignItemTwo{1.12in}
  {If   $\relSynSub{p,\Gamma}{\varUnTyp_1}{\varUnTyp_2}{n}$,}
  {then $\relSynSub{  \Gamma}{\varUnTyp_1}{\varUnTyp_2}{n}$.}
\alignItemTwo{1.12in}
  {If   $\relSub{p,\Gamma}{\varScm_1}{\varScm_2}{n}$,}
  {then $\relSub{  \Gamma}{\varScm_1}{\varScm_2}{n}$.}
\alignItemTwo{1.12in}
  {If   $\relTyp{p,\Gamma}{e}{\varScm}{n}$,}
  {then $\relTyp{  \Gamma}{e}{\varScm}{n}$.}
\end{enumerate}
Furthermore, for each property, the size of the output derivation is the
same size as the original.
\end{lemma}

\begin{proof}
By mutual induction.

\begin{proof}[Proof of (1)]\noqedsymbol
\mbox{}

\spaceCase
\caseHead{C-Valid-n}
  {$\inferrule*
    {\formulaSat
      {n}
      {\embed{\varFormOne,\Gamma}\wedge\varFormTwo_1\Rightarrow\varFormTwo_2}}
    {\relImpl{\varFormOne,\Gamma}{\varFormTwo_1}{\varFormTwo_2}{n}}$}
\spaceCase

By expanding the embedding,
  $\formulaSat
     {n}
     {\varFormOne\wedge\embed{\Gamma}\wedge\varFormTwo_1\Rightarrow\varFormTwo_2}$.

Thus,
  $\formulaSat
     {n}
     {\varFormOne\Rightarrow
      \embed{\Gamma}\wedge\varFormTwo_1\Rightarrow\varFormTwo_2}$.

Because of the assumption,
  $\formulaSat
     {n}
     {\embed{\Gamma}\wedge\varFormTwo_1\Rightarrow\varFormTwo_2}$.

By $\ruleName{C-Valid-n}$,
  $\relImpl{\Gamma}{\varFormTwo_1}{\varFormTwo_2}{n}$.

\spaceCase
\caseHead{C-Valid}
  {$\inferrule*
    {\valid
      {\embed{\varFormOne,\Gamma}\wedge\varFormTwo_1\Rightarrow\varFormTwo_2}}
    {\relImpl{\varFormOne,\Gamma}{\varFormTwo_1}{\varFormTwo_2}{n}}$}
\spaceCase

By Validity,
  $\formulaSat
    {n}
    {\embed{\varFormOne,\Gamma}\wedge\varFormTwo_1\Rightarrow\varFormTwo_2}$.

The rest of the reasoning in this case follows the previous case.

\spaceCase
\caseHead{C-ImpSyn}
  {$\inferrule*
    {\exists j.\ \valid{\embed{\varFormOne,\Gamma} \wedge \varFormTwo \Rightarrow
                        \synHastyp{\varLogVal_j}{\varUnTyp}}}
    {\relImpl{\varFormOne,\Gamma}
             {\varFormTwo}
             {\vee_i\ \synHastyp{\varLogVal_i}{\varUnTyp_i}}{n}}$}
\spaceCase

By \ruleName{C-Valid},
           $\relImpl{\Gamma}
                    {\varFormTwo}
                    {\synHastyp{\varLogVal_j}{\varUnTyp}}{n}$.

By IH (2), $\relSynSub{\Gamma,\varFormTwo}
                      {\varUnTyp}
                      {\varUnTyp_j}{n}$.

By $\ruleName{C-ImpSyn}$,
  $\relImpl{\Gamma}
           {\varFormTwo}
           {\vee_i\ \synHastyp{\varLogVal_i}{\varUnTyp_i}}{n}$.
\end{proof}

\begin{proof}[Proof of (2), (3), and (4)]\noqedsymbol
Straightforward induction.
\end{proof}

\end{proof}

\separatingLine


\noindent
The following lemma intuitively captures the relationship between the type
system and the underlying refinement logic: if a closed value $\varVal$ can be
given the type $\varTyp$ with a derivation at level $n$, then the formula
$\embed{\varTyp}(\varVal)$ is true in the System D Interpretation at level
$n+1$. This property plays a crucial role in the proof of Value Substitution.
Notice that nothing is said about values that are assigned polytypes.

Because the following lemma works only with the empty environment, the
Strengthening lemma is helpful for proving the $\ruleName{C-ImpSyn}$ and
$\ruleName{S-Mono}$ cases, which have premises that use non-empty
environments.

\begin{mainlemma}[Satisfiable Typing]
\mbox{}
\begin{enumerate}
\alignItemTwo{1.30in}
  {If   $\relImpl{\ }{\varFormOne}{\varFormTwo}{n}$,}
  {then $\formulaSat{n+1}{\varFormOne\Rightarrow\varFormTwo}$.}
\alignItemTwo{1.30in}
  {If   $\relSynSub{\ }{\varUnTyp_1}{\varUnTyp_2}{n}$,}
  {then $\formulaSat{n+1}{
                          \synHastyp{\theV}{\varUnTyp_1}\Rightarrow
                          \synHastyp{\theV}{\varUnTyp_2}}$.}
\alignItemTwo{1.30in}
  {If   $\relSub{\ }{\refTyp{\varFormOne}}{\refTyp{\varFormTwo}}{n}$,}
  {then $\formulaSat{n+1}{\varFormOne\Rightarrow\varFormTwo}$.}
\alignItemTwo{1.30in}
  {If   $\relTyp{\ }{\varVal}{\varTyp}{n}$,}
  {then $\formulaSat{n+1}{\embed{\varTyp}(\varVal)}$}.
\end{enumerate}
\end{mainlemma}

\noindent
In the first three properties, the variable $\theV$ appears
free in the implication. Thus, they are implicitly quantified over
\emph{all} values.

\begin{proof}
By mutual induction on the size of derivations, not by structural
induction. The reason for this induction principle is that in
the $\ruleName{C-ImpSyn}$ and $\ruleName{S-Mono}$ cases, subderivations
are manipulated by Lifting and Strengthening (which preserve derivation
size) before appealing to the induction hypothesis.

\begin{proof}[Proof of (1)]\noqedsymbol
\mbox{}

\spaceCase
\caseHead{C-Valid}
  {$\inferrule*
    {\valid{\formTrue\wedge\varFormOne\Rightarrow\varFormTwo}}
    {\relImpl{}{\varFormOne}{\varFormTwo}{n}}$}
\spaceCase

By Validity, $\formulaSat{n}{\formTrue\wedge\varFormOne\Rightarrow\varFormTwo}$,
and thus, $\formulaSat{n}{\varFormOne\Rightarrow\varFormTwo}$.

By Lifting, $\formulaSat{n+1}{\varFormOne\Rightarrow\varFormTwo}$.

\spaceCase
\caseHead{C-Valid-n}
  {$\inferrule*
    {\formulaSat{n}{\formTrue\wedge\varFormOne\Rightarrow\varFormTwo}}
    {\relImpl{}{\varFormOne}{\varFormTwo}{n}}$}
\spaceCase

By Validity and Lifting, $\formulaSat{n+1}{\varFormOne\Rightarrow\varFormTwo}$.

\spaceCase
\caseHead{C-ImpSyn}
  {$\inferrule*
    {\exists j.\ 
       \valid{\embed{\emptyset} \wedge \varFormOne \Rightarrow
              \synHastyp{\varLogVal_j}{\varUnTyp}} \sep\sep
       \relSynSub{\varFormOne}{\varUnTyp}{\varUnTyp_j}{n}}
    {\relImpl{}{\varFormOne}
               {\vee_i\ \synHastyp{\varLogVal_i}{\varUnTyp_i}}{n}}$}
\spaceCase

We assume $\formulaSat{n+1}{\varFormOne}$
and will prove $\formulaSat{n+1}{\vee_i\ \synHastyp{\varLogVal_i}{\varUnTyp_i}}$.

By $\ruleName{C-Valid}$,
  $\relImpl{\ }{\varFormOne}{\synHastyp{\varLogVal_j}{\varUnTyp}}{n}$.

By IH (1),
  $\formulaSat{n+1}
              {\varFormOne\Rightarrow\synHastyp{\varLogVal_j}{\varUnTyp}}$.

Thus, $\formulaSat{n+1}{\synHastyp{\varLogVal_j}{\varUnTyp}}$.

By Lifting, $\relSynSub{p}{\varUnTyp}{\varUnTyp_j}{n+1}$.

By Strengthening, $\relSynSub{\ }{\varUnTyp}{\varUnTyp_j}{n+1}$.

This last derivation is the same size as
  $\relSynSub{\varFormOne}{\varUnTyp}{\varUnTyp_j}{n}$

  \indent\indent
  since Lifting and Strengthening preserve derivation size.

Thus, we can apply the induction hypothesis.

By IH (2),
  $\formulaSat{n+1}
     {\synHastyp{\theV}{\varUnTyp}\Rightarrow\
      \synHastyp{\theV}{\varUnTyp_j}}$.

Thus, $\formulaSat{n+1}{\synHastyp{\varLogVal_j}{\varUnTyp_j}}$.

Thus, $\formulaSat{n+1}{\vee_i\ \synHastyp{\varLogVal_i}{\varUnTyp_i}}$.
%
\end{proof}

\begin{proof}[Proof of (2)]\noqedsymbol
\mbox{}

\spaceCase
\caseHead{U-Arrow}
  {$\inferrule*
    {\relSub{}{\varTyp_{21}}{\varTyp_{11}}{n} \sepPremise
     \relSub{\tyBind{x}{\varTyp_{21}}}{\varTyp_{12}}{\varTyp_{22}}{n}}
    {\relSynSub{}{\utArrow{x}{\varTyp_{11}}\varTyp_{12}{}}
                 {\utArrow{x}{\varTyp_{21}}\varTyp_{22}{}}{n}}$}
\spaceCase

Let $\varUnTyp_1=\utArrow{x}{\varTyp_{11}}{\varTyp_{12}}$
and $\varUnTyp_2=\utArrow{x}{\varTyp_{21}}{\varTyp_{22}}$.

We assume $\formulaSat{n+1}{\synHastyp{\theV}{\varUnTyp_1}}$
and will prove $\formulaSat{n+1}{\synHastyp{\theV}{\varUnTyp_2}}$.

By Type Predicate Interpretation, there are two cases.

  \spaceCase
  \subCaseHead{$\theV=\funBare{x}{e}$ and
               $\relTyp{\tyBind{x}{\varTyp_{11}}}{e}{\varTyp_{12}}{n-1}$}
  \spaceCase

  {\subCaseIndentation

  By Lifting,
    $\relTyp{\tyBind{x}{\varTyp_{11}}}{e}{\varTyp_{12}}{n}$.

  By Narrowing,
    $\relTyp{\tyBind{x}{\varTyp_{21}}}{e}{\varTyp_{11}}{n}$.

  By $\ruleName{T-Sub}$,
    $\relTyp{\tyBind{x}{\varTyp_{21}}}{e}{\varTyp_{22}}{n}$.

  Thus, by Type Predicate Interpretation,
    $\formulaSat{n+1}{\synHastyp{\funBare{x}{e}}{\varUnTyp_2}}$.
  }

  \spaceCase
  \subCaseHeadEmpty

    \indent\indent
    $\varVal=c$,
    $\typConst{c}=
        \refTyp{\theV=c\wedge
                \synHastyp{\theV}{\utArrow{x}{\varTyp_{01}}{\varTyp_{02}}}}$, and

    \indent\indent
    $\relSynSub{}{\utArrow{x}{\varTyp_{01}}{\varTyp_{02}}}
                 {\utArrow{x}{\varTyp_{11}}{\varTyp_{12}}}{n-1}$.

  \spaceCase

  {\subCaseIndentation

  By Lifting,
    $\relSynSub{\ }{\utArrow{x}{\varTyp_{01}}{\varTyp_{02}}}
                   {\utArrow{x}{\varTyp_{11}}{\varTyp_{12}}}{n}$.

  By Inversion,
    $\relSub{\ }{\varTyp_{11}}{\varTyp_{01}}{n}$ and
    $\relSub{\tyBind{x}{\varTyp_{11}}}{\varTyp_{02}}{\varTyp_{12}}{n}$.

  By Transitive Subtyping,
    $\relSub{\ }{\varTyp_{21}}{\varTyp_{01}}{n}$.

  By Narrowing,
    $\relSub{\tyBind{x}{\varTyp_{21}}}{\varTyp_{02}}{\varTyp_{12}}{n}$.

  By Transitive Subtyping,
    $\relSub{\tyBind{x}{\varTyp_{21}}}{\varTyp_{02}}{\varTyp_{22}}{n}$.

  By $\ruleName{U-Arrow}$,
    $\relSynSub{\ }{\utArrow{x}{\varTyp_{01}}{\varTyp_{02}}}
                   {\utArrow{x}{\varTyp_{21}}{\varTyp_{22}}}{n}$.

  By Type Predicate Interpretation,
    $\formulaSat{n+1}{\synHastyp{c}{\utArrow{x}{\varTyp_{21}}{\varTyp_{22}}}}$.

  }

\spaceCase
\caseHead{U-Var}{Trivial.}

\spaceCase
\caseHead{U-Null}{By Type Predicate Interpretation.}
\spaceCase

\spaceCase
\caseHead{U-Datatype}
  {$\inferrule*
    {\lookupCon{\varCon}{\seq{\varPole\tyVar}}{f}{\varTyp'} \\\\
     \forall i.\ 
       \textrm{if } \varPole_i\in\set{\poleCo,\poleBi}
       \textrm{ then } \relSub{}{\varTyp_{1i}}{\varTyp_{2i}}{n} \\\\
     \forall i.\ 
       \textrm{if } \varPole_i\in\set{\poleContra,\poleBi}
       \textrm{ then } \relSub{}{\varTyp_{2i}}{\varTyp_{1i}}{n}}
    {\relSynSub{}{\typeCon{\varCon}{\seq{\varTyp_1}}}
                 {\typeCon{\varCon}{\seq{\varTyp_2}}}{n}}$}
\spaceCase

We consider the special case when there is exactly one type parameter
$\tyVar$ with variance annotation $\varPole$.
The type actuals are, therefore, labeled $\varTyp_{11}$ and $\varTyp_{21}$.
The reasoning extends to an arbitrary number of type parameters by a strong
induction on the length of the sequence.

\spaceCase
\subCaseHead{\varPole\ =\ \poleCo}
\spaceCase

  {\subCaseIndentation
  Consider an arbitrary $\varVal_0$ such that
  $\formulaSat{n+1}{\synHastyp{\varVal_0}{\typeCon{\varCon}{\varTyp_{11}}}}$.

  By Type Predicate Interpretation, there are two cases.

  In one case, $\varVal_0=\vNull$, and trivially

  \indent\indent
  $\formulaSat{n+1}{\synHastyp{\vNull}{\typeCon{\varCon}{\varTyp_{21}}}}$.

  In the other case, $\varVal_0=\newDataBare{\varCon}{\seq{\varVal}}$ and

  \indent\indent
  for all $j$,
  $\relTyp{\ }{\varVal_j}{\inst{\varTyp'_j}{\tyVar}{\varTyp_{11}}}{n}$.

  By well-formedness of the type definition, $\tyVar$ appears only

    \indent\indent
    positively in every $\varTyp'_j$.

  By Sound Variance (1),

    \indent\indent
    $\relSub{\ }{\inst{\varTyp'_j}{\tyVar}{\varTyp_{11}}}
                {\inst{\varTyp'_j}{\tyVar}{\varTyp_{21}}}{n}$.

  By $\ruleName{T-Sub}$,
  $\relTyp{\ }{\varVal_j}{\inst{\varTyp'_j}{\tyVar}{\varTyp_{21}}}{n}$.

  By Type Predicate Interpretation,
  $\formulaSat{n+1}{\synHastyp{\newDataBare{\varCon}{\seq{\varVal}}}
                              {\typeCon{\varCon}{\varTyp_{21}}}}$.
  }

\spaceCase
\subCaseHead{ \varPole\ =\ \poleContra.\sepPremise Similar, using Sound Variance (2)}
\spaceCase

\subCaseHead{ \varPole\ =\ \poleBi.\sepPremise Similar, using Sound Variance (3)}
\end{proof}

\begin{proof}[Proof of (3)]\noqedsymbol
Only the rule for monotypes applies.

\spaceCase
\caseHead{S-Mono}
  {$\inferrule*
    {x \text{ fresh} \sepPremise
     \varFormOne' = \subst{\varFormOne}{\theV}{x} \sepPremise
     \varFormTwo' = \subst{\varFormTwo}{\theV}{x} \\\\
     \forall(\varFormTwo_{1i},\varFormTwo_{2i})\in\cnf{\varFormTwo'}.\
       \relImpl{\varFormOne'}{\varFormTwo_{1i}}{\varFormTwo_{2i}}{n}
    }
    {\relSub{}{\refTyp{\varFormOne}}{\refTyp{\varFormTwo}}{n}}$}
\spaceCase

Note that the alpha-renaming preserves satisfiability.

So we assume $\formulaSat{n+1}{\varFormOne'}$
and then prove $\formulaSat{n+1}{\varFormTwo'}$.

By Strengthening on each premise,
  $\relImpl{}{\varFormTwo_{1i}}{\varFormTwo_{2i}}{n}$.

Each of these derivations has the same size as the original.

Thus, by IH (1) on each,
  $\formulaSat{n+1}{\varFormTwo_{1i}\Rightarrow\varFormTwo_{2i}}$.

Thus, $\formulaSat{n+1}{\wedge_i\ \varFormTwo_{1i}\Rightarrow\varFormTwo_{2i}}$.

Thus, by equivalence of normalized formulas, $\formulaSat{n+1}{\varFormTwo'}$.
\end{proof}

\begin{proof}[Proof of (4)]\noqedsymbol
We only need to consider the rules that can derive a monotype
$\varTyp$ for a value $\varVal$ in the empty environment.

\spaceCase
\caseHead{T-Const}{By Constant Types (Valid)}.

\spaceCase
\caseHead{T-Extend}{Trivially, since
                    $\subst{(\theV=\varVal)}{\theV}{\varVal}=\varVal=\varVal$}.

\spaceCase
\caseHead{T-Fun}
  {$\inferrule*
    {\varUnTyp=\utArrow{x}{\varTyp_{1}}{\varTyp_{2}} \sepPremise
     \relTyp{\tyBind{x}{\varTyp_{1}}}{e}{\varTyp_{2}}{n}
    }
    {\relTyp{}{\funBare{x}{e}}
            {\refTypShort{\theV=\funBare{x}{e}\wedge
                          \synHastyp{\theV}{\varUnTyp}}}{n}}$}
\spaceCase


By Type Predicate Interpretation,
  $\formulaSat{n+1}{\synHastyp{\funBare{x}{e}}{\varUnTyp}}$.

Furthermore, by Validity,
  $\formulaSat{n+1}{\funBare{x}{e}=\funBare{x}{e}}$.

\spaceCase
\caseHead{T-Fold}{
  $\inferrule*
    {\lookupCon{\varCon}{\seq{\varPole\tyvar}}{f}{\varTyp'} \\\\
     \forall j.\ \relTyp{}{\varVal_j}
                        {\inst{\varTyp_j}{\seq{\tyvar}}{\seq{\varTyp}}}{n}
    }
    {\relTyp{}{\newDataBare{\varCon}{\seq{\varVal}}}
            {\refTyp{\fold{\varCon}{\seq{\varTyp}}{\seq{\varVal}}}}{n}}$}
\spaceCase

We consider each of the components of the formula from $\foldName$.

By Validity, $\formulaSat{n+1}{\newDataBare{\varCon}{\seq{\varVal}}\neq\vNull}$.

By Type Predicate Interpretation,
$\formulaSat{n+1}{\synHastyp{\newDataBare{\varCon}{\seq{\varVal}}}
                            {\typeCon{\varCon}{\seq{\varTyp}}}}$.

By Datatype Representation,
$\formulaSat{n+1}{\tagof{\newDataBare{\varCon}{\seq{\varVal}}}=\tagRecd}$

  \indent\indent
  and
  $\formulaSat
     {n+1}
     {\wedge_j \ \selR{\newDataBare{\varCon}{\seq{\varVal}}}{f_j}=\varVal_j}$.

\spaceCase
\caseHead{T-Unfold}{
  $\inferrule*[right=\ruleNameFig{T-Unfold}]
    {\relTyp{}{\varVal}{\refTyp{\synHastyp{\theV}{\typeCon{\varCon}{\seq{\varTyp}}}}}{n}}
    {\relTyp{}{\varVal}{\refTyp{\unfold{\varCon}{\seq{\varTyp}}}}{n}}$}
\spaceCase

By IH (3), $\formulaSat{n+1}{\synHastyp{\varVal}{\typeCon{\varCon}{\seq{\varTyp}}}}$.

The goal follows by Type Predicate Interpretation and

  \indent\indent
  Datatype Representation.

\spaceCase
\caseHead{T-Sub}
  {$\inferrule*
    {\relTyp{}{\varVal}{\varTyp'}{n} \sepPremise
     \relSub{}{\varTyp'}{\varTyp}{n}}
    {\relTyp{}{\varVal}{\varTyp}{n}}$}
\spaceCase

By IH (3), $\formulaSat{n+1}{\embed{\varTyp'}(\varVal)\Rightarrow
                             \embed{\varTyp}(\varVal)}$.

By IH (4), $\formulaSat{n+1}{\embed{\varTyp'}(\varVal)}$.

Thus, $\formulaSat{n+1}{\embed{\varTyp'}(\varVal)}$.
\end{proof}
\end{proof}


\separatingLine

\noindent
In the following lemma we lift substitution
to judgments in the obvious way. For example, we write
$(\relTyp{\Gamma}{e}{\varScm}{n})[w/x]$
to mean
$\relTyp{\Gamma[w/x]}{e[w/x]}{\varScm[w/x]}{n}$.


\begin{mainlemma}[Stratified Value Substitution]
Let $\relTyp{\ }{\varVal}{\varScm}{n}$.
\begin{enumerate}
\alignItemTwo{1.25in}
  {If   $\relImpl{\tyBind{x}{\varScm},\Gamma}{\varFormOne}{\varFormTwo}{n}$,}
  {then $(\relImpl{\Gamma}{\varFormOne}{\varFormTwo}{n+1})[w/x]$.}
\alignItemTwo{1.25in}
  {If   $\relSynSub{\tyBind{x}{\varScm},\Gamma}{\varUnTyp_1}{\varUnTyp_2}{n}$,}
  {then $(\relSynSub{\Gamma}{\varUnTyp_1}{\varUnTyp_2}{n+1})[w/x]$.}
\alignItemTwo{1.25in}
  {If   $\relSub{\tyBind{x}{\varScm},\Gamma}{\varScm_1}{\varScm_2}{n}$,}
  {then $(\relSub{\Gamma}{\varScm_1}{\varScm_2}{n+1})[w/x]$.}
\alignItemTwo{1.25in}
  {If   $\relTyp{\tyBind{x}{\varScm},\Gamma}{e}{\varScm'}{n}$,}
  {then $(\relTyp{\Gamma}{e}{\varScm'}{n+1})[w/x]$.}
\end{enumerate}
\end{mainlemma}

\begin{proof} By mutual induction.
In the $\ruleName{C-Valid}$ and $\ruleName{C-Valid-n}$ cases, we will
distinguish between whether $\varScm$ is a monotype or a polymorphic
type scheme. In all other cases, this difference will not affect the
reasoning. The $\ruleName{T-Var}$ case is interesting because
singleton types must be preserved after substitution.

\begin{proof}[Proof of (1)]\noqedsymbol
Recall that we use the notation $\varFormOne(x)$ to mean
$\subst{\varFormOne}{\theV}{x}$ and $\embed{\varTyp}(x)$ to mean
$\subst{\embed{\varTyp}}{\theV}{x}$.
Furthermore, we lift this to $\embed{\Gamma}(x)$ in the obvious way.


\spaceCase
\caseHead{C-Valid-n}
  {$\inferrule*
    {\formulaSat{n}{\embed{\tyBind{x}{\varScm},\Gamma}
                    \wedge\varFormOne\Rightarrow\varFormTwo}}
    {\relImpl{\tyBind{x}{\varScm},\Gamma}{\varFormOne}{\varFormTwo}{n}}$}
\spaceCase

  \subCaseHead{$\varScm=\varTyp$}
  \spaceCase

  {\subCaseIndentation

  Thus,
    $\formulaSat{n}{\embed{\varTyp}(x)\wedge\embed{\Gamma}(x)\wedge
                    \varFormOne(x)\Rightarrow\varFormTwo}(x)$.

  Thus,
    $\formulaSat{n}{\embed{\varTyp}(x)\Rightarrow\embed{\Gamma}(x)\wedge
                    \varFormOne(x)\Rightarrow\varFormTwo}(x)$.

  By Lifting,
    $\formulaSat{n+1}{\embed{\varTyp}(x)\Rightarrow\embed{\Gamma}(x)\wedge
                      \varFormOne(x)\Rightarrow\varFormTwo}(x)$.

  By Satisfiable Typing, $\formulaSat{n+1}{\embed{\varTyp}(\varVal)}$.

  Thus, 
    $\formulaSat{n+1}{\substwx{\embed{\Gamma}}\wedge
                      \substwx{\varFormOne}\Rightarrow
                      \substwx{\varFormTwo}}$.

  By $\ruleName{C-Valid-n}$,
    $\relImpl{\substwx{\Gamma}}
             {\substwx{\varFormOne}}
             {\substwx{\varFormTwo}}{n+1}$.

  }

  \spaceCase
  \subCaseHead{$\varScm=\typAll{\tyVar}{\varScm'}$}
  \spaceCase

  {\subCaseIndentation

  Thus,
    $\formulaSat{n}{\formTrue\wedge\embed{\Gamma}(x)\wedge
                    \varFormOne(x)\Rightarrow\varFormTwo(x)}$.

  Thus,
    $\formulaSat{n}{\embed{\Gamma}(x)\wedge
                    \varFormOne(x)\Rightarrow\varFormTwo(x)}$.

  Thus,
    $\formulaSat{n}{\substwx{\embed{\Gamma}}\wedge
                    \substwx{\varFormOne}\Rightarrow
                    \substwx{\varFormTwo}}$.

  By Lifting,
    $\formulaSat{n+1}{\substwx{\embed{\Gamma}}\wedge
                      \substwx{\varFormOne}\Rightarrow
                      \substwx{\varFormTwo}}$.

  By $\ruleName{C-Valid-n}$,
    $\relImpl{\substwx{\Gamma}}
             {\substwx{\varFormOne}}
             {\substwx{\varFormTwo}}{n+1}$.
  }

\spaceCase
\caseHead{C-Valid}
  {$\inferrule*
    {\valid{\embed{\tyBind{x}{\varScm},\Gamma}
            \wedge\varFormOne\Rightarrow\varFormTwo}}
    {\relImpl{\tyBind{x}{\varScm},\Gamma}{\varFormOne}{\varFormTwo}{n}}$}
\spaceCase

  \subCaseHead{$\varScm=\varTyp$}
  \spaceCase

  {\subCaseIndentation

  Thus,
    $\valid{\embed{\varTyp}(x)\wedge\embed{\Gamma}(x)\wedge
            \varFormOne(x)\Rightarrow\varFormTwo(x)}$.

  Thus,
    $\formulaSat{n}{\embed{\varTyp}(x)\wedge\embed{\Gamma}(x)\wedge
                    \varFormOne(x)\Rightarrow\varFormTwo}(x)$.

  The rest of the reasoning follows the $\ruleName{C-Valid-n}$ subcase.
  }

  \spaceCase
  \subCaseHead{$\varScm=\typAll{\tyVar}{\varScm'}$}
  \spaceCase

  {\subCaseIndentation

  Thus,
    $\valid{\formTrue\wedge\embed{\Gamma}(x)\wedge
            \varFormOne(x)\Rightarrow\varFormTwo(x)}$.

  Thus,
    $\formulaSat{n}{\formTrue\wedge\embed{\Gamma}(x)\wedge
                    \varFormOne(x)\Rightarrow\varFormTwo(x)}$.

  The rest of the reasoning follows the $\ruleName{C-Valid-n}$ subcase.
  }

\spaceCase
\caseHead{C-ImpSyn}
  {$\inferrule*
    {\exists j.\
      \valid{\embed{\tyBind{x}{\varScm}, \Gamma} \wedge \varFormOne \Rightarrow
             \synHastyp{\varLogVal_j}{\varUnTyp}} \\\\
      \relSynSub{\tyBind{x}{\varScm},\Gamma,\varFormOne}
                {\varUnTyp}{\varUnTyp_j}{n}}
    {\relImpl{\tyBind{x}{\varScm},\Gamma}{\varFormOne}
             {\vee_i\ \synHastyp{\varLogVal_i}{\varUnTyp_i}}{n}}$}
\spaceCase

By \ruleName{C-Valid},
  $\relImpl{\tyBind{x}{\varScm},\Gamma}
           {\varFormOne}
           {\synHastyp{\varLogVal_j}{\varUnTyp}}{n+1}$.

By IH (1),
  $\relImpl{\substwx{\Gamma}}
           {\substwx{\varFormOne}}
           {\synHastyp{\substwx{\varLogVal_j}}{\substwx{\varUnTyp}}}{n+1}$.

By IH (2),
  $\relSynSub{\substwx{\Gamma},\substwx{\varFormOne}}
             {\substwx{\varUnTyp}}
             {\substwx{\varUnTyp_j}}{n+1}$.

By $\ruleName{C-ImpSyn}$,
  $\relImpl{\substwx{\Gamma}}
           {\substwx{\varFormOne}}
           {\substwx{(\vee_i\ \synHastyp{\varLogVal_j}{\varUnTyp_j})}}{n+1}$.
\end{proof}

\begin{proof}[Proof of (2)]\noqedsymbol
Straightforward induction.
\end{proof}

\begin{proof}[Proof of (3)]\noqedsymbol
Straightforward induction, appealing to the equisatisfiability of
normalized formulas in the
$\ruleName{S-Mono}$ case.
\end{proof}

\begin{proof}[Proof of (4)]\noqedsymbol
\mbox{}

\spaceCase
\caseHead{T-Const}{}
\spaceCase

By $\ruleName{T-Const}$, $\relTyp{\substwx{\Gamma}}{c}{\typConst{c}}{n+1}$.

By Constant Types (Well-formed), $\relWf{\ }{\typConst{c}}$,

  \indent\indent
  so $\typConst{c}$ has no free variables.

Thus, $\substwx{\typConst{c}}=\typConst{c}$.

Also, $\substwx{c}=c$, which concludes the case.

\spaceCase
\caseHead{T-Var}
  {$\inferrule*
    {(\tyBind{x}{\varScm},\Gamma)(y)=\varTyp}
    {\relTyp{\tyBind{x}{\varScm},\Gamma}{y}{\refTyp{\theV=y}}{n}}$}
\spaceCase

  \subCaseHead{$x\neq y$}
  \spaceCase

  {\subCaseIndentation

  By substitution on environments, schemes, types and

  \indent\indent
    formulas,
    $(\substwx{\Gamma})(y)=\substwx{\varTyp}$.

  By $\ruleName{T-Var}$, $\relTyp{\substwx{\Gamma}}{y}{\refTyp{\theV=y}}{n+1}$.

  This concludes the subcase since

    \indent\indent
    $\substwx{y}=y$ and
    $\substwx{\refTypShort{\theV=y}}=\refTypShort{\theV=y}$.
  }

  \spaceCase
  \subCaseHead{$x=y$}
  \spaceCase

  {\subCaseIndentation

  Note that
    $\substwx{x}=\varVal$ and
    $\substwx{\refTypShort{\theV=x}}=\refTypShort{\theV=\varVal}$.

    \spaceCase
    \subSubCaseHead{$\varVal=z$}
    \spaceCase
    
    {\subSubCaseIndentation
      Impossible, since the typing environment is empty.
    }

    \spaceCase
    \subSubCaseHead{$\varVal=\dictextend{\varVal_1}{\varVal_2}{\varVal_3}$}
    \spaceCase
    
    {\subSubCaseIndentation
      Trivial, by $\ruleName{T-Extend}$.
    }

    \spaceCase
    \subSubCaseHead{$\varVal=c$}
    \spaceCase
    
    {\subSubCaseIndentation

      By $\ruleName{T-Const}$,
        $\relTyp{\substwx{\Gamma}}{c}{\typConst{c}}{n+1}$.

      By Constant Types (Normal),
        $\typConst{c}=\refTypShort{\theV=c\wedge\varFormOne}$.

      By $\ruleName{C-Valid}$ and $\ruleName{S-Mono}$,

        \indent\indent
        $\relSub{\substwx{\Gamma}}
                         {\refTypShort{\theV=c\wedge\varFormOne}}
                         {\refTypShort{\theV=c}}{n+1}$.

      By $\ruleName{T-Sub}$,
        $\relTyp{\substwx{\Gamma}}{c}{\refTypShort{\theV=c}}{n+1}$.
    }

    \spaceCase
    \subSubCaseHead{$\varVal=\funBare{z}{e_0}$}
    \spaceCase
    
    {\subSubCaseIndentation

      By Inversion,
        $\relTyp{\ }
                {\varVal}
                {\refTypShort{\theV=\varVal \wedge
                              \synHastyp{\theV}{\varUnTyp}}}{n+1}$.

      By $\ruleName{C-Valid}$ and $\ruleName{S-Mono}$,

        \indent\indent
        $\relSub{\ }
                {\refTypShort{\theV=\varVal \wedge
                              \synHastyp{\theV}{\varUnTyp}}}
                {\refTypShort{\theV=\varVal}}{n+1}$.

      By $\ruleName{T-Sub}$,
        $\relTyp{\ }{\varVal}{\refTypShort{\theV=\varVal}}{n+1}$.
        
      By Weakening,
        $\relTyp{\substwx{\Gamma}}{\varVal}{\refTypShort{\theV=\varVal}}{n+1}$.
    }

    \spaceCase
    \subSubCaseHead{$\varVal=\typFun{\tyVar}{e_0}$}
    \spaceCase
    
    {\subSubCaseIndentation
      Impossible, since $\varTyp$ is a monotype.
    }
  }

\spaceCase
\caseHead{T-VarPoly}
  {$\inferrule*
    {(\tyBind{x}{\varScm},\Gamma)(y)=\typAll{\tyVar}{\varScm_0}}
    {\relTyp{\tyBind{x}{\varScm},\Gamma}{y}{\typAll{\tyVar}{\varScm_0}}{n}}$}
\spaceCase

By substitution on environments, schemes, types and

\indent\indent
  formulas,
  $(\substwx{\Gamma})(y)=\substwx{(\typAll{\tyVar}{\varScm_0})}$.

  \spaceCase
  \subCaseHead{$x\neq y$}
  \spaceCase

  {\subCaseIndentation

  Since $x$ is a term variable,
    $\substwx{(\typAll{\tyVar}{\varScm_0})}$ is a polytype.

  By $\ruleName{T-VarPoly}$,
    $\relTyp{\substwx{\Gamma}}{y}{\substwx{(\typAll{\tyVar}{\varScm_0})}}{n+1}$.

  This concludes the subcase, since $\substwx{y}=y$.
  }

  \spaceCase
  \subCaseHead{$x=y$}
  \spaceCase

  {\subCaseIndentation

  Thus, $\varScm=\typAll{\tyVar}{\varScm_0}$.

  Since $\relWf{\ }{\varScm}$, $x$ does not appear in $\varScm$.

  Thus, $\substwx{\varScm_0}=\varScm_0$.

  The goal follows from $\ruleName{T-VarPoly}$.

  }

\spaceCase
\caseHead{T-Fun}
  {$\inferrule*
    {\relTyp{\tyBind{x}{\varScm},\Gamma,\tyBind{y}{\varTyp_1}}
            {e_0}
            {\varTyp_2}{n} \\\\
     \varUnTyp = \utArrow{y}{\varTyp_1}{\varTyp_2}
    }
    {\relTyp{\tyBind{x}{\varScm},\Gamma}
            {e}
            {\refTypShort{\theV=e \wedge \synHastyp{\theV}{\varUnTyp}}}{n}}$}
\spaceCase

Note that in this case, $e=\funBare{y}{e_0}$.

By IH (4),
  $\relTyp{\substwx{\Gamma},\tyBind{y}{\substwx{\varTyp_1}}}
          {\substwx{e_0}}
          {\substwx{\varTyp_2}}{n+1}$.

By $\ruleName{T-Fun}$,

  \indent\indent
  $\relTyp{\substwx{\Gamma}}
          {\substwx{e}}
          {\refTypShort{\theV=\substwx{e} \wedge
                        \synHastyp{\theV}{\substwx{\varUnTyp}}}}{n+1}$.

Thus,
  $\relTyp{\substwx{\Gamma}}
          {\substwx{e}}
          {\substwx{\refTypShort{\theV=e \wedge
                                 \synHastyp{\theV}{\varUnTyp}}}}{n+1}$.

\spaceCase
\caseHead{T-App}
  {$\inferrule*
    {\relTyp{\tyBind{x}{\varScm},\Gamma}
            {\varVal_1}
            {\refTypShort{\synHastyp{\theV}{\utArrow{x}
                                           {\varTyp_{11}}
                                           {\varTyp_{12}}}}}{n} \\\\
     \relTyp{\tyBind{x}{\varScm},\Gamma}{\varVal_2}{\varTyp_{11}}{n}
    }
    {\relTyp{\tyBind{x}{\varScm},\Gamma}
            {\varVal_1\ \varVal_2}{\subst{\varTyp_{12}}{y}{\varVal_2}}{n}}$}
\spaceCase

Let $\Gamma'=\substwx{\Gamma}$,
    $\varVal_1'=\substwx{\varVal_1}$,
    $\varVal_2'=\substwx{\varVal_2}$,

  \indent\indent
    $\varTyp_{11}'=\substwx{\varTyp_{11}}$, and
    $\varTyp_{12}'=\substwx{\varTyp_{12}}$,

By IH (4),
  $\relTyp{\Gamma'}
          {\varVal_1'}
          {\substwx{\refTypShort{\synHastyp{\theV}
                                           {\utArrow{y}
                                                    {\varTyp_{11}}
                                                    {\varTyp_{12}}}}}}{n+1}$.

Thus,
  $\relTyp{\Gamma'}
          {\varVal_1'}
          {\refTypShort{\synHastyp{\theV}
                                  {\utArrow{y}
                                           {\varTyp_{11}'}
                                           {\varTyp_{12}'}}}}{n+1}$.

By IH (4),
  $\relTyp{\Gamma'}{\varVal_2'}{\varTyp_{11}'}{n+1}$.

By $\ruleName{T-App}$,
  $\relTyp{\Gamma'}
          {\varVal_1'\ \varVal_2'}
          {\subst{\varTyp_{12}'}{y}{\varVal_2'}}{n+1}$.


Now we expand
$\subst{\varTyp_{12}'}{y}{\varVal_2'}$ to
$\substwx{\subst{\substwx{\varTyp_{12}}}{y}{\varVal_2}}$.

Since $\varVal$ and $\varVal_2$ are closed values, and $x$ and $y$ are
distinct, this

  \indent\indent
  is the same as
  $\substwx{\substwx{\subst{\varTyp_{12}}{y}{\varVal_2}}}$.

Furthermore, this is
  $\substwx{(\subst{\varTyp_{12}}{y}{\varVal_2})}$.

Finally, we note that $\varVal_1'\ \varVal_2'=\substwx{(\varVal_1\ \varVal_2)}$.

Thus, the derivation from $\ruleName{T-App}$ does indeed satisfy the goal.

\spaceCase
\caseHead{T-Sub}
  {$\inferrule*
    {\relTyp{\tyBind{x}{\varScm},\Gamma}{e}{\varScm''}{n} \sep
     \relSub{\tyBind{x}{\varScm},\Gamma}{\varScm''}{\varScm'}{n}}
    {\relTyp{\tyBind{x}{\varScm},\Gamma}{e}{\varScm'}{n}}$}
\spaceCase

By IH (4),
  $\relTyp{\substwx{\Gamma}}
          {\substwx{e}}
          {\substwx{\varScm''}}{n+1}$.

By IH (3),
  $\relTyp{\substwx{\Gamma}}
          {\substwx{\varScm''}}
          {\substwx{\varScm'}}{n+1}$.

By $\ruleName{T-Sub}$,
  $\relTyp{\substwx{\Gamma}}
          {\substwx{e}}
          {\substwx{\varScm'}}{n+1}$.

\spaceCase
\multiCaseHead{T-Let, T-If, T-TFun, T-TApp, T-Extend}

By IH on the premises and original rule to conclude.

\spaceCase
\multiCaseHead{T-Fold, T-Unfold}
\spaceCase

By IH on the premises and original rule to conclude.
\end{proof}

\end{proof}

\separatingLine


\noindent
In the following lemma we lift instantiation
to judgments in the obvious way. For example, we write
$\inst{(\relTyp{\Gamma}{e}{\varScm}{n})}{\tyVar}{\varTyp}$
to mean
$\relTyp{\inst{\Gamma}{\tyVar}{\varTyp}}
        {e}
        {\inst{\varScm}{\tyVar}{\varTyp}}{n}$.

\begin{lemma}[Type Substitution]
Let $\relWf{\ }{\varTyp}$.
\begin{enumerate}
\alignItemTwo{1.15in}
  {If   $\relImpl{\tyVar,\Gamma}{\varFormOne}{\varFormTwo}{n}$,}
  {then $\inst{(\relImpl{\Gamma}{\varFormOne}{\varFormTwo}{n})}
               {\tyVar}{\varTyp}$.}
\alignItemTwo{1.15in}
  {If   $\relSynSub{\tyVar,\Gamma}{\varUnTyp_1}{\varUnTyp_2}{n}$,}
  {then $\inst{(\relSynSub{\Gamma}{\varUnTyp_1}{\varUnTyp_2}{n})}
              {\tyVar}{\varTyp}$.}
\alignItemTwo{1.15in}
  {If   $\relSub{\tyVar,\Gamma}{\varScm_1}{\varScm_2}{n}$,}
  {then $\inst{(\relSub{\Gamma}{\varScm_1}{\varScm_2}{n})}
              {\tyVar}{\varTyp}$.}
\alignItemTwo{1.15in}
  {If   $\relTyp{\tyVar,\Gamma}{e}{\varScm}{n}$,}
  {then $\inst{(\relTyp{\Gamma}{e}{\varScm}{n})}
              {\tyVar}{\varTyp}$.}
\end{enumerate}
\end{lemma}

\begin{proof}
By mutual induction. Even (1) is straightforward, since type variables
in the environment play no rule in the embedding of formulas into the logic
(they are embedded as $\formTrue$).
\end{proof}


\begin{lemma}[Canonical Forms]
Suppose $\relTyp{\ }{\varVal}{\varScm}{n}$.
\begin{enumerate}
\item If $\varScm=\tyBool$,
      then $\varVal=\vTrue$ or $\varVal=\vFalse$.

\item If $\varScm=\refTyp{\synHastyp{\theV}
                                    {\utArrow{x}{\varTyp_{1}}{\varTyp_{2}}}}$,
      then either
      \begin{enumerate}
        \item $\varVal=\funBare{x}{e}$ and
              $\relTyp{\tyBind{x}{\varTyp_{1}}}{e}{\varTyp_{2}}{n}$, or


        \item $\varVal=c$ and
              for all $\varVal'$ such that $\relTyp{\ }{\varVal'}{\varTyp_{1}}{n}$,
              $\deltaApp{c}{\varVal'}$ is defined and
              $\relTyp{\ }{\deltaApp{c}{\varVal'}}
                          {\subst{\varTyp_{2}}{x}{\varVal'}}{n}$.
      \end{enumerate}

\item If $\varScm=\typAll{\tyVar}{\varScm'}$,
      then $\varVal=\typFun{\tyVar}{e}$ and $\relTyp{\tyVar}{e}{\varScm'}{n}$.
\end{enumerate}
\end{lemma}

\begin{proof}[Proof of (1)]
By Satisfiable Typing, $\formulaSat{n+1}{\tagof{w}=\tagBool}$.
By Boolean Values, $w$ is either $\vTrue$ or $\vFalse$.
\end{proof}

\begin{proof}[Proof of (2)]
By Satisfiable Typing,
  $\formulaSat{n+1}{\synHastyp{w}{\utArrow{x}{\varTyp_{1}}{\varTyp_{2}}}}$.
The goal follows by Type Predicate Interpretation and Constant Types (App).
\end{proof}

\begin{proof}[Proof of (3)]
By induction on the derivation. We consider only the rules that
can derive a polytype.

\spaceCase
\caseHead{T-TFun}{Immediate.}

\spaceCase
\caseHead{T-TApp}{Impossible, since $\varVal$ is a value.}

\spaceCase
\caseHead{T-VarPoly}{Impossible, since the environment is empty.}

\spaceCase
\caseHead{T-Sub}
  {$\inferrule*
    {\relTyp{}{\varVal}{\varScm_0}{n} \sepPremise
     \relSub{}{\varScm_0}{\typAll{\tyVar}{\varScm'}}{n}}
    {\relTyp{}{\varVal}{\typAll{\tyVar}{\varScm'}}{n}}$}
\spaceCase

The subtyping derivation can only conclude by $\ruleName{S-Poly}$.

From its premises, $\varScm_0=\typAll{\tyVar}{\varScm''}$
where $\relSub{\tyVar}{\varScm''}{\varScm'}{n}$.

By IH, $\varVal=\typFun{\tyVar}{e}$ and $\relTyp{\tyVar}{e}{\varScm''}{n}$.

By $\ruleName{T-Sub}$, $\relTyp{\tyVar}{e}{\varScm'}{n}$.

\end{proof}

\separatingLine


\noindent
We are now ready to prove the following type soundness theorem
that combines progress and preservation. Soundness of the basic
type system, which is the system at level zero and is used
for typechecking source programs, follows as a corollary.

\begin{theorem}[\dntypes Type Soundness]
\mbox{}
\vspace{0.07in}

\noindent
If $\relTyp{\ }{e}{\varScm}{n}$, then either $e$ is a value or
$\reducesTo{e}{e'}$ and $\relTyp{\ }{e'}{\varScm}{n+1}$.
\end{theorem}

\begin{proof} By induction on the typing derivation.

\spaceCase
\multiCaseHead{T-Var, T-VarPoly}
\spaceCase

Impossible, since the typing environment is empty.

\spaceCase
\multiCaseHead{T-Const, T-Extend, T-Fun, T-TFun, T-Fold}
\spaceCase

Immediate, since $e$ is a value.

\spaceCase
\multiCaseHead{T-Unfold}
\spaceCase

By Satisfiable Typing and Type Predicate Interpretation, $e$ is a

  \indent\indent
  value.

\spaceCase
\caseHead{T-If}
  {$\inferrule*
    {\relTyp{}{\varVal}{\tyBool}{n} \\\\
     \relTyp{\varVal=\vTrue}{e_1}{\varScm}{n} \\\\
     \relTyp{\varVal=\vFalse}{e_2}{\varScm}{n}}
    {\relTyp{}{\ite{\varVal}{e_1}{e_2}}{\varScm}{n}}$}
\spaceCase

By Canonical Forms, there are two cases.

  \spaceCase
  \subCaseHead{$w=\vTrue$}
  \spaceCase

  {\subCaseIndentation

  By $\ruleName{E-IfTrue}$, $e'=e_1$.

  $\valid{\vTrue=\vTrue}$,
    so by Strengthening, $\relTyp{\ }{e_1}{\varScm}{n}$.

  By Lifting, $\relTyp{\ }{e_1}{\varScm}{n+1}$.

  }

  \spaceCase
  \subCaseHead{$w=\vFalse$}
  \spaceCase

  {\subCaseIndentation

    By $\ruleName{E-IfFalse}$, $e'=e_2$.

    $\valid{\vFalse=\vFalse}$,
      so by Strengthening, $\relTyp{\ }{e_2}{\varScm}{n}$.

    By Lifting, $\relTyp{\ }{e_2}{\varScm}{n+1}$.
  }

\spaceCase
\caseHead{T-App}
  {$\inferrule*
    {\relTyp{}{\varVal_1}
              {\refTypShort{\synHastyp{\theV}
                                      {\utArrow{x}{\varTyp_{11}}
                                                  {\varTyp_{12}}}}}{n} \\\\
     \relTyp{}{\varVal_2}{\varTyp_{11}}{n}}
    {\relTyp{}{\varVal_1\ \varVal_2}{\subst{\varTyp_{12}}{x}{\varVal_2}}{n}}$}
\spaceCase

By Canonical Forms, there are two cases.

  \spaceCase
  \subCaseHead{
    $\varVal_1=\funBare{x}{e_0}$ and
    $\relTyp{\tyBind{x}{\varTyp_{11}}}{e_0}{\varTyp_{12}}{n}$.
  }
  \spaceCase

  {\subCaseIndentation

  By Value Substitution,
    $\relTyp{}{\subst{e_0}{x}{\varVal_2}}
              {\subst{\varTyp_{12}}{x}{\varVal_2}}{n+1}$.

  This concludes the subcase, since by $\ruleName{E-App}$,
    $e'=\subst{e_0}{x}{\varVal_2}$.

  }

  \spaceCase
  \subCaseHead{$\varVal_1=c$}
  \spaceCase

  {\subCaseIndentation

  Since $\relTyp{\ }{\varVal_2}{\varTyp_{11}}{n}$,
  we are also given that $\deltaApp{c}{\varVal_2}$

    \indent\indent
    is defined and
    $\relTyp{\ }{\deltaApp{c}{\varVal_2}}{\subst{\varTyp_{12}}{x}{\varVal_2}}{n}$.

  By Lifting,
    $\relTyp{\ }{\deltaApp{c}{\varVal_2}}{\subst{\varTyp_{12}}{x}{\varVal_2}}{n+1}$.

  This concludes the subcase, since by $\ruleName{E-Delta}$,
    $e'=\deltaApp{c}{\varVal_2}$.

  }

\spaceCase
\caseHead{T-TApp}
  {$\inferrule*
    {\relTyp{}{\varVal'}{\typAll{\tyVar}{\varScm'}}{n}}
    {\relTyp{}{\typInst{\varVal'}{\varTyp}}
              {\inst{\varScm'}{\tyVar}{\varTyp}}{n}}$}
\spaceCase

By Canonical Forms, $\varVal'=\typFun{\tyVar}{e_0}$
  and $\relTyp{\tyVar}{e_0}{\varScm'}{n}$.

By Type Substitution,
  $\relTyp{\ }{e_0}{\inst{\varScm'}{\tyVar}{\varTyp}}{n}$.

By Lifting,
  $\relTyp{\ }{e_0}{\inst{\varScm'}{\tyVar}{\varTyp}}{n+1}$.

This concludes the case, since by $\ruleName{T-TApp}$,
  $e'=e_0$.

\spaceCase
\caseHead{T-Let}
  {$\inferrule*
    {\relWf{}{\varScm_1} \sepPremise
     \relTyp{}{e_1}{\varScm_1}{n} \\\\
     \relWf{}{\varScm_2} \sepPremise
     \relTyp{\tyBind{x}{\varScm_1}}{e_2}{\varScm_2}{n}}
    {\relTyp{}{\letinBare{x}{e_1}{e_2}}{\varScm_2}{n}}$}
\spaceCase

By the IH, there are two cases.

  \spaceCase
  \subCaseHead{$e_1$ is a value $\varVal$}
  \spaceCase

  {\subCaseIndentation

  By $\ruleName{E-Let}$, $e'=\subst{e_2}{x}{\varVal}$.

  By Value Substitution,
    $\relTyp{}{\subst{e_2}{x}{\varVal}}{\subst{\varScm_2}{x}{\varVal}}{n+1}$.

  Since $\relWf{\ }{\varScm_2}$, $x$ does not appear free in $\varScm_2$,
    so $\subst{\varScm_2}{x}{\varVal} = \varScm_2$.

  }

  \spaceCase
  \subCaseHead{$\reducesTo{e_1}{e_1'}$ and $\relTyp{}{e_1'}{\varScm}{n+1}$}
  \spaceCase

  {\subCaseIndentation

  By $\ruleName{E-Compat}$, $e'=\letinBare{x}{e_1'}{e_2}$. 

  By Lifting, $\relTyp{\tyBind{x}{\varScm}}{e_2}{\varScm_2}{n+1}$.

  By $\ruleName{T-Let}$, $\relTyp{}{\letinBare{x}{e_1'}{e_2}}{\varScm_2}{n+1}$.

  }

\spaceCase
\caseHead{T-Sub}
  {$\inferrule*
    {\relTyp{}{\varVal}{\varScm'}{n} \sepPremise
     \relSub{}{\varScm'}{\varScm}{n}}
    {\relTyp{}{\varVal}{\varScm}{n}}$}
\spaceCase

By IH, Lifting, and $\ruleName{T-Sub}$.

\end{proof}


\begin{corollary}[System D Type Soundness]
\mbox{}

If $\relTyp{\ }{e}{\varScm}{0}$,
then either $e$ diverges or
$\reducesToMulti{e}{\varVal}$ and $\relTyp{\ }{\varVal}{\varScm}{*}$.
\end{corollary}

\begin{proof}
Follows from \dntypes Type Soundness.
\end{proof}


  \clearpage

\section{Algorithmic Typing}
\label{sec:app-algorithmic}

A type checker for \dtypes cannot directly implement the declarative
type system for a couple of reasons.
First, the typing rules are not
syntax-directed because of \ruleName{T-Sub} and \ruleName{T-Unfold},
which can apply to any expression $e$, and \ruleName{C-ImpSyn},
which non-deterministically refers to a type term $\varUnTyp$.
Second, the syntax of values lacks type annotations,
so the premises of rules like \ruleName{T-Fun}, \ruleName{T-Let},
and \ruleName{T-If} manipulate types that cannot
be inferred by the syntax of the expression being checked.

In this section, we define an algorithmic version of the type system.
First, we extend the syntax of the language with optional type
annotations for binding constructs and for constructed data.
Next, we show how to implement the non-deterministic
\ruleName{C-ImpSyn} rule.
Then, we define an algorithmic type system without the
non-deterministic \ruleName{T-Sub} and \ruleName{T-Unfold} rules.
To eliminate the
former, we derive unique types and then add explicit subtyping checks
in the typing rules that require them. To eliminate the latter, we
eagerly attempt to unfold the types of bindings in anticipation of where
\ruleName{T-Unfold} might be needed.
Furthermore, although we could require that all binding constructs
and constructed data be annotated with types, this would lead to
redundant and tedious type annotations.
Instead, we define a \emph{bidirectional type system}
in the style of \cite{pierce-turner} that locally infers type
annotations where possible.


\subsection{Syntax}

We extend the syntax of \dtypes as follows.
\[
\begin{array}{rcll}
\varVal
  & ::= & \cdots                            & \textbf{Values} \\
  &  |  & \funAnn{x}{\varTyp}{e}            & \text{annotated function} \\
  &  |  & \newDataAnn{\varCon}{\seq{\varTyp}}{\seq{\varVal}}
                                            & \text{annotated constructed data} \\
\blankrow{4} \\
e & ::= & \cdots                            & \textbf{Expressions} \\
  &  |  & \letinAnn{x}{\varScm}{e_1}{e_2}   & \text{annotated let-binding}
\end{array}
\]


\subsection{Subtyping}

\begin{figure}[t]
\centering

\judgementHead{Algorithmic Subtyping}{$\relASub{\Gamma}{\varScm_1}{\varScm_2}{\usedBoxes}$}

\vsepRule



$\inferrule*[right=\ruleNameFig{SA-Mono}]
  {x \text{ fresh} \sepPremise
   \varFormOne_1' = \subst{\varFormOne_1}{\theV}{x} \sepPremise
   \varFormOne_2' = \subst{\varFormOne_2}{\theV}{x} \\\\
   \cnf{\varFormOne_2'} = \wedge_i(\clause{\varFormTwo_i}{\varFormThr_i}) \\\\
   \forall i.\
     \relAImpl{\Gamma,\varFormOne_1'}{\varFormTwo_i}{\varFormThr_i}{\usedBoxes}}
  {\relASub{\Gamma}{\refTyp{\varFormOne_1}}{\refTyp{\varFormOne_2}}{\usedBoxes}}
$

\vsepRule

$\inferrule*[right=\ruleNameFig{SA-Poly}]
  {\relASub{\Gamma}{\varScm_1}{\varScm_2}{\usedBoxes}}
  {\relASub{\Gamma}{\typAll{\tyVar}{\varScm_1}}{\typAll{\tyVar}{\varScm_2}}{\usedBoxes}}
$

\vsepRule

\judgementHead{Algorithmic Clause Implication}
              {$\relAImpl{\Gamma}{\varFormTwo}{\varFormThr}{\usedBoxes}$}

\vsepRule

$\inferrule*[right=\ruleNameFig{CA-Valid}]
  {\valid{\embed{\Gamma} \land \varFormTwo \Rightarrow \varFormThr}}
  {\relAImpl{\Gamma}{\varFormTwo}{\varFormThr}{\usedBoxes}}
$

\vsepRule


$\inferrule*[right=\ruleNameFig{CA-ImpSyn}]
  {\exists\ j. \sep
   \usedBoxes' = \extract{\Gamma}{\refTyp{\theV=\varLogVal_j}}{\usedBoxes} \\\\
   \exists\ \varUnTyp \in \usedBoxes'. \sep
   \relASynSub{\Gamma, \varFormTwo}{\varUnTyp}{\varUnTyp_j}{\usedBoxes\cup\usedBoxes'}
  }
  {\relAImpl{\Gamma}
            {\varFormTwo}
            {\vee_i\ \synHastyp{\varLogVal_i}{\varUnTyp_i}}
            {\usedBoxes}}
$

\vsepRule

\judgementHead{Algorithmic Syntactic Subtyping}{$\relASynSub{\Gamma}{\varUnTyp_1}{\varUnTyp_2}{\usedBoxes}$}

\vsepRule

$\inferrule*[right=\ruleNameFig{UA-Arrow}]
  {
   \relASub{\Gamma}{\varTyp_{21}}{\varTyp_{11}}{\usedBoxes} \\\\
   \relASub{\extendGamma{\Gamma}{x_1}{\varTyp_{21}}}
           {\varTyp_{12}}{\varTyp_{22}}{\usedBoxes}
  }
  {\relASynSub{\Gamma}{\utArrow{x_1}{\varTyp_{11}}{\varTyp_{12}}}
                      {\utArrow{x_2}{\varTyp_{21}}{\varTyp_{22}}}{\usedBoxes}}
$

\vsepRule

$\inferrule*[right=\ruleNameFig{UA-Var}]
  { }
  {\relASynSub{\Gamma}{\tyvar}{\tyvar}{\usedBoxes}}
$
\hsepRule
$\inferrule*[right=\ruleNameFig{UA-Null}]
  { }
  {\relASynSub{\Gamma}{\nullCon}{\typeCon{\varCon}{\seq{\varTyp}}}{\usedBoxes}}
$

\vsepRule

$\inferrule*[right=\ruleNameFig{UA-Datatype}]
  {\lookupCon{\varCon}{\seq{\varPole\tyvar}}{f}{\varTyp} \\\\
   \forall i.\
      \text{if } \varPole_i\in\set{\poleCo,\poleBi}
      \text{ then } \relASub{\Gamma}{\varTyp_{1i}}{\varTyp_{2i}}{\usedBoxes} \\\\
   \forall i.\ 
      \text{if } \varPole_i\in\set{\poleContra,\poleBi}
      \text{ then } \relASub{\Gamma}{\varTyp_{2i}}{\varTyp_{1i}}{\usedBoxes}
  }
  {\relASynSub{\Gamma}{\typeCon{\varCon}{\seq{\varTyp_1}}}
                      {\typeCon{\varCon}{\seq{\varTyp_2}}}{\usedBoxes}}
$

\vsepRule


\caption{Algorithmic subtyping for \dtypes}
\label{fig:alg-subtyping}
\end{figure}

The algorithmic subtyping rules for \dtypes are shown in
\autoref{fig:alg-subtyping}.
The derivation rules of the algorithmic subtyping, clause implication,
and syntactic subtyping relations are analagous to their counterparts in
in the declarative system,
except that they include an additional input $\usedBoxes$, which is a set
of type terms $\varUnTyp$.
To begin the discussion, this additional input $\usedBoxes$ should be ignored,
and the procedure $\extendGamma{\Gamma}{x}{\varScm}$ can be assumed
to extend a type environment in the usual way,
that is, $\Gamma,\tyBind{x}{\varScm}$;
we will return to both of these issues shortly.


\parahead{Type Extraction}
We now show how \ruleName{CA-ImpSyn} implements the non-deterministic
$\ruleName{C-ImpSyn}$ rule.
First, we define the procedure $\boxesName$ that
traverses the environment $\Gamma$ and syntactically collect all of its
type terms $\varUnTyp$.
\begin{align*}
\boxes{\Gamma,\tyBind{x}{\refTyp{\varFormOne}}} &= 
    \boxes{\Gamma} \cup \boxes{\varFormOne} \\
\intertext{
The interesting case for formulas is for type predicates:
}
\boxes{\synHastyp{\varLogVal}{\varUnTyp}} &= \set{\varUnTyp}
\end{align*}
Notice that types contained within $\varUnTyp$ are not
collected, only ``top-level type terms" are.

The \ruleName{CA-ImpSyn} rule then uses the following $\extractName$
procedure to compute which type terms $\varUnTyp$ out of all possible
type terms in the environment 
(ignoring the $``\setminus\usedBoxes"$ part for now)
are such that
the solver can prove $\synHastyp{\varVal}{\varUnTyp}$ is true for all
values $\varVal$ of type $\varTyp$
\begin{align*}
  \extract{\Gamma}{\varTyp}{\usedBoxes} =\ &
     \setComp{\varUnTyp \in \usedBoxes'}
             {\valid{\embed{\Gamma, \tyBind{x}{\varTyp}} \Rightarrow
                     \synHastyp{x}{\varUnTyp}}}
\\[-2pt] & \hspace{0.10in} 
  \textrm{ where }\ \usedBoxes' = \boxes{\Gamma}\ \setminus\ \usedBoxes
\\[-2pt] & \hspace{0.10in} 
  \textrm{ and $x$ is fresh}
\end{align*}
That is, \ruleName{CA-ImpSyn} tries all type terms $\varUnTyp$ that
\ruleName{C-ImpSyn} might possibly refer to.


\parahead{Termination}
We now turn to the question of whether algorithmic subtyping terminates.
Because the subtyping, implication, and syntactic subtyping relations are
mutually defined, we may worry that it is possible to construct an implication
query (and hence a subtyping obligation) which is non-terminating.
Indeed, a na\"ive approach to deciding implications over type predicates 
using the above strategies (without considering the $\usedBoxes$ parameters)
may not terminate.
In the following, we write judgments without the $\usedBoxes$ parameters
to see what goes wrong when they are not considered.

Consider the environment
\begin{align*}
  \Gamma \defeq\  &
     y: \tyTop,\ 
     x: \refTyp{\theV = y \land \synHastyp{\theV}{\varUnTyp}} \\
  \text{where } \varUnTyp \defeq\ &
     \utArrow{a}
             {\refTyp{\synHastyp{\theV}{\utArrow{b}{\refTyp{\theV = y}}{\tyTop}}}}
             {\tyTop}
\end{align*}
and suppose we wish to check that
\begin{equation}
  \relImpl{\Gamma}
          {\formTrue}
          {\synHastyp{y}{\utArrow{x}{\refTyp{\theV = y}}{\tyTop}}}{}.
\label{eq:nontermination:start}
\end{equation}
\ruleName{CA-Valid} cannot derive this judgment, since the
implication
\[
  \embed{\Gamma} \wedge \formTrue \Rightarrow
  \synHastyp{y}{\utArrow{x}{\refTyp{\theV = y}}{\tyTop}}
\]
is not valid.
Thus, we must try to derive \autoref{eq:nontermination:start}
by \ruleName{CA-ImpSyn}.
Type extraction derives that $\synHastyp{y}{\varUnTyp}$ in $\Gamma$,
so the remaining obligation is
\[
  \relSynSub{\Gamma}{\varUnTyp}{\utArrow{x}{\refTyp{\theV = y}}{\tyTop}}{}.
\]
Because of the contravariance of function subtyping on the left-hand side
of the arrow, the following judgment must be derivable:
$$\relSub{\Gamma}
         {\refTyp{\theV = y}}
         {\refTyp{\synHastyp{\theV}{\utArrow{b}{\refTyp{\theV = y}}{\tyTop}}}}{}.$$
After \ruleName{SA-Mono} substitutes a fresh variable, say $\theV'$, for
$\theV$ in both types, this reduces to the clause implication obligation
$$\relImpl{\Gamma, \theV' = y}
          {\formTrue}
          {\synHastyp{\theV'}{\utArrow{b}{\refTyp{\theV' = y}}{\tyTop}}}
          {}.$$
Alas, this is essentially \autoref{eq:nontermination:start}, so we
are stuck in an infinite loop! We will again extract the type $\varUnTyp$
for $y$ (aliased to $\theV'$ here) and repeat the process \emph{ad inifinitum}.

This situation arises because we are allowed to invoke
the rule \ruleName{CA-ImpSyn} infinitely many times.
Then it must also be the case that \ruleName{CA-ImpSyn} extracts a single
type term from the environment infinitely often, since there are only
finitely many in the environment.
Thus, to ensure termination, we make the restriction that
along any branch of a subtyping derivation, a type term may be
extracted from the environment at most once.
This is the purpose of the set $\usedBoxes$ that is propagated through
subtyping judgments;
the $\extractName$ procedure excludes from consideration any type terms
in the set $\usedBoxes$ of already-used type terms.
Notice that in the \ruleName{CA-ImpSyn} rule, the results of the call to
$\extractName$ are included in the already-used set of the syntactic
subtyping judgment.


\subsection{Bidirectional Type Checking}

In this section, we define an algorithm for type checking programs where
type annotations for binding constructs and constructed data expressions
may or may not be provided.
Following work on local type inference \cite{pierce-turner},
our type checking algorithm is split into two mutually-dependent parts:
%
a \emph{type synthesis} relation $\relSynth{\Gamma}{e}{\varScm}$
that given an expression $e$, a type environment $\Gamma$, and no information
about the expected type of $e$ attempts to synthesize, or derive, a
well-formed type $\varScm$; and
%
a \emph{type conversion} relation $\relConvert{\Gamma}{e}{\varScm}$
that, in addition to $e$ and $\Gamma$, takes a type $\varScm$ that is
required of $e$, and checks whether or not $e$ can indeed
be given type $\varScm$.
Thus, $\varScm$ is an output of a synthesis judgment but an input to a
conversion judgment.
We will highlight some of the more interesting cases of type checking
relations after dealing with two issues.


\parahead{Inconsistent Type Environments}
Recall that the type extraction procedure collects the type terms
$\varUnTyp$ such that
$\valid{\embed{\Gamma,\tyBind{x}{\varTyp}}\Rightarrow
        \synHastyp{x}{\varUnTyp}}$.
If the environment $\Gamma,\tyBind{x}{\varTyp}$ happens to
be inconsistent, then all such implications will be valid.
As we will see, our typing rules for function application will depend
on type extraction returning exactly one syntactic arrow,
which will not be the case in an inconsistent environment.
This is a precision issue that we avoid by simply not performing
type extraction when in an inconsistent environment.
To this end, both the synthesis and conversion algorithms start off
by checking whether the environment is inconsistent, and if it
is, they trivially succeed.
\[
  \inferrule*[lab=\ruleNameFig{TS-False}]
    {\valid{\embed{\Gamma}\Rightarrow\formFalse}}
    {\relSynth{\Gamma}{e}{\refTypShort{\formFalse}}}
\hsepRule
  \inferrule*[lab=\ruleNameFig{TC-False}]
    {\valid{\embed{\Gamma}\Rightarrow\formFalse}}
    {\relConvert{\Gamma}{e}{\varScm}}
\]
These rules are sound because when the environment is inconsistent,
the underlying implications can be discharged by \ruleName{CA-Valid}
anyway.


\parahead{Unfolding}
Unlike \ruleName{T-Sub}, uses of
\ruleName{T-Unfold} cannot be factored into other typing rules, since
we cannot syntactically predict where it is needed.
It is not sufficient, for example, to unfold type definitions only
at uses of variables (that is, in the typing rule for variables).
To demonstrate, consider the function 
\begin{verbatim}
    let get_hd x = get x "hd"
\end{verbatim}
and an attempt to assign it the type
\[
  \tyDec{get\_hd}
        {\utArrow{\ttx}{\refTypShort{\theV\neq\vNull\wedge
                                     \synHastyp{\theV}{\tyList{\tyTop}}}}
                       {\refTypShort{\theV=\selR{\ttx}{\ttfld{hd}}}}}.
\]
Say we unfold the type $\tyList{\tyTop}$ at the use of $\ttx$,
when it is passed to the $\vGet$ function.
By the definition of $\unfold{\listCon}{\tyTop}$, we obtain
\[
  \ttx\neq\vNull \Rightarrow
  (\tagof{\ttx}=\tagRecd \wedge
  \hasR{\ttx}{\ttfld{hd}} \wedge
  \hasR{\ttx}{\ttfld{tl}})
\]
which, together with the assumption that $\ttx\neq\vNull$,
allows the call to $\vGet$ to typecheck.
Then, to check the subsequent call with argument $\ttfld{hd}$, we require
that $\hasR{\ttx}{\ttfld{hd}}$. The unfolded formula is sufficient
to prove this, but it is no longer in the environment
of logical assumptions, since it was not recorded in the type environment.

Languages like ML leverage pattern matching to determine exactly
when to unfold type definitions.
We do not have this option, however, since our core language does
not include a syntactic form for unpacking constructed data.
Instead, we eagerly try to unfold type definitions every time a
variable is added to the environment. 
We define a procedure $\extendGammaName$ that, in addition to
extending a type environment as usual,
uses type extraction to determine whether the variable has a
constructed type and, if it does, unfolds and records its
type definition.
\begin{align*}
  \extendGamma{\Gamma}{x}{\varTyp} =\ &
    \Gamma, \tyBind{x}{\varTyp},
      \wedge_{\typeCon{\varCon}{\seq{\varTyp'}}\in\usedBoxes}\ 
        \subst{\unfold{\varCon}{\seq{\varTyp'}}}{\theV}{x}
  \\[-2pt] 
  & \hspace{0.10in} \textrm{where }
      \usedBoxes = \extract{\Gamma}{\refTypShort{\theV=x}}{\emptyset}
  \\
  \extendGamma{\Gamma}{x}{\typAll{\tyvar}{\varScm}} =\ &
    \Gamma, \tyBind{x}{\typAll{\tyvar}{\varScm}}
\end{align*}
%


\parahead{Constants and Variables}
We now consider some of the algorithmic typing rules.
For non-function values, the synthesis rules are similar to the declarative
typing rules, whereas the conversion rules invoke synthesis and then call
into subtyping to check the synthesized type against the goal.

\[
  \inferrule*[lab=\ruleNameFig{TS-Const}]
    { }
    {\relSynth{\Gamma}{c}{\typConst{c}}{}}
\hsepRule
  \inferrule*[lab=\ruleNameFig{TC-Const}]
    {\relSynth{\Gamma}{c}{\varScm'}{} \sepPremise
     \relASub{\Gamma}{\varScm'}{\varScm}{\emptyset}}
    {\relConvert{\Gamma}{c}{\varScm}{}}
\]
\[
  \inferrule*[lab=\ruleNameFig{TS-Var}]
    {\Gamma(x)=\varTyp}
    {\relSynth{\Gamma}{x}{\refTypShort{\theV=x}}{}}
\hsepRule
  \inferrule*[lab=\ruleNameFig{TC-Var}]
    {\relSynth{\Gamma}{x}{\varTyp'}{} \sepPremise
     \relASub{\Gamma}{\varTyp'}{\varTyp}{\emptyset}}
    {\relConvert{\Gamma}{x}{\varTyp}{}}
\]


\parahead{Functions}
The synthesis rule for annotated functions is straightforward.
The best we can do when the function binder $x$ is not annotated is
try to typecheck the body assuming that $x$ has type $\tyTop$.

\[
  \inferrule*[right=\ruleNameFig{TS-FunAnn}]
    {\relWf{\Gamma}{\varTyp_1} \sepPremise
     \relSynth{\extendGamma{\Gamma}{x}{\varTyp_1}}{e}{\varTyp_2}}
    {\relSynth{\Gamma}
              {\funAnn{x}{\varTyp_1}{e}}
              {\refTypShort{\synHastyp{\theV}{\utArrow{x}{\varTyp_1}{\varTyp_2}}}}}
\] 

\[
  \inferrule*[right=\ruleNameFig{TS-FunBare}]
    {\relSynth{\extendGamma{\Gamma}{x}{\tyTop}}{e}{\varTyp_2}}
    {\relSynth{\Gamma}
              {\funBare{x}{e}}
              {\refTypShort{\synHastyp{\theV}{\utArrow{x}{\tyTop}{\varTyp_2}}}}}
\] 

\vsepRule
\noindent
When checking whether a function, annotated or not, can be converted to a
particular type $\varTyp$, we require that $\varTyp$ syntactically have the
form $\refTypShort{\synHastyp{\theV}{\varUnTyp}}$ where $\varUnTyp$ is
an arrow.
This seems to be a reasonable source-level requirement, but it could be
loosened if needed.

\[
  \inferrule*[right=\ruleNameFig{TC-FunAnn}]
    {\relWf{\Gamma}{\varTyp} \sepPremise
     \relASub{\Gamma}{\varTyp_1}{\varTyp}{\emptyset} \\\\
     \relSynth{\extendGamma{\Gamma}{x}{\varTyp_1}}{e}{\varTyp_2}}
    {\relConvert{\Gamma}
                {\funAnn{x}{\varTyp}{e}}
                {\refTypShort{\synHastyp{\theV}{\utArrow{x}{\varTyp_1}{\varTyp_2}}}}}
\]

\[
  \inferrule*[right=\ruleNameFig{TC-FunBare}]
    {\relSynth{\extendGamma{\Gamma}{x}{\varTyp_1}}{e}{\varTyp_2}}
    {\relConvert{\Gamma}
                {\funBare{x}{e}}
                {\refTypShort{\synHastyp{\theV}{\utArrow{x}{\varTyp_1}{\varTyp_2}}}}}
\]


\parahead{Function Applications}
The cases for application are the most unique to our setting.
To synthesize an application, we must be able to synthesize a type
$\varTyp_1$ for the function $\varVal_1$ and use type extraction to
convert $\varTyp_1$ to a syntactic arrow.
The procedure $\extractName$ can return an arbitrary number of
syntactic type terms, so we must decide how to proceed in the event
that $\varTyp_1$ can be extracted to multiple different arrow types.
To avoid the need for backtracking in the type checker, and to
provide a semantics
that is simple for the programmer to understand and use, we
consider an application $\varVal_1\ \varVal_2$ to be well-typed
if there is \emph{exactly one} syntactic arrow that is applicable
for the given argument $\varVal_2$.

Determining what is ``applicable'' separates into two cases.
In the case that we can synthesize a type $\varTyp_2$ for $\varVal_2$,
we use the following procedure that succeeds if there is
exactly one arrow in the set $\usedBoxes$ of type terms
with a domain that is a supertype of $\varTyp_2$.
\begin{align*}
&
  \filterAppOne{\Gamma}{\usedBoxes}{\varTyp_2} = \\
& \hspace{0.20in}
  \left\{
    \begin{array}{ll}
      \utArrow{x}{\varTyp_{11}}{\varTyp_{12}} & \textrm{if }
        \utArrow{x}{\varTyp_{11}}{\varTyp_{12}}\ \textrm {is the only }
          \varUnTyp\in\usedBoxes
      \\
      & \hspace{0.10in} \textrm{such that }
          \relASub{\Gamma}{\varTyp_2}{\varTyp_{11}}{\emptyset}
      \\
      \textit{fail} & \textrm{otherwise}
    \end{array}
  \right.
\end{align*}

The first synthesis rule for application uses this procedure to
derive an output type for the call. (We write parentheses
around the last premise, because it is not needed; it is implied
by the successful $\filterAppOneName$ call. We include the
premise in the rule for clarity.)
\[
  \inferrule*[right=\ruleNameFig{TS-App1}]
    {\relSynth{\Gamma}{\varVal_1}{\varTyp_1} \sepPremise
     \relSynth{\Gamma}{\varVal_2}{\varTyp_2} \\\\
     \usedBoxes = \extract{\Gamma}{\varTyp_1}{\emptyset} \\\\
     \utArrow{x}{\varTyp_{11}}{\varTyp_{12}} =
       \filterAppOne{\Gamma}{\usedBoxes}{\varTyp_2} \\\\
     (\relASub{\Gamma}{\varTyp_2}{\varTyp_{11}}{\emptyset})
    }
    {\relSynth{\Gamma}{\varVal_1\ \varVal_2}{\varTyp_{12}[\varVal_2/x]}}
\]

\noindent
In the case that we cannot synthesize a type for $\varVal_2$,
we use the following procedure that succeeds if there is
exactly one arrow in $\usedBoxes$ with a domain type that $\varVal_2$
can be converted to.
\begin{align*}
&
  \filterAppTwo{\Gamma}{\usedBoxes}{\varVal_2} = \\
& \hspace{0.20in}
  \left\{
    \begin{array}{ll}
      \utArrow{x}{\varTyp_{11}}{\varTyp_{12}} & \textrm{if }
        \utArrow{x}{\varTyp_{11}}{\varTyp_{12}}\ \textrm {is the only }
          \varUnTyp \in \usedBoxes
      \\
      & \hspace{0.10in} \textrm{such that }
          \relConvert{\Gamma}{\varVal_2}{\varTyp_{11}}
      \\
      \textit{fail} & \textrm{otherwise}
    \end{array}
  \right.
\end{align*}

\noindent
The second synthesis rule for application uses this procedure to
derive an output type for the call.
\[
  \inferrule*[right=\ruleNameFig{TS-App2}]
    {\relSynth{\Gamma}{\varVal_1}{\varTyp_1} \\\\
     \usedBoxes = \extract{\Gamma}{\varTyp_1}{\emptyset} \\\\
     \utArrow{x}{\varTyp_{11}}{\varTyp_{12}} =
       \filterAppTwo{\Gamma}{\usedBoxes}{\varVal_2} \\\\
     (\relConvert{\Gamma}{\varVal_2}{\varTyp_{11}})
    }
    {\relSynth{\Gamma}{\varVal_1\ \varVal_2}{\varTyp_{12}[\varVal_2/x]}}
\]

Type conversion for an application can proceed in two ways,
if either the type of the function or argument can be synthesized.
The first case, when the function type can be synthesized to an arrow,
is similar to \ruleName{TS-App2} with an additional subtyping check.

\[
  \inferrule*[right=\ruleNameFig{TC-App1}]
    {\relSynth{\Gamma}{\varVal_1}{\varTyp_1} \\\\
     \usedBoxes = \extract{\Gamma}{\varTyp_1}{\emptyset} \\\\
     \utArrow{x}{\varTyp_{11}}{\varTyp_{12}} =
       \filterAppTwo{\Gamma}{\usedBoxes}{\varVal_2} \\\\
     (\relConvert{\Gamma}{\varVal_2}{\varTyp_{11}}) \\\\
     \relASub{\Gamma}{\varTyp_{12}[\varVal_2/x]}{\varTyp}{\emptyset}
    }
    {\relConvert{\Gamma}{\varVal_1\ \varVal_2}{\varTyp}}
\]

\noindent
In the second case, when we can synthesize a type $\varTyp_2$ for the
argument, we combine $\varTyp_2$ with the goal $\varTyp$ to infer a
plausible arrow type for the function. Notice that we use a
dummy formal parameter $x$, since we have no (reasonable) way of computing
where $x$ might have appeared in $\varTyp$ before substituting
$\varVal_2$ for $x$.

\[
  \inferrule*[right=\ruleNameFig{TC-App2}]
    {\relSynth{\Gamma}{\varVal_2}{\varTyp_2} \sepPremise
     x \textrm{ fresh} \\\\
     \relConvert{\Gamma}
                {\varVal_1}
                {\refTyp{\synHastyp{\theV}{\utArrow{x}{\varTyp_2}{\varTyp}}}}
    }
    {\relConvert{\Gamma}{\varVal_1\ \varVal_2}{\varTyp}}
\]


\parahead{If-expressions}
We can synthesize a precise type for if-expressions by tracking the
guard predicates in the output type.
Type conversion for if-expressions is straightforward.

\[
  \inferrule*[right=\ruleNameFig{TS-If}]
    {\relConvert{\Gamma}{\varVal}{\tyBool} \\\\
     \relSynth{\Gamma,\varVal=\vTrue}{e_1}{\refTyp{\varFormOne_1}} \\\\
     \relSynth{\Gamma,\varVal=\vFalse}{e_2}{\refTyp{\varFormOne_2}} \\\\
     \varFormTwo \defeq (\varVal=\vTrue\Rightarrow\varFormOne_1 \wedge
                         \varVal=\vFalse\Rightarrow\varFormOne_2)
    }
    {\relSynth{\Gamma}{\ite{\varVal}{e_1}{e_2}}{\refTyp{\varFormTwo}}}
\]

\[
  \inferrule*[right=\ruleNameFig{TC-If}]
    {\relConvert{\Gamma}{\varVal}{\tyBool} \\\\
     \relConvert{\Gamma,\varVal=\vTrue}{e_1}{\varTyp} \\\\
     \relConvert{\Gamma,\varVal=\vFalse}{e_2}{\varTyp}
    }
    {\relConvert{\Gamma}{\ite{\varVal}{e_1}{e_2}}{\varTyp}}
\]


\parahead{Let-expressions}
The rules for let-expressions share a similar structure.
The choice whether to use synthesis or conversion on the equation
expression $e_1$ depends on whether there is an annotation $\varScm$ or not.
The choice for the body expression $e_2$ depends on the kind of derivation
for the overall let-expression.
Whenever a let-binding contains an annotation $\varScm$, we
must check that it is well-formed.
The synthesis rules for both kinds of let-bindings must also
check that the synthesized type $\varTyp$ is well-formed in
$\Gamma$, since we need to ensure that synthesized types are
always well-formed in their environment.

\[
  \inferrule*[lab=\ruleNameFig{TS-LetAnn-1}]
    {\relWf{\Gamma}{\varScm} \sepPremise
     \relConvert{\Gamma}{e_1}{\varScm} \sepPremise
     \relSynth{\extendGamma{\Gamma}{x}{\varScm}}{e_2}{\varTyp} \sepPremise
     \relWf{\Gamma}{\varTyp}}
    {\relSynth{\Gamma}{\letinAnn{x}{\varScm}{e_1}{e_2}}{\varTyp}}
\]
\[
  \inferrule*[lab=\ruleNameFig{TC-LetAnn}]
    {\relWf{\Gamma}{\varScm} \sepPremise
     \relConvert{\Gamma}{e_1}{\varScm} \sepPremise
     \relConvert{\extendGamma{\Gamma}{x}{\varScm}}{e_2}{\varTyp}}
    {\relConvert{\Gamma}{\letinAnn{x}{\varScm}{e_1}{e_2}}{\varTyp}}
\]
\[
  \inferrule*[lab=\ruleNameFig{TS-LetBare-1}]
    {\relSynth{\Gamma}{e_1}{\varScm} \sepPremise
     \relSynth{\extendGamma{\Gamma}{x}{\varScm}}{e_2}{\varTyp} \sepPremise
     \relWf{\Gamma}{\varTyp}}
    {\relSynth{\Gamma}{\letinBare{x}{e_1}{e_2}}{\varTyp}}
\]
\[
  \inferrule*[lab=\ruleNameFig{TC-LetBare}]
    {\relSynth{\Gamma}{e_1}{\varScm} \sepPremise
     \relConvert{\extendGamma{\Gamma}{x}{\varScm}}{e_2}{\varTyp}}
    {\relConvert{\Gamma}{\letinBare{x}{e_1}{e_2}}{\varTyp}}
\]

\vsepRule
Because the syntax of \dtypes is A-normal form,
programs will contain many let-expressions.
Ideally, our algorithmic type rules will deal well with
bare let-expressions well to avoid an overwhelming
and redundant annotation burden.
The \ruleName{TS-LetBare-1} rule does not, however, successfully
synthesize types in common situations where we would expect it to.
We will show three problematic examples and then incorporate
a simple technique that supports them.

First, consider the function
\begin{verbatim}
let get_f (x:{tag(v)="Dict" /\ has(v,"f")}) =
  get x "f"
\end{verbatim}
In A-normal form, this function might be written as

\begin{verbatim}
let get_f (x:{tag(v)="Dict" /\ has(v,"f")}) =
  let a = get x in
  let b = a "f" in
    b
\end{verbatim}

\noindent
Notice that the function binder is annotated but the let-binders are not.
It seems reasonable to expect that the annotation on $\ttx$ would be
sufficient for type synthesis to derive the type
\[
  \tyDec{\texttt{get\_f}}
        {\utArrow{\ttx}
          {\refTypShort{\tyRecd(\theV)\wedge\hasR{\theV}{\ttfld{f}}}}
          {\refTypShort{\selR{\ttx}{\ttfld{f}}}}}
\]
but it does not. Consider an attempt to apply \ruleName{TS-LetBare-1} for
the let-expression that binds $\ttb$. At that point, type synthesis
can derive the type $\varTyp=\refTypShort{\theV=\selR{\ttx}{\ttfld{f}}}$
for the equation expression \verb+a "f"+. Then, in the type environment
extended with \tyBind{\ttb}{\varTyp}, \ruleName{TS-Var} synthesizes
the singleton type \refTypShort{\theV=\ttb} for the body expression.
But this type is, of course, not-well formed in the type environment without
the binding for $\ttb$, so the \ruleName{TS-LetBare-1} rule fails.
This is quite unfortunate, since the \ruleName{TS-Var} rule will be
used extensively, and clearly there is a type that we could have
used instead of $\refTypShort{\theV=\ttb}$, namely, the type stored for
$\ttb$ in the environment!

As a second problematic situation, consider the following variation
of the previous example.

\begin{verbatim}
let maybe_get_f (x:Dict) =
  if mem x "f" then get x "f" else 0
\end{verbatim}
In A-normal form, this function might be written as

\begin{verbatim}
let maybe_get_f (x:Dict) =
  let a = mem x in
  let b = a "f" in
  if b then
    let c = get x in
    c "f"
  else
    0
\end{verbatim}

\noindent
Again, we have a problem applying the \ruleName{TC-LetBare-1} rule
to the let-expression that binds $\ttb$.
The type synthesized for the equation \verb+a "f"+ is
$\varTyp=\refTypShort{\tyBool(\theV)\wedge
                      (\theV=\vTrue\Leftrightarrow\hasR{\ttx}{\ttfld{f}})}$.
To synthesize the type of the body,
the culprit this time is the \ruleName{TS-If} rule, which derives the type
$\refTypShort{\ttb=\vTrue\Rightarrow\theV=\selR{x}{\ttfld{f}}\wedge
              \ttb=\vFalse\Rightarrow\theV=0}$
that refers to $\ttb$.
We observe that the type $\varTyp$ indicates that it is a boolean flag
that records the property $\hasR{\ttx}{\ttfld{f}}$,
so in this case, we would like to replace the problematic body type with
$\refTypShort{\hasR{\ttx}{\ttfld{f}}=\vTrue\Rightarrow\theV=\selR{x}{\ttfld{f}}\wedge
              \hasR{\ttx}{\ttfld{f}}=\vFalse\Rightarrow\theV=0}$.
Furthermore, we might expect to be able to play this trick quite often,
since the shape of $\varTyp$ --
$\refTypShort{\varFormOne=\vTrue\Rightarrow\theV=\selR{x}{\ttfld{f}}\wedge
              \varFormOne=\vFalse\Rightarrow\theV=0}$
for some formula $\varFormOne$ --
is the same as the return type of several common primitive functions,
including $\vHas$ and \texttt{=}.

The third and final problematic situation that we consider originates with
a small twist on the previous example.

\begin{verbatim}
let another_maybe_get_f (x:Dict) =
  let a = mem x in
  let b = a "f" in
  let b' = b in
  if b' then
    let c = get x in
    c "f"
  else
    0
\end{verbatim}

\noindent
This time, the boolean condition used in the if-expression goes
through one more level of indirection, namely, the variable $\texttt{b'}$.
Thus, when processing the $\texttt{b'}$ let-expression,
the type synthesized by \ruleName{TS-If} for the body expression is
$\refTypShort{\texttt{b'}=\vTrue\Rightarrow\theV=\selR{x}{\ttfld{f}}\wedge
              \texttt{b'}=\vFalse\Rightarrow\theV=0}$
The type for $\texttt{b'}$, which is $\refTypShort{\theV=\texttt{b}}$,
does not, however, match the special shape of boolean flags from before.
The trick we can play this time is to simply replace $\texttt{b'}$ with
$\ttb$, and derive
$\refTypShort{\texttt{b}=\vTrue\Rightarrow\theV=\selR{x}{\ttfld{f}}\wedge
              \texttt{b}=\vFalse\Rightarrow\theV=0}$
for the body expression.
This type is well-formed, and when considered as the body expression for
the enclosing let-expression that binds $\ttb$, will be further rewritten using the
technique for eliminating singletons to the type
$\refTypShort{\hasR{\ttx}{\ttfld{f}}=\vTrue\Rightarrow\theV=\selR{x}{\ttfld{f}}\wedge
              \hasR{\ttx}{\ttfld{f}}=\vFalse\Rightarrow\theV=0}$

We encapsulate these three simple heuristics in a procedure $\elimName$
and use it to define the following more precise synthesis rule for bare
let-bindings.

\[
  \inferrule*[right=\ruleNameFig{TS-LetBare-2}]
    {\relSynth{\Gamma}{e_1}{\varScm} \sepPremise
     \relSynth{\extendGamma{\Gamma}{x}{\varScm}}{e_2}{\varTyp} \\\\
     \varTyp'=\elim{\Gamma}{x}{\varScm}{\varTyp} \sepPremise
     (\relWf{\Gamma}{\varTyp'})
    }
    {\relSynth{\Gamma}{\letinBare{x}{e_1}{e_2}}{\varTyp'}}
\]

\noindent
The procedure $\elim{blah}{x}{\varScm}{\varTyp}$ procedure takes a variable
$x$ whose equation expression has been synthesized to type $\varScm$, and the
type $\varTyp$ for the body expression, and attempts to remove occurrences
of $x$.
When the procedure succeeds, the resulting type is guaranteed to be
well-formed in the environment without $x$.
It starts by processing the top-level refinement predicate.
\begin{align*}
  \elim{\Gamma}{x}{\varScm}{\refTyp{\varFormOne}} &=
    \refTyp{\elim{\Gamma}{x}{\varScm}{\varFormOne}}
\end{align*}

\noindent
The first non-trivial case is for equality predicates that
correspond to the singleton types synthesized by \ruleName{TS-Var}.
\begin{align*}
  \elim{\Gamma}{x}{\varScm}{\theV=x} &=
    \left\{
      \begin{array}{ll}
        \varFormOne    & \textrm{if } \varScm=\refTyp{\varFormOne} \\
        \mathit{fail}  & \textrm{otherwise}
      \end{array}
    \right.
\end{align*}
The other non-trivial case is for equality predicates
that equate variables with boolean values, as the \ruleName{TS-If} rule
does. The two cases correspond to whether $\varScm$ matches the canonical
shape of boolean flags or whether $\varScm$ is a singleton type.
\begin{align*}
& \elim{\Gamma}{x}{\varScm}{x=\vTrue} = \\
& \hspace{0.20in}
    \left\{
      \begin{array}{ll}
        \varFormOne   & \textrm{ if }
           \varScm=\refTypShort{\tyBool(\theV)\wedge
                                (\theV=\vTrue\Leftrightarrow\varFormOne}) \\
        y = \vTrue\ \ & \textrm{ if } \varScm=\refTypShort{\theV=y} \\
        \mathit{fail} & \textrm{ otherwise}
      \end{array}
    \right. \\
& \elim{\Gamma}{x}{\varScm}{x=\vFalse} = \\
& \hspace{0.20in}
    \left\{
      \begin{array}{ll}
        \neg\varFormOne  & \textrm{ if }
           \varScm=\refTypShort{\tyBool(\theV)\wedge
                                (\theV=\vTrue\Leftrightarrow\varFormOne}) \\
        y = \vFalse   & \textrm{ if } \varScm=\refTypShort{\theV=y} \\
        \mathit{fail} & \textrm{ otherwise}
      \end{array}
    \right.
\end{align*}
The rest of the cases recursively process the formula.
\begin{align*}
  \elim{\Gamma}{x}{\varScm}{\lprimF(\seq{\varLogVal})} &=
     \lprimF(\seq{\elim{\Gamma}{x}{\varScm}{\varLogVal}}) \\
  \elim{\Gamma}{x}{\varScm}{\synHastyp{\varLogVal}{\varUnTyp}} &=
     \synHastyp{\elim{\Gamma}{x}{\varScm}{\varLogVal}}
               {\elim{\Gamma}{x}{\varScm}{\varUnTyp}} \\
  \elim{\Gamma}{x}{\varScm}{\varFormOne\wedge\varFormTwo} &=
     \elim{\Gamma}{x}{\varScm}{\varFormOne} \wedge
     \elim{\Gamma}{x}{\varScm}{\varFormTwo}
\intertext{As one final heuristic, we attempt to rewrite occurrences of $x$
that do not appear in the two kinds of equality predicates that we have
built support for. The following is the non-trivial case for logical values
that replaces the variable $x$ when its type is a singleton.}
  \elim{\Gamma}{x}{\varScm}{x} &=
    \left\{
      \begin{array}{ll}
        y              & \textrm{if } \varScm=\refTypShort{\theV=y} \\
        \mathit{fail}  & \textrm{otherwise}
      \end{array}
    \right. \\
  \elim{\Gamma}{x}{\varScm}{y} &= y
\end{align*}

\noindent
If variable elimination fails, we can synthesize $\tyTop$ as a last resort.

\[
  \inferrule*[right=\ruleNameFig{TS-LetBare-3}]
    {\relSynth{\Gamma}{e_1}{\varScm} \sepPremise
     \relSynth{\extendGamma{\Gamma}{x}{\varScm}}{e_2}{\varTyp}
    }
    {\relSynth{\Gamma}{\letinBare{x}{e_1}{e_2}}{\tyTop}}
\]

\noindent
Since synthesis annotated let-expressions must also check
that the output type is well-formed, we define two additional rules
\ruleName{TS-LetAnn-2} and \ruleName{TS-LetAnn-3} that are analagous
to the conversion rules.


\parahead{Constructed Data}
We briefly discuss how we infer type parameters that are omitted in
constructed data expressions.
We extend the syntax of type definitions as follows.
For every type variable $\tyVar$ of a type definition for constructor
$\varCon$, we allow exactly one occurrence of $\tyVar$ to be \emph{marked},
written $\inferenceMarker\tyVar$, in the definition of $\varCon$.
When attempting to synthesize a type for unannotated constructed data, we use
the positions of marked type variables to match the corresponding positions in
the types of the value arguments that are used to construct the record.
For simplicity, we infer omitted type parameters for constructed data
only when \emph{all} type parameters are omitted. Therefore, we require that
either zero or all of the type parameters in a definition are marked.

For example, we update the $\listCon$ definition as follows
to use the type of the \ttfld{tl} field to infer the type parameter:
\[
  \typeConDefOneTwo
    {\listCon}
    {\poleCo\tyVar}
    {\ttfld{hd}}{\refTypShort{\synHastyp{\theV}{\tyVar}}}
    {\ttfld{tl}}{\refTypShort{\synHastyp{\theV}
                                        {\tyList{\inferenceMarker\tyVar}}}}
\]
Therefore, if the variable $\texttt{xs}$ has type $\tyList{\tyInt}$, then
$\newDataBare{\listCon}{1,\texttt{xs}}$
is well-typed; we infer the type argument $\tyInt$, which is
a supertype of $\refTypShort{\theV=1}$.
Notice that putting the marker for $\tyVar$ in the type of the \ttfld{hd}
field would lead to less successful inference, since the type of an element
added to a list will often be more specific than the type of the rest of the
list, and so the inferred type parameter would be too specific.
For example, $\newDataBare{\listCon}{1,\texttt{xs}}$ would \emph{not} be
well-typed, since the type $\refTypShort{\theV=\texttt{xs}}$
is a subtype of $\tyList{\tyInt}$,
but is not a subtype of $\tyList{\refTypShort{\theV=1}}$.


\parahead{Remaining Rules}
We omit the definition of the remaining synthesis and conversion rules
since they do not illuminate any new concerns.
Although the techniques that we have employed so far
would allow us to, we do not synthesize type instantiations.


\subsection{Soundness}

We now consider how derivations in the algorithmic type system relate to
derivations in the declarative type system.
We use a procedure $\eraseName$ to remove type annotations from functions,
let-bindings, and constructed data because the syntax of the declarative
system does not permit them.

\begin{prop}[Sound Algorithmic Typing]
\mbox{}
\begin{enumerate}
\alignItemTwo{1.10in}
  {If   $\relAImpl{\Gamma}{\varFormOne}{\varFormTwo}{\usedBoxes}$,}
  {then $\relImpl{\Gamma}{\varFormOne}{\varFormTwo}{}$.}
\alignItemTwo{1.10in}
  {If   $\relASynSub{\Gamma}{\varUnTyp_1}{\varUnTyp_2}{\usedBoxes}$,}
  {then $\relSynSub{\Gamma}{\varUnTyp_1}{\varUnTyp_2}{}$.}
\alignItemTwo{1.10in}
  {If   $\relASub{\Gamma}{\varScm_1}{\varScm_2}{\usedBoxes}$,}
  {then $\relSub{\Gamma}{\varScm_1}{\varScm_2}{}$.}
\alignItemTwo{1.10in}
  {If   $\relSynth{\Gamma}{e}{\varScm}$,}
  {then $\relTyp{\Gamma}{\erase{e}}{\varScm}{}$.}
\alignItemTwo{1.10in}
  {If   $\relConvert{\Gamma}{e}{\varScm}$,}
  {then $\relTyp{\Gamma}{\erase{e}}{\varScm}{}$.}
\end{enumerate}
\end{prop}

\noindent
\textit{Proof sketch.} 
We consider the key aspects of the development of the algorithmic type
system and provide an intuition for why they are sound.
To prove that algorithmic clause implication is sound with respect
to declarative clause implication, we must consider \ruleName{CA-ImpSyn}
and its use of the type extraction procedure. It is easy to see
that uses of $\extractName$ can be converted into derivations by
$\ruleName{C-Valid}$, since it depends on the validity of logical
implications.
Proving that algorithmic subtyping and syntactic subtyping are sound
with respect to their declarative counterparts goes by induction on
their derivation rules, which correspond one-to-one.

To prove that type synthesis and type conversion are sound with respect
to declarative typing, there are a few points to consider.
%
The first is the initial check for an inconsistent type environment that
\ruleName{TS-False} and \ruleName{TC-False} perform. It is simple to
show that in the declarative system any judgment is derivable when the
type environment is inconsistent. The proof is a straightforward induction,
using the \ruleName{C-Valid} rule to check that an inconsistent environment
means all clause implications can be proven valid.
Second, we can show that the $\extendGammaName$ procedure, which uses
type extraction to unfold type definitions, can be replaced with uses of
\ruleName{T-Unfold}.
Third, we can show that in the \ruleName{TC-LetBare-2} rule, when
$\elim{blah}{x}{\varScm}{\varTyp}$ successfully returns a type $\varTyp'$,
it is well-formed in $\Gamma$ and, furthermore, since the heuristics
employed soundly replace equality predicates,
$\relTyp{\Gamma}{e_2}{\varTyp'}{}$.
Finally, we can show that the subtyping premises used in the algorithmic
rules can be replaced with uses of \ruleName{T-Sub}.
%


\subsection{Implementation}

We have implemented a prototype checker for \dtypes in approximately 2,000
lines of OCaml, using Z3 \cite{z3} to discharge SMT queries.
A noteworthy, but unsurprising, optimization in our implementation
compared to the algorithmic system presented here is that the environment of
logical assumptions is maintained incrementally. We add and remove assertions
to and from the logical environment whenever the type system manipulates
the type environment, so that by the time the \ruleName{CA-Valid} rules
needs to check $\valid{\embed{\Gamma}\wedge\varFormOne\Rightarrow\varFormTwo}$, 
the formula $\embed{\Gamma}$ is already in the background assumptions of
the environment; only $\valid{\varFormOne\Rightarrow\varFormTwo}$ needs to be
discharged.
%



  \clearpage

\section{Examples}
\label{sec:examples}

In this section, we present the original, unadapted source code corresponding
to the noted examples in \ref{sec:intro} and \ref{sec:overview}.

\subsection{Introduction}

The introduction references the following function from the
Dojo Javascript library, version 1.6.1 \cite{dojo-js}:

\begin{lstlisting}[firstnumber=193, title="\_base/\_loader/loader.js", language=Caml, morekeywords={typeof}]
d._onto = function(arr, obj, fn){
  if(!fn){
   arr.push(obj);
 }else if(fn){
   var func = (typeof fn == "string") ? obj[fn] : fn;
   arr.push(function(){ func.call(obj); });
 }
}
\end{lstlisting}

\subsection{Overview}

The \verb+toXML+ example is adapted from the Python 3.2 standard
library:

\begin{lstlisting}[title=Lib/plistlib.py, firstnumber=111]
class DumbXMLWriter:
    def __init__(self, file, indentLevel=0, indent="\t"):
        self.file = file
        self.stack = []
        self.indentLevel = indentLevel
        self.indent = indent

    def beginElement(self, element):
        self.stack.append(element)
        self.writeln("<%s>" % element)
        self.indentLevel += 1

    def endElement(self, element):
        assert self.indentLevel > 0
        assert self.stack.pop() == element
        self.indentLevel -= 1
        self.writeln("</%s>" % element)

    def simpleElement(self, element, value=None):
        if value is not None:
            value = _escape(value)
            self.writeln("<%s>%s</%s>" % (element, value, element))
        else:
            self.writeln("<%s/>" % element)

    def writeln(self, line):
        if line:
            # plist has fixed encoding of utf-8
            if isinstance(line, str):
                line = line.encode('utf-8')
            self.file.write(self.indentLevel * self.indent)
            self.file.write(line)
        self.file.write(b'\n')

class PlistWriter(DumbXMLWriter):

    def __init__(self, file, indentLevel=0, indent=b"\t", writeHeader=1):
        if writeHeader:
            file.write(PLISTHEADER)
        DumbXMLWriter.__init__(self, file, indentLevel, indent)

    def writeValue(self, value):
        if isinstance(value, str):
            self.simpleElement("string", value)
        elif isinstance(value, bool):
            # must switch for bool before int, as bool is a
            # subclass of int...
            if value:
                self.simpleElement("true")
            else:
                self.simpleElement("false")
        elif isinstance(value, int):
            self.simpleElement("integer", "%d" % value)
        elif isinstance(value, float):
            self.simpleElement("real", repr(value))
        elif isinstance(value, dict):
            self.writeDict(value)
        elif isinstance(value, Data):
            self.writeData(value)
        elif isinstance(value, datetime.datetime):
            self.simpleElement("date", _dateToString(value))
        elif isinstance(value, (tuple, list)):
            self.writeArray(value)
        else:
            raise TypeError("unsupported type: %s" % type(value))

    def writeData(self, data):
        self.beginElement("data")
        self.indentLevel -= 1
        maxlinelength = 76 - len(self.indent.replace(b"\t", b" " * 8) *
                                 self.indentLevel)
        for line in data.asBase64(maxlinelength).split(b"\n"):
            if line:
                self.writeln(line)
        self.indentLevel += 1
        self.endElement("data")

    def writeDict(self, d):
        self.beginElement("dict")
        items = sorted(d.items())
        for key, value in items:
            if not isinstance(key, str):
                raise TypeError("keys must be strings")
            self.simpleElement("key", key)
            self.writeValue(value)
        self.endElement("dict")

    def writeArray(self, array):
        self.beginElement("array")
        for value in array:
            self.writeValue(value)
        self.endElement("array")
\end{lstlisting}


}
\fi

\end{document}